\documentclass{aa}
\usepackage{graphicx}
\usepackage{xspace}
\usepackage{txfonts}
\usepackage{hyperref}
\hypersetup{
    colorlinks=true,
    linkcolor=red,
    urlcolor=magenta,
}

\def\deg{\ensuremath{^\circ}}

\providecommand{\mas}{\ensuremath{\textrm{mas}}}

\providecommand{\deg}{\ensuremath{^\circ}}
\providecommand{\gmag}{\ensuremath{G}}

\newcommand\gdr[1]{\gaia~DR#1}

\newcommand{\sersic}{S\'ersic\xspace}
\newcommand{\gaia}{\textit{Gaia}\xspace}

\begin{document}

\title{\gaia Data Release 3: Surface brightness profiles of galaxies and host galaxies of quasars. \thanks{This paper is dedicated to the memory of Dimitri Pourbaix who supported the CU4-EO group over the years and with whom we shared very nice moments.}}

\author{
C. Ducourant\inst{1}\thanks{Corresponding author: C. Ducourant, e-mail: christine.ducourant@u-bordeaux.fr},
A. Krone-Martins\inst{2,3},
L. Galluccio\inst{4},
R. Teixeira\inst{5},
J.-F. Le Campion\inst{1},\\
E. Slezak\inst{4},
R. de Souza\inst{5},
P. Gavras\inst{6},
F. Mignard\inst{4},
J. Guiraud\inst{7}, 
W. Roux\inst{7},\\ 
S. Managau\inst{9}, 
D. Semeux\inst{10}, 
A. Blazere\inst{10}, 
A. Helmer\inst{9},
and D. Pourbaix\inst{8}}
  \institute{
        Laboratoire d'Astrophysique de Bordeaux, Univ. Bordeaux, CNRS, B18N, all{\'e}e Geoffroy Saint-Hilaire, 33615 Pessac, France
\and
                 CENTRA, Faculdade de Ci\^encias, Universidade de Lisboa, Ed. C8, Campo Grande, 1749-016 Lisboa, Portugal
\and
         Donald Bren School of Information and Computer Sciences, University of California, Irvine, CA 92697, USA
\and         
        Université Côte d’Azur, Observatoire de la Côte d’Azur, CNRS, Laboratoire Lagrange, Bd de l’Observatoire, CS 34229, 06304 Nice Cedex 4, France.
\and
         Instituto de Astronomia, Geof\'isica e Ci\^encias Atmosf\'ericas, Universidade de S\~{a}o Paulo, Rua do Mat\~{a}o, 1226, Cidade Universit\'aria, 05508-          900 S\~{a}o Paulo, SP, Brazil
\and 
          RHEA for European Space Agency (ESA), Camino bajo del Castillo,
s/n, Urbanizacion Villafranca del Castillo, Villanueva de la Cañada,
28692 Madrid, Spain
\and CNES Centre Spatial de Toulouse, 18 avenue Edouard Belin, 31401
Toulouse Cedex 9, France
\and FNRS, Institut d’Astronomie et d’Astrophysique, Université Libre de Bruxelles, boulevard du Triomphe, 1050 Bruxelles, Belgium
\and Thales Services for CNES Centre Spatial de Toulouse, 18 avenue Edouard Belin, 31401 Toulouse Cedex
9, France
\and ATOS for CNES Centre Spatial de Toulouse, 18 avenue Edouard Belin, 31401 Toulouse Cedex 9, France
}

\date{Received 15 April 2022; accepted 12 May 2022}
\abstract
   {Since July 2014, the \gaia space mission has been continuously scanning the sky and observing the extragalactic Universe with unprecedented spatial resolution in the optical domain ($\sim$ 180 mas by the end of the mission). \gaia provides an opportunity to study the morphology of the galaxies of the local Universe (z<0.45) with much higher resolution than has ever been attained from the ground. It also allows us to provide the first morphological all-sky space catalogue of nearby galaxies and galaxies that host quasars in the visible spectrum.}
   {We present the Data Processing and Analysis Consortium CU4-Surface Brightness Profile fitting pipeline, which aims to recover the light profile of nearby galaxies and galaxies hosting quasars. } 
   {The pipeline uses a direct model based on the Radon transform to measure the two-dimensional surface brightness profile of the extended sources. It simulates a large set of 2D light profiles and iteratively looks for the one that best reproduces the 1D observations by means of a Bayesian exploration of the parameters space. We also present our method for setting up the input lists of galaxies and quasars to be processed.}
   {We successfully analysed 1\,103\,691 known quasars and detected a host galaxy around 64\,498 of them ($\sim$6\%). We publish the surface brightness profiles of the host for a subset of 15\,867 quasars with robust solutions. The distribution of the S\'ersic index describing the light profile of the host galaxies peaks at $\sim$ 0.8 with a mean value of $\sim$ 1.9, indicating that these galaxies hosting a quasar are consistent with disc-like galaxies. The pipeline also analysed 940\,887 galaxies with both a \sersic and a de Vaucouleurs profile and derived robust solutions for 914\,837 of them. The distribution of the S\'ersic indices confirms that \gaia mostly detects elliptical galaxies and that very few discs are measured.}
{}

\titlerunning{Surface brightness profiles of extragalactic sources in  {\it Gaia} DR3}
\authorrunning{C. Ducourant et al.}
\maketitle

\section{Introduction}\label{intro}
The \gaia space mission, \cite{2016Prusti} is one of the most ambitious and spectacular projects in astronomy from the last 20 years. The observations carried out by this satellite have radically transformed the base of knowledge upon which the astronomy community relies to explore and understand the Milky Way as well as the extragalactic Universe. The primary target of \gaia is the stellar content of the Milky Way but the satellite also observes extragalactic sources like quasars and galaxies. In this way, \gaia provides the first opportunity to analyse extragalactic sources from space, with the instrument reaching an exquisite spatial resolution of $\sim$180 $\mas$ (by the end of the mission), and to produce the first all-sky and space-based catalogues of extragalactic objects and of their properties in optical wavelengths. \gaia  allows us to gain insight into a population of the local Universe that is mostly unresolved by ground-based facilities and has not been studied on such scale.\\

As explained in \cite{2016Prusti}, all sources brighter than the detector sensitivity and with a 2D surface brightness profile complying with the rules of the sophisticated onboard video processing algorithm (VPA) are accepted and observed. The VPA has a strong impact on the type of sources observed, as it filters out most disc-like galaxies \citep{2014deSouza, 2015deBruijne}. Observation windows are transmitted to the data processing centres. These windows are one dimensional for most astrometric fields (AFs 1-9) and two dimensional for the Sky Mapper (SM). In order to derive extremely accurate astrometric measurements of the objects, \gaia repeatedly scans the full sky. The scanning law of the satellite determines the  frequency at which a given object is observed. By the end of the mission,  approximately
$140$ observations should have been made for   each source, with various transit angles on the sky (as illustrated in Figure \ref{cov}) so that a large fraction of the selected extragalactic extended objects are completely covered by different observations at different transit angles. From these, it is possible to extract information about their morphology by running the CU4-Surface Brightness pipeline which is designed to reproduce the observed data by means of a direct model. 

Quasars are key objects for the \gaia astrometric mission. One of the main scientific goals of Gaia is to provide the first ever realisation of a rotation-free celestial reference frame (CRF) at sub-mas level in the visible wavelength domain, matching the International Celestial Reference System (ICRS) specifications. The axes of the resulting Gaia inertial optical frame have been aligned on the third realisation of ICRS (ICRF3) based on ~4300 radio-loud quasars observed by the very long baseline interferometry (VLBI). Among those objects in common between the ICRF3 and Gaia-CRF2, some tens of sources exhibit large (up to 10 mas) angular positional differences which could be real offsets between the centres of emission at optical and radio wavelengths. These spatial offsets could be linked to various effects: active galactic nucleus (AGN) activity triggering star formation, dual AGN, or recoiling super massive black holes; these offsets are the subject of an active field of research \citep[e.g.][]{2018Skipper, 2019Suh}.

Galaxies are not the primary targets of the Gaia mission but the compact and bulge-dominated ones are quite easily detected by Gaia, which enables us to obtain valuable information on the morphological characteristics of this population, provided a dedicated processing exists. Galaxy morphology is a fundamental tracer in observational cosmology. Indeed, it provides clues as to how galaxies form and evolve over the Hubble time by a combination of minor and major mergers, interaction with the neighbourhood, gas accretion, and secular evolution, thereby allowing a better understanding of the relations between this morphology, mass assembly, and star formation. Having a large sample of galaxies with a shape classification is therefore necessary in order to address, for example, the issue of the formation history of the Hubble sequence, to discriminate among the inside-out and outside-in scenarios \citep{2013Perez}, to study the relative role of (major) mergers and AGN feedback in quenching star formation, to measure the time-delay between this quenching and the colour and morphological transformations, and so on. These topics motivate our methodological efforts on the subject.

The \gaia Data Release 3 \citep{2022Brown} is based on data collected during the first 34 months of the nominal mission (between 25 July 2014 and 28 May 2017) and provides an astrometric and photometric catalogue for more than 1.8 billion sources with an apparent G magnitude down to 21 mag. It also provides, for the first time, a comprehensive set of results for the extragalactic sources observed by the satellite. Several coordination units
(CU3, CU4, CU7, and CU8) within the data processing and analysis consortium (DPAC) have developed specific pipelines to analyse and classify these sources. The results of these CUs (surface brightness profiles, redshifts, variability, \gaia-CRF3, and classifications) are gathered in two separate tables, namely \href{https://gea.esac.esa.int/archive/documentation/GDR3/Gaia_archive/chap_datamodel/sec_dm_extra--galactic_tables/ssec_dm_qso_candidates.html}{\textbf{qso\_candidates}} and \href{https://gea.esac.esa.int/archive/documentation/GDR3/Gaia_archive/chap_datamodel/sec_dm_extra--galactic_tables/ssec_dm_galaxy_candidates.html}{\textbf{galaxy\_candidates,}} which are provided alongside \gaia DR3. An overview of these tables and their main properties is presented by \cite{2022Bailer-Jones}.

The \gaia CU4-Surface brightness profile fitting includes several modules dedicated to different tasks: preparing, transforming and organising the observations; performing the fitting; filtering out the solutions according to their reliability; and attributing quality flags to the sources. This article presents the method used to fit the surface brightness profile of the extragalactic sources detected by \gaia and presents the results with relevant indications for their use. No scientific exploitation of these results is made in this paper. We describe the data processing behind the surface brightness profiles of extragalactic sources. Section~\ref{lists} presents the construction of the lists of quasars and galaxies that we processed and Section~\ref{filtering} presents the filtering applied to these lists. An overview of the pipeline is given in Section~\ref{pipeline}. The specificity of the \gaia data is presented in Section~\ref{data} together with a description of how extended sources are seen by the satellite. The algorithms used to derive the surface brightness profiles are presented in Section~\ref{fitting}. A post-processing step to eliminate and flag non-robust solutions has been applied to the results and is presented in Section~\ref{postproc}. Results and their validation are presented in Section~\ref{results} and the data product is given in Section \ref{catalogue}. Finally, we summarise our findings and briefly describe our plans for future improvement of the methods in Section~\ref{conclusion}.

\section{Input lists of quasars and galaxies}\label{lists}
In this section, we summarise the creation of the input lists of objects that were processed by our pipeline. \gaia DR3 will release a probabilistic classification of the sources into five classes: star, galaxy, quasar, binary star, or white dwarf \citep[see][]{CreeveyDR3-DPACP-157}. Nevertheless, due to the complex and long processing plan within the \gaia DPAC, we were not able to benefit from this classification before publication of the present paper. Indeed, we had to set up input lists of sources using surveys and literature studies as of early 2018.
The list of quasars was set up by compiling major AGN and quasars catalogues. The list of galaxies was established from a previous paper \citep{2022Krone-Martins} classifying sources from the \gaia DR2 catalogue with entry in the allWISE catalogue \citep{2013Cutri} into point-like or extended sources. Due to long processing cycles within DPAC, these input lists had to be delivered by early 2018, preventing the inclusion of more recent catalogues in our compilation of quasars.

\subsection{Input list of quasars}\label{qso_input}
We set up the input list of quasars by merging the major catalogues of candidate quasars and candidate AGN published before 2018. We considered the following catalogues: AllWISE R90 \citep{2018Assef}, HMQ (Half Million Quasars catalogue) \citep{2015Flesch}, AllWISE \citep{2015Secrest}, LQAC3 \citep{2015Souchay}, SDSS-DR12Q \citep{2017Paris}, and the ICRF2 \citep{2009Ma}. Most of these catalogues include stellar contaminants, except for ICRF2, which provides spectroscopically confirmed quasars. A selection of unpublished classifications of \gaia DR2 quasars based on photometric variability (designated CU7  hereafter) shared within the \gaia DPAC was also appended to the compiled list \citep{2019Rimoldini}.

In order to cross-match the seven catalogues, we first estimated their astrometric precision by cross-matching them with the \gdr{2} using a search radius of 1 $\arcsec$. The mean distance between the catalogue positions and the \gaia positions for the matched sources is adopted as an estimation of the astrometric precision of the catalogues, assuming that \gaia is error-free. 

The catalogues were then cross-matched one with another using a search radius of three times the precision before merging. This compilation contains $6\,166\,355$ sources of which $1\,996\,597$ have a match in \gdr{2}. 

This list was cleaned of stellar contaminants by applying an astrometric filter that rejects sources with a two-parameter solution in \gdr{2,} a parallax of $\varpi$ $\ge$ 7 mas, or total proper motion of $\lvert\mu\rvert\ge$7 mas/yr. This filter was derived from the \gdr2 astrometric properties of multiply imaged quasars by gravitational lensing \citep[][Fig. 2]{2018ducourant}. No additional constraint on the galactic latitude has been added because several ICRF2  validated quasars are located at very low Galactic latitude. This filter is intentionally not severe because most quasars are in the faint luminosity regime of \gaia (see Figure~\ref{G}) where the astrometry is less accurate and the potential presence of a surrounding host galaxy can perturb the astrometry of the central nucleus. 

The final list of quasars encompasses $1\,392\,788$ sources with an entry in \gdr{3}. In this list, more than 1 million sources have an entry in at least two catalogues (see Table~\ref{count_qso}). We keep track of the catalogue where the sources are identified in the \href{https://gea.esac.esa.int/archive/documentation/GDR3/Gaia_archive/chap_datamodel/sec_dm_extra--galactic_tables/ssec_dm_qso_catalogue_name.html}{\textbf{qso\_catalogue\_name}} table that is provided alongside \gdr{3}.

\begin{table}[h]
\centering
\caption{Number of sources from the input list of quasars that are present in the various original catalogues.}
\label{count_qso}
\begin{tabular}{lr}
\hline
Catalogue & Nb of sources \\
\hline
\hline
HMQ                     & 946\,984 \\
AllWISE Assef+2018 R90  & 893\,622 \\
CU7 GaiaVariQso2018     & 645\,175 \\
AllWISE Secrest+2015    & 550\,905 \\
LQAC3                   & 191\,987 \\
SDSS-DR12Q              & 144\,531 \\
ICRF2                   &   2\,186 \\
\hline
Input quasar list & $1\,392\,788$ \\
\hline
\end{tabular}
\end{table}

\begin{figure*}
    \centering
    \includegraphics[width=1\textwidth]{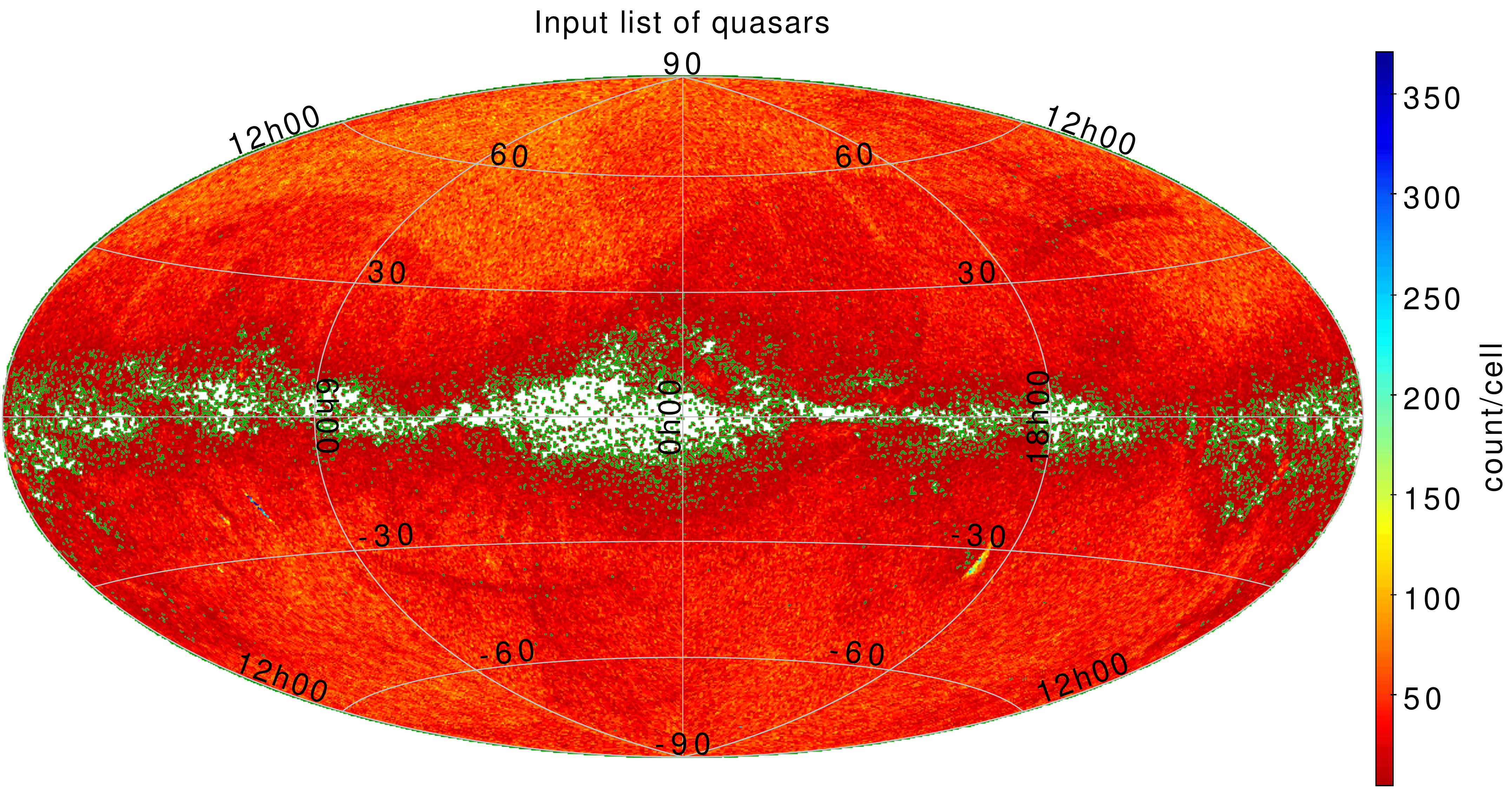} 
    \caption{Sky distribution in galactic coordinates of the quasars included in the input list. The cell of this map is approximately 0.2 deg$^2$, and the colour indicates the number of sources in each cell.}
    \label{qso_lb}
\end{figure*}

The sky distribution of the quasars from the compiled list is given in Figure~\ref{qso_lb} in galactic coordinates. The sky coverage of each of the merged catalogues is heterogeneous, as is the resulting input list. One notices a light over-density in the region of the Large Magellanic clouds (LMC: l = 280.4652$\deg$, b = $-32.8884\deg$), which corresponds to contamination by stars that were not filtered out by the astrometric filter described above. Similarly, some of the sources lying near the Galactic plane are probably stellar contaminants. The yellow overdensity in the Northern hemisphere corresponds to the coverage of the SDSS DR12Q catalogue.

\subsection{Input list of galaxies}
The list of galaxies analysed by our pipeline was established by \cite{2022Krone-Martins}. This catalogue was compiled using a fully unsupervised method based on the use of a stochastic iterative scheme specifically tailored to the star--galaxy separation problem. It sets up a catalogue of extended extragalactic sources. This approach first relies on a random sampling of the data points then on a random selection of a dimension in the analysed data space at each iteration. This method applies a hierarchical density-based clustering method \citep[HDBSCAN][]{Campello10.1145/2733381} and an automatically optimised supervised method \citep[a Radial Basis Support Vector Machine] []{andrew_2000} trained on the initial unsupervised solution obtained at each iteration. 
The method analyses \gdr{2} combined with the AllWISE survey \citep{2013Cutri}. 

\begin{figure*}
    \centering
    \includegraphics[width=1\textwidth]{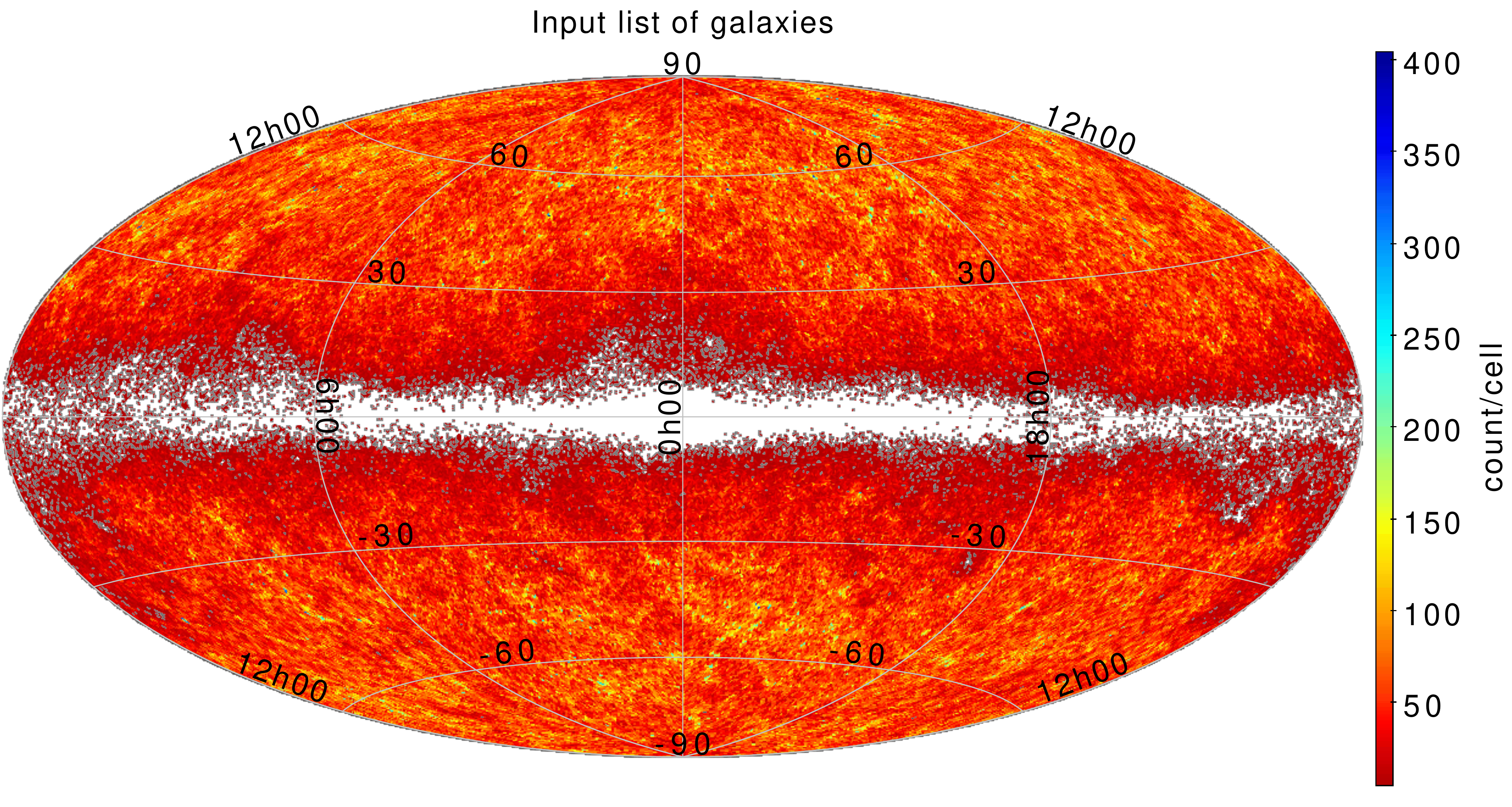} 
    \caption{Sky distribution in galactic coordinates of the galaxies included in the input list. The cell of this map is approximately 0.2 deg$^2$, and the colour indicates the number of sources in each cell.}
    \label{gal_lb}
\end{figure*}

The resulting list of extended sources contains $1\,742\,933$ galaxy candidates with an entry in \gdr{3}. The distribution of these sources on the sky is given in Figure~\ref{gal_lb} where one can observe a homogeneous spread of the sources, except in the Galactic plane which appears mostly empty. 

\section{Filtering}\label{filtering}
\subsection{On-board filtering of extragalactic sources}
While \gaia is scanning the entire sky, all the sources observed are not sent to Earth because the flow of data would exceed the capabilities of the telemetry. The VPA \citep[]{2016Prusti} is implemented on board to select the observations to be transferred. A windowing scheme is used that selects part of the CCD detectors centred on the source (hereafter windows of observation). The VPA takes the decision of whether or not to send a window in order to filter out a large number of contaminants (e.g. cosmics) while preserving as many real sources as possible. Schematically, the decision is based on the shape of the central light profile of the source as seen by the  SM and the astrometric field 1 (AF1) detector . If the profile is steep enough to be similar to a star-like source then the observation is transmitted, but if the profile is too flat then it is rejected. This selection function of the VPA was analysed by \cite{2014deSouza, 2015deBruijne}. These authors showed that the majority of disc-like galaxies are rejected by the VPA except when they encompass a bright bulge, while most elliptical galaxies are detected in the limit of the sensitivity of the detectors.

This VPA filtering mostly affects the galaxies to be analysed and we therefore produce an incomplete and Hubble-type biased catalogue of galaxies. 

\subsection{Filtering on angular coverage}
\begin{figure}[h]
    \includegraphics[width=0.48\textwidth]{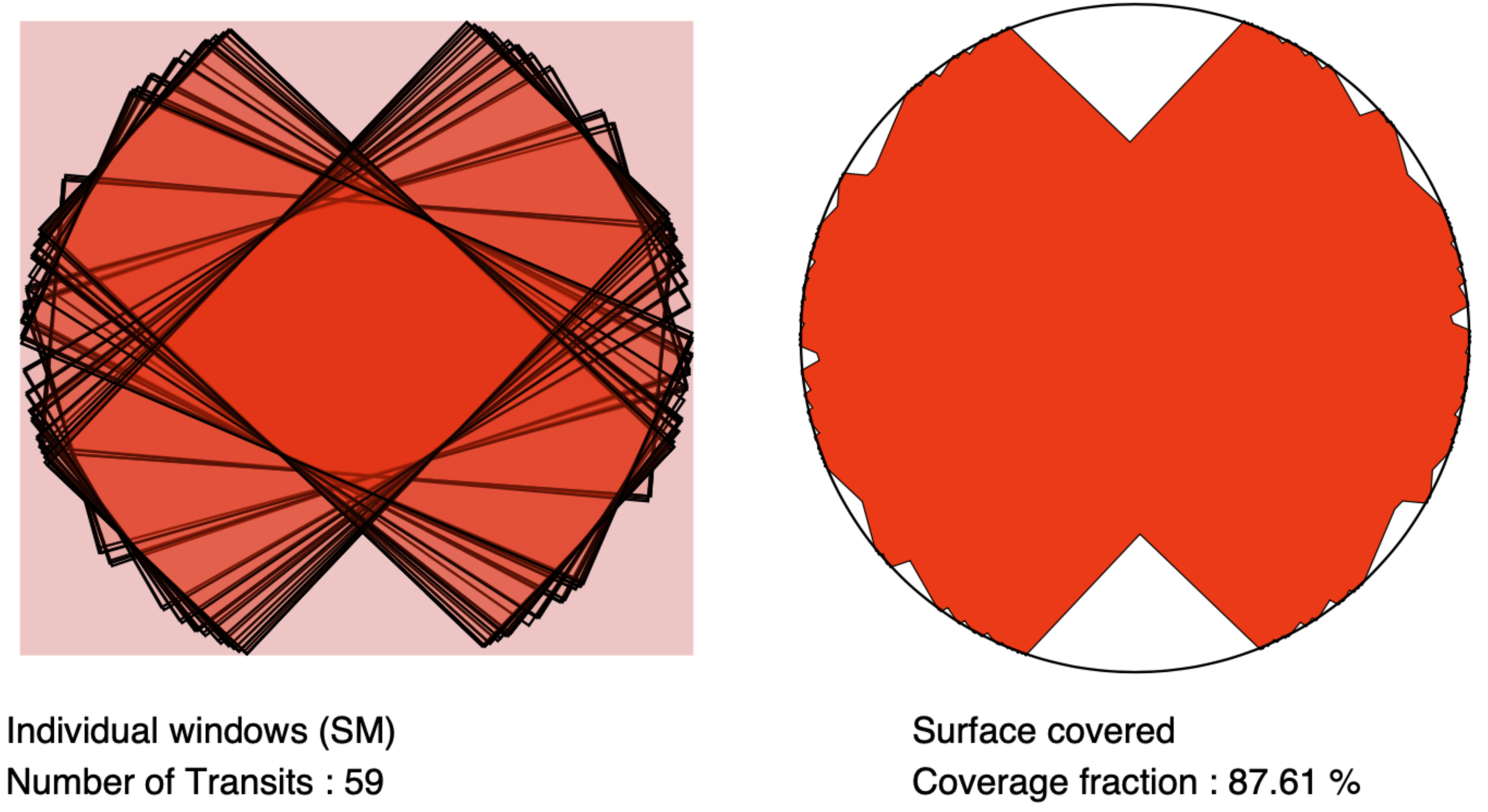} 
    \caption{Typical angular coverage by the SM windows of a source located at the Galactic coordinates (l,b)=($0\deg$,0$\deg$) with 59 transits and a resulting 87.61\% angular coverage.}
    \label{cov}
\end{figure}

The sources are repetitively scanned along the mission through various transit angles that are determined by the nominal scanning law of \gaia \citep{2016Prusti}. To recover the morphology of the extended sources, one must have a sufficient number of transits whose angles are uniformly spread over the source as illustrated in Figure~\ref{cov}. 

The \textit{angular coverage} of a source is estimated as the ratio between the area of the polygon created from the union of all the observed windows and transit angles over the area of a circle with a diameter equal to the diagonal of the largest observed window. The coverage corresponds to the fraction of the surface of the source covered by the observations \citep[see][for more details]{2011Krone-Martins}.

\begin{figure}
    \includegraphics[width=0.44\textwidth]{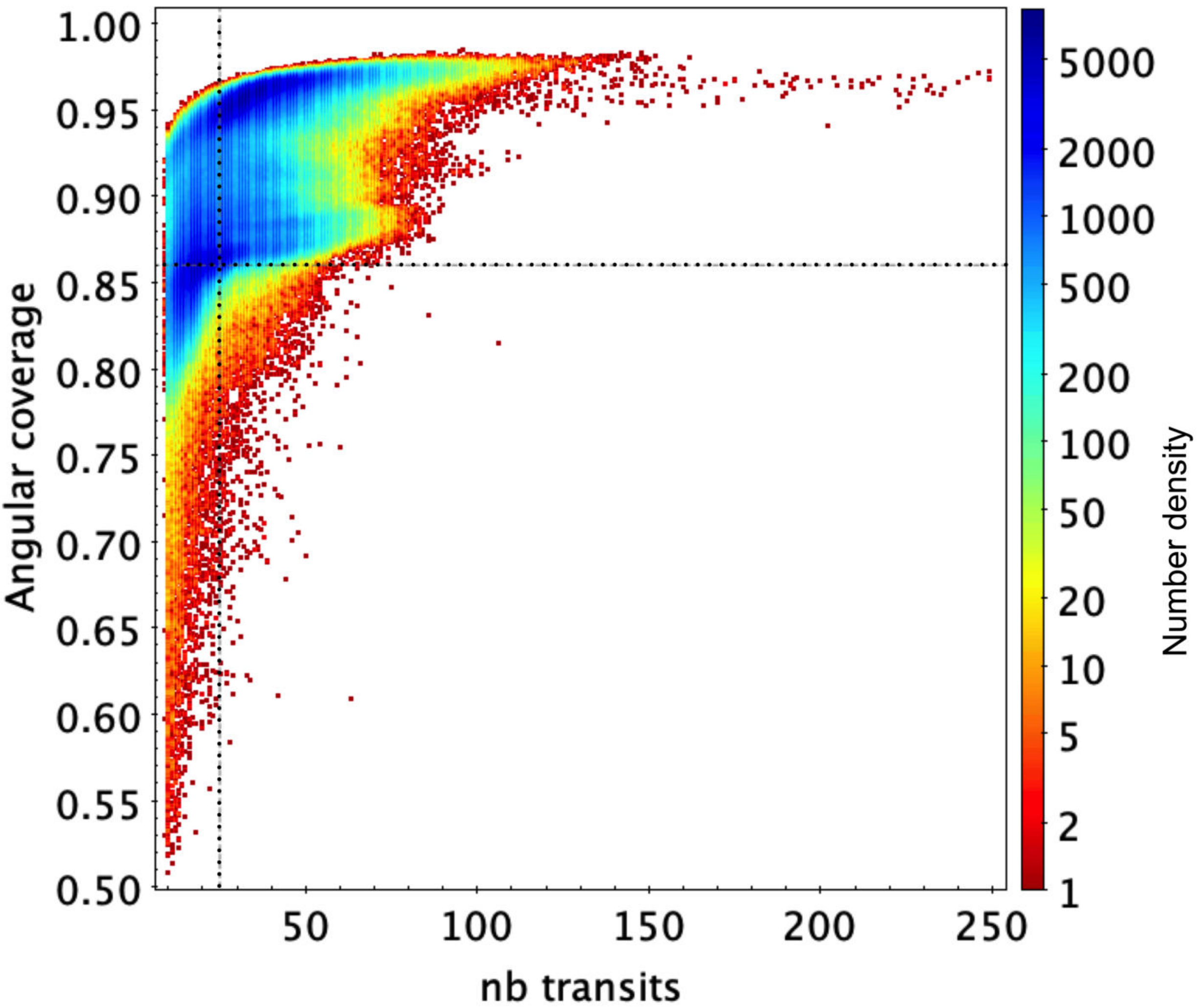} 
    \caption{Density plot of the distribution of angular coverage of quasars and galaxies as a function of their number of transits. Sources with less than 25 transits or with a coverage $< 86\%$ are discarded from the analysis. These limits are indicated by the dotted lines.}
    \label{angularCoverage}
\end{figure}

The angular coverage of the quasars and galaxies from the input lists is presented in Figure~\ref{angularCoverage} as a function of the number of transits over the sources. When the number of transits over a source is too small or the angular coverage low, it is then impossible to properly
recover the morphology of the source \citep{2013Krone-Martins}. The first action of the pipeline is then to filter out sources with less than 25 transits or with an angular coverage $< 86\%$. 

\begin{figure}
    \includegraphics[width=0.5\textwidth]{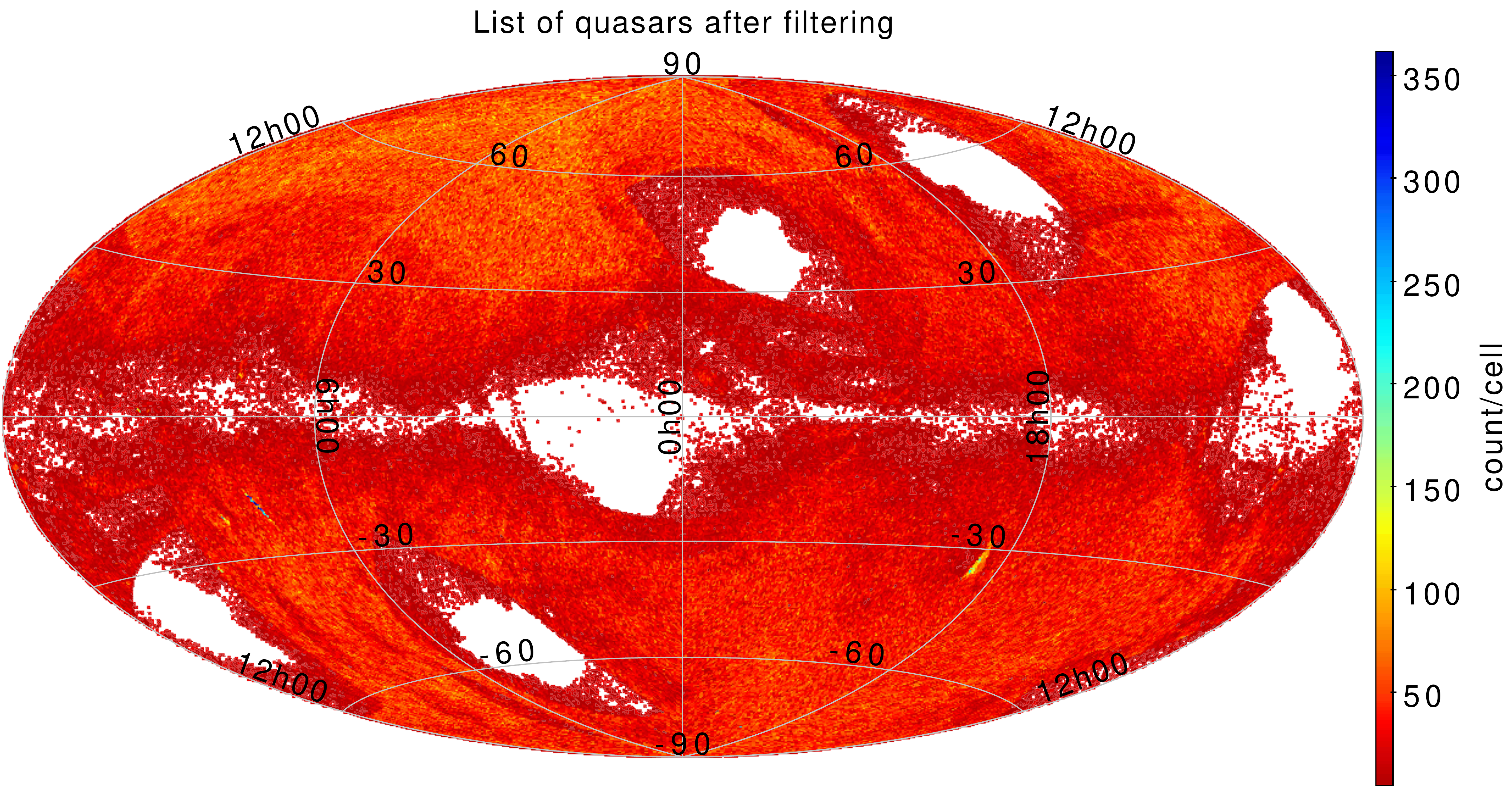} 
    \includegraphics[width=0.5\textwidth]{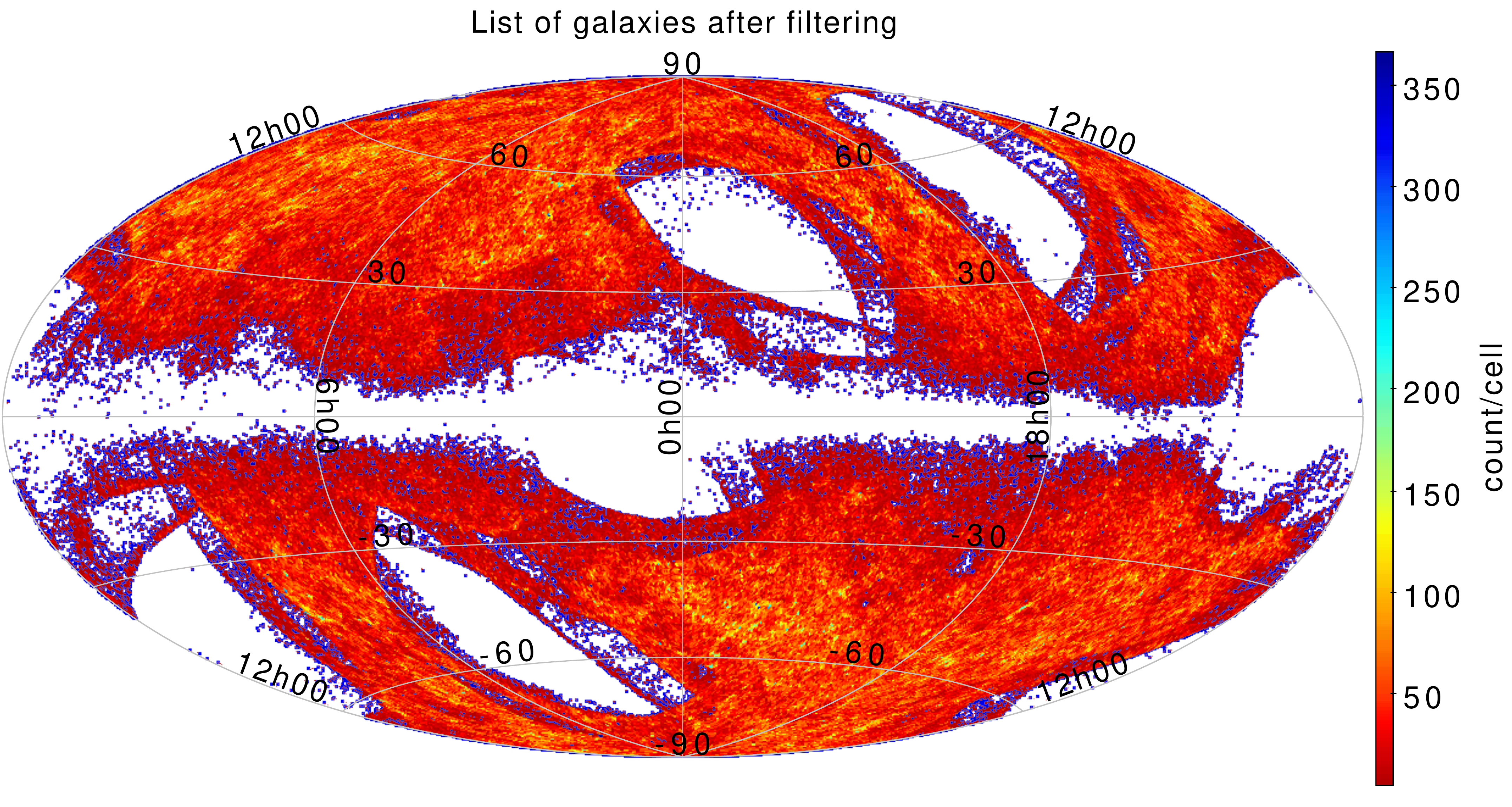} 
    \caption{Sky distribution in Galactic coordinates of the final lists of $1\,103\,691$ quasars and of $940\,887$ galaxies retained after applying a filtering on the angular coverage and on the number of transits. The cell of these maps is approximately 0.2 deg$^2$, and the colours indicate the number of sources in each cell.}
    \label{proc_lb}
\end{figure}

The filtering on the number of transits and on the angular coverage leaves us with \textbf{$1\,103\,691$} quasars and \textbf{$940\,887$} galaxies to be analysed. Their distribution on the sky is presented in Figure~\ref{proc_lb} where the signature of the \gaia scanning law unambiguously appears: the depletion area corresponds to regions where the number of transits is too small or where the coverage is insufficient. 

\begin{figure}
    \includegraphics[width=0.45\textwidth]{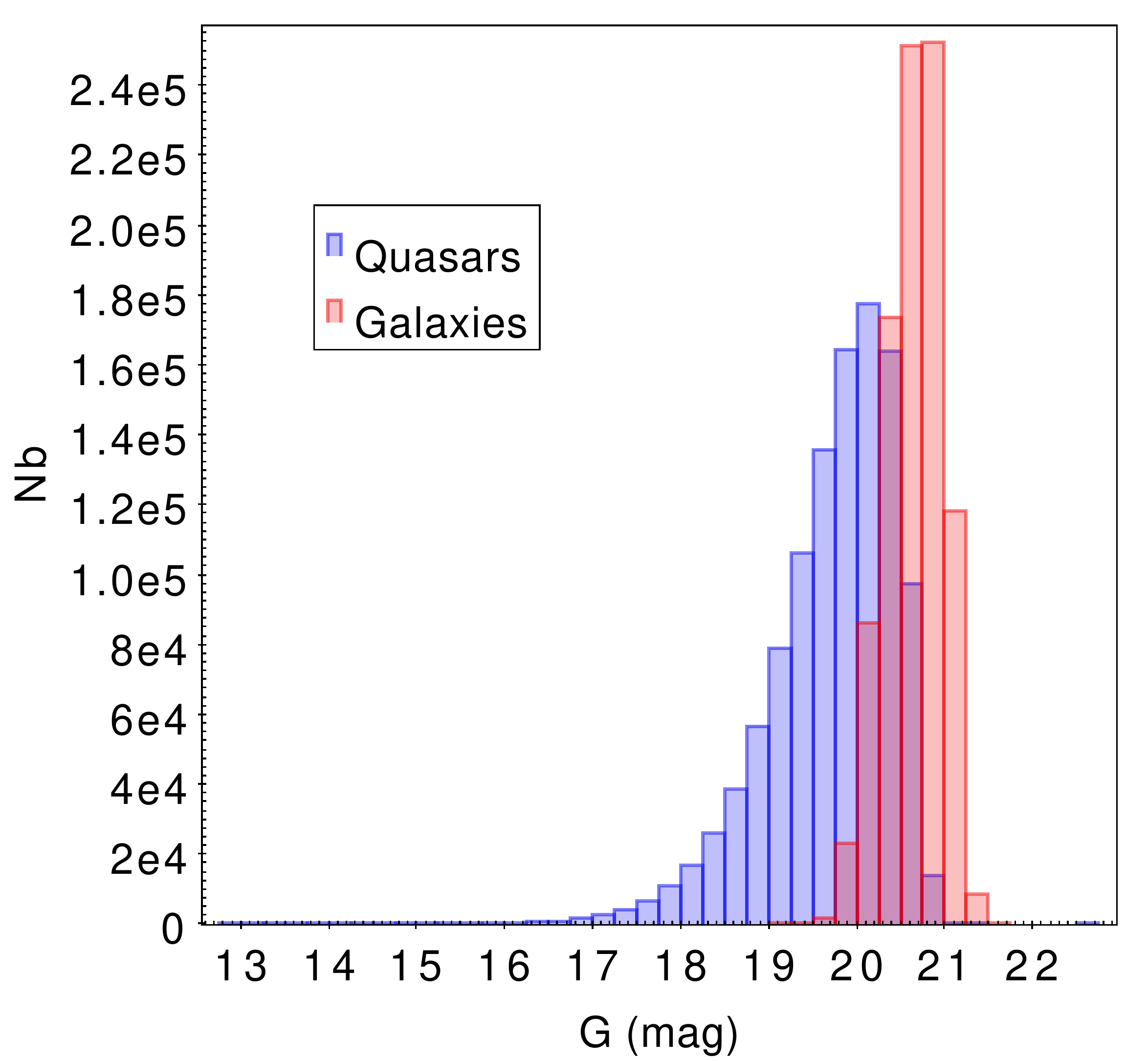} 
    \includegraphics[width=0.45\textwidth]{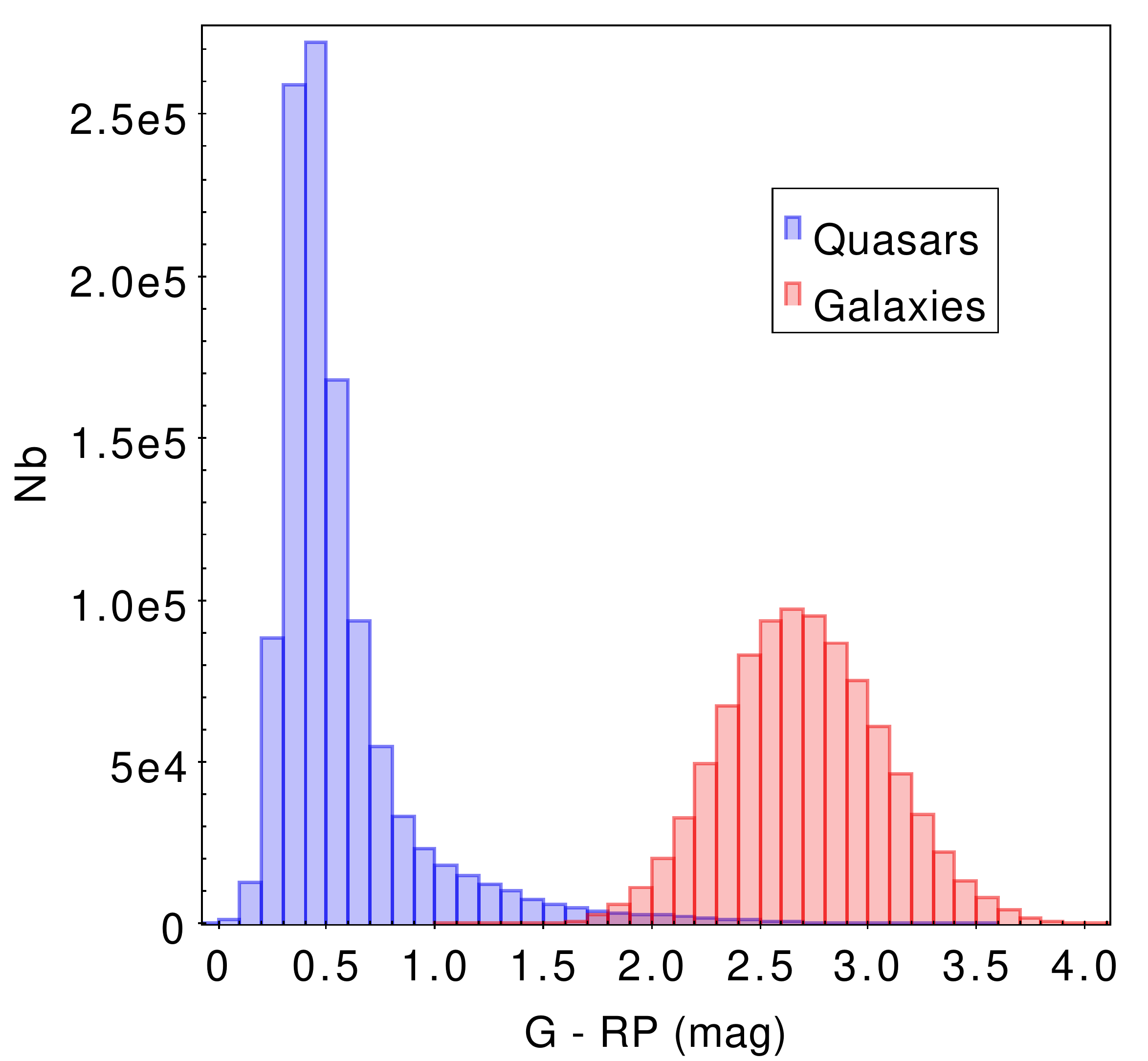} 
    \caption{Distribution of G magnitudes (phot\_g\_mean\_mag parameter, bin size=0.25 mag) and G-RP colour (phot\_g\_mean\_mag - phot\_rp\_mean\_mag, bin width=0.1 mag) of the final lists of $1\,103\,691$ quasars and of $940\,887$ galaxies analysed in terms of surface brightness profiles.}
    \label{G}
\end{figure}

The distribution of the \gaia magnitudes of the final list of quasars and of galaxies is presented in Figure~\ref{G} along with their colour G-RP. Most galaxies appear fainter than G=20 mag, whereas quasars appear brighter. This is a consequence of the way \gaia measures magnitudes (\textit{phot\_g\_mean\_mag}) which is tuned for point-like sources (and most quasars) and not suited for extended objects. All galaxies are redder than quasars. There is a slight overlap between two distributions  that corresponds to quasars with host galaxies and for which the host modifies the mean colour.

\section{Pipeline overview}\label{pipeline}
\begin{figure*}[h]
    \centering
        \includegraphics[width=1\textwidth]{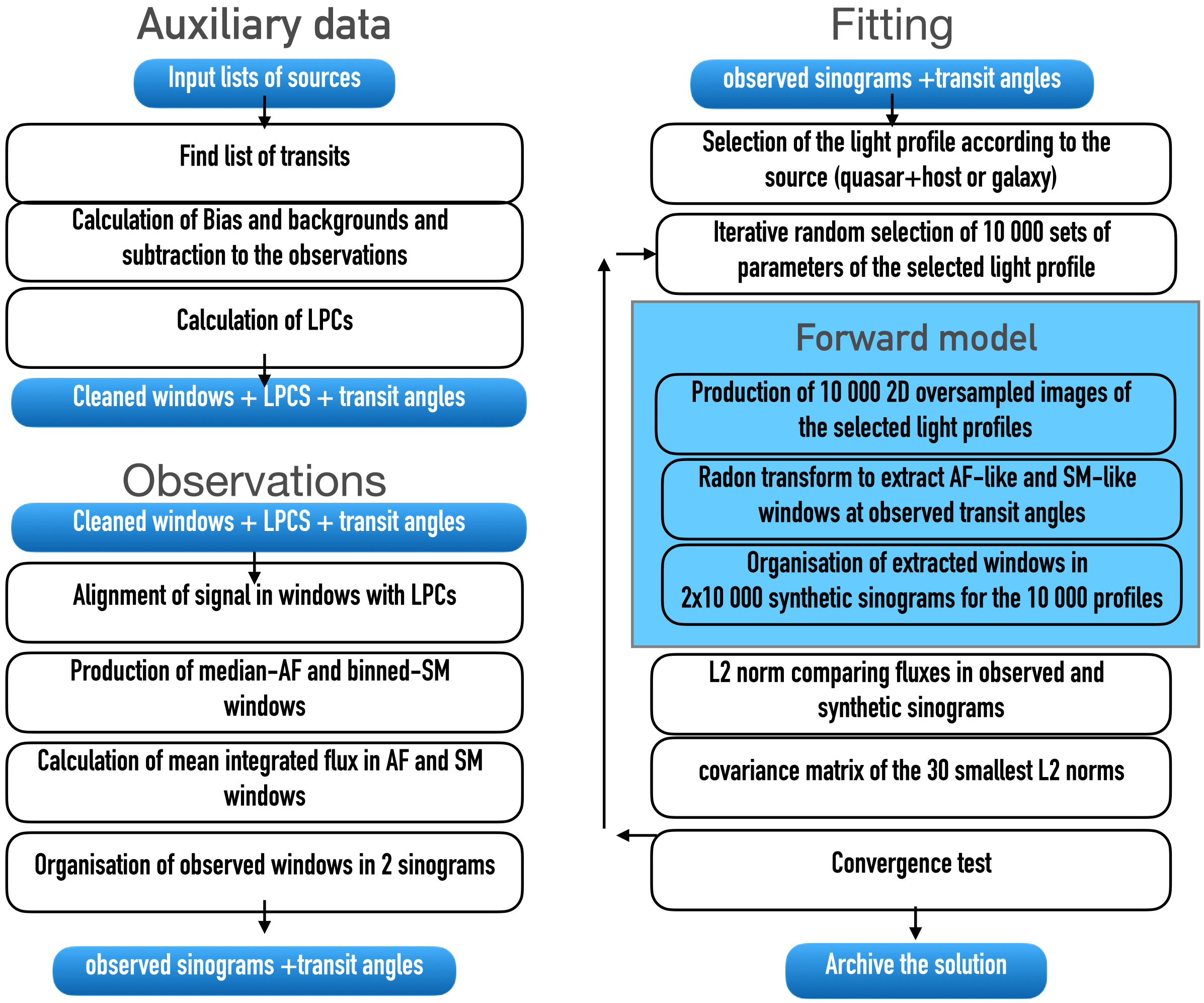}
    \caption {Pipeline steps: (1) production of auxiliary data, (2) treatment and organisation of observed windows, (3) fitting of surface brightness profiles.}
    \label{fit}
\end{figure*}

Figure~\ref{fit} presents a flowchart of the CU4\textit{-Surface Brightness Profile fitting} pipeline, illustrating the different tasks of the pipeline. 
It presents the three major steps and their details: preparation of auxiliary data needed by the chain, organisation of observations, and fitting of light profiles. The pipeline is operated at the Centre d'Etudes Spatiales (CNES). The details of the steps are presented in the following sections.

\section{Gaia data}\label{data}
The observations processed by the CU4 \textit{Surface Brightness Profile} pipeline for quasars and galaxy candidates were acquired between 25 July 2014 and 28 May 2017 and represent a total of $\sim$116 million transits over the sources. Observations suspected to contain a corrupted signal were removed from these data (e.g. observation gaps, decontamination, basic angle gaps).

\subsection{AF/SM observed windows}\label{afsmobservations}
\begin{figure}[h]
    \includegraphics[width=0.45\textwidth]{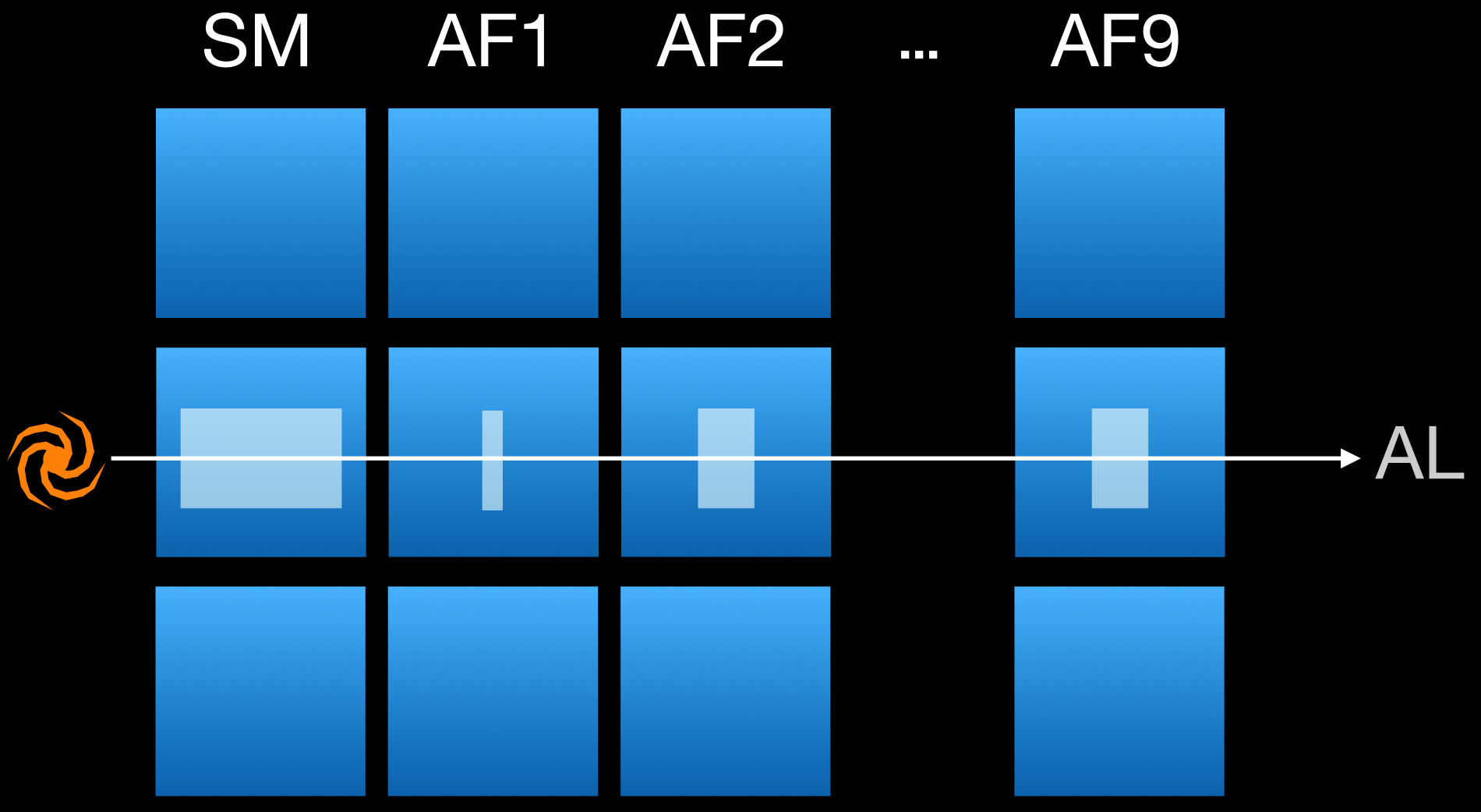} 
    \caption{Observation of a galaxy passing through the focal plane of \gaia in the AL direction, entering the SM then the AFs (AF1-AF9). Light blue windows represent the observed windows that are transmitted to the ground.}
    \label{focal}
\end{figure}

\begin{figure}[h]
    \includegraphics[width=0.45\textwidth]{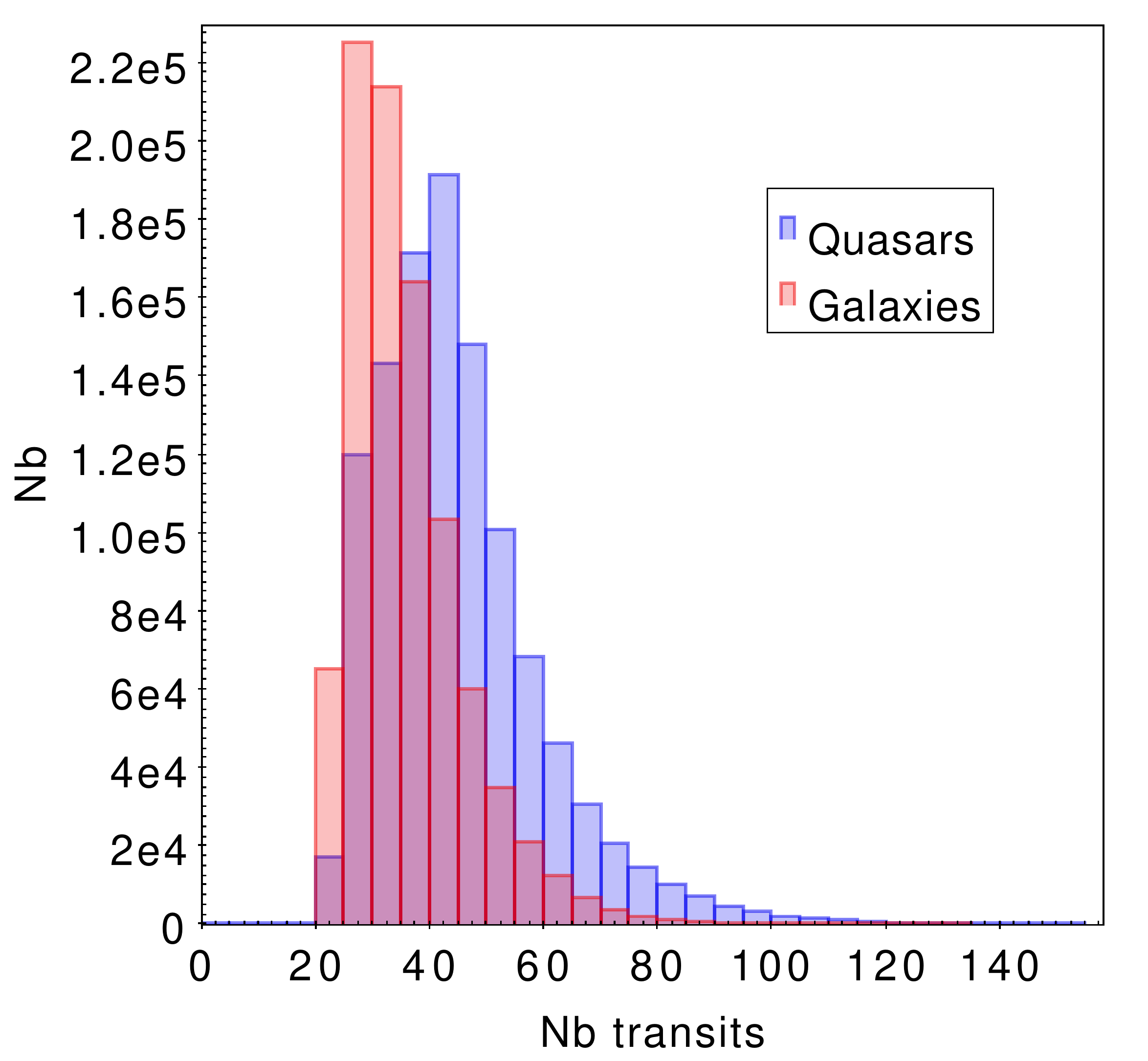} 
    \caption{Histogram of the number of valid transits for the quasars and the galaxies to be processed  after the  elimination of abnormal sample values using a 3$\sigma$ clipping rejection criterion (bin width=5).}
    \label{ntransits}
\end{figure}

The satellite scans sources with various transit angles as imposed by the scanning law. One transit over a source is illustrated in Figure~\ref{focal} where the source passes through the focal plane of \gaia in the along scan (AL) direction, entering first the SM and then the nine AFs (AF1-AF9). The transmitted windows of observation, as selected by the VPA, are represented in light blue. Figure~\ref{ntransits} presents the distribution of the number of transits of the quasars and galaxies. Quasars have an average number of 40 transits while galaxies have 35 transits. Quasars have slightly more transits because of the presence of a bright nucleus (the quasar) while all galaxies are diffuse and therefore less frequently detected.

The size of the transmitted windows depends on the magnitude of the source. There are three regimes of magnitudes: G$<$13 mag, 13 mag $< G <$ 16 mag, and G$>$16 mag, corresponding to different window sizes. Extragalactic objects are generally fainter than G=16 mag. Table~\ref{sample} presents the characteristics of these windows. 

\begin{table}[h]
\caption{Characteristics of the observed windows for sources fainter than G=16 mag. The dimensions are given in the AL and AC directions and expressed in \textit{sample} units (binned physical pixels) and in \mas.}
\label{sample}
\centering
\begin{tabular}{lcrr}
\hline
Detector & \multicolumn{2}{c}{Window size (AL,AC)} & Sample size \\
         &  [sample]    & [mas] & [mas]\\
\hline
SM       &  20 x 3    & 4\,715 x 2\,121 & 236 x 707\\
AF 1     &   6 x 1    &  354 x 2\,121 &  59 x 2\,121 \\
AF 2-9   &  12 x 1    &  707 x 2\,121 &  59 x 2\,121 \\
\hline
\end{tabular}
\end{table}

The nine AF windows are 1D with rectangular samples narrower in the along scan (AL) direction and larger in the across scan (AC) direction. The SM window is two-dimensional, covering a surface that is approximately seven times larger than that of the AF. Its resolution in the AL direction is much poorer than that  in the AF direction. SM catches a wide-angle view of the objects with low resolution and AF provides an extremely accurate view of their inner part. 

\begin{figure}
        \centering
        \includegraphics[height=0.4\textwidth]{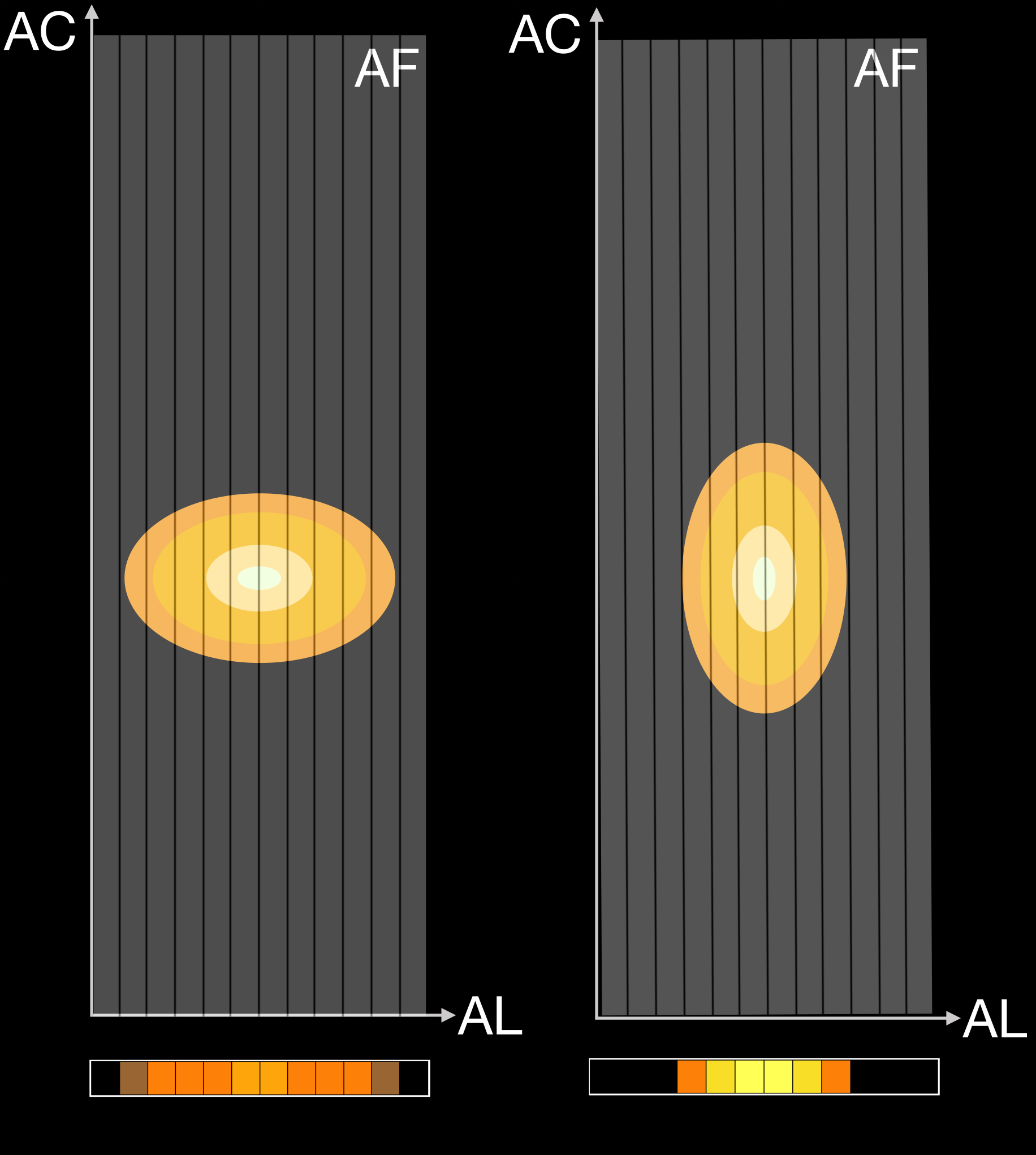} 
    \caption{Illustration of the observation by AF of an extended and elongated source when the scan direction (AL) is along the major axis (left) or along the minor axis (right) of the source. The resulting observed windows are given at the bottom of the figure. The colours illustrate the amount of flux collected in the different samples. When scanned along minor axis, the observations present an over-brightness at the centre. This effect is also seen in the SM detector.}
    \label{principle}
\end{figure}

For extended objects, the rectangular shape of the samples is responsible for a very specific light distribution in the observed windows, as illustrated in Figure~\ref{principle}. When an elongated object is scanned along its minor axis, the flux collected by the central samples is much larger than when it is scanned along its major axis. This property is one of the specificities of the observations that our algorithm exploits in particular to extract the position angles of the objects. 

The small overall angular size of the \gaia windows (2\,121 mas for AF (AC), 4\,715 mas for SM) is a limitation to the analysis of surface brightness profiles of large extended sources. Large sources with an effective radius (encompassing half of their total flux) of larger than $\sim$2 \arcsec \ have less than 50\% of their total flux collected in the observed windows. The surface brightness profile algorithm is then forced to use extrapolation of the flux outside the observation windows. Accordingly, a minimum of one effective radius should be encompassed within the AF windows for a reliable analysis of the light profiles. 


\subsection{Bias and background correction}
The raw observations (AF and SM observed windows) have to be corrected for bias and background before being analysed in terms of surface brightness profile fitting. A dedicated chain has been developed to handle this task. The bias and background are calculated using routines provided by the CU5 DPAC group in charge of the photometric calibration of the observations \citep[see][for a detailed description of the bias and background treatment within DPAC]{Castaneda2021gdr3.reptE...3C}\footnote{The user can also read the online documentation and the sections \url{https://gea.esac.esa.int/archive/documentation/GDR3/Data_processing/chap_cu3pre/sec_cu3pre_cali/ssec_cu3pre_cali_ccdbias.html} for details on the bias calibration and \url{
https://gea.esac.esa.int/archive/documentation/GDR3/Data_processing/chap_cu3pre/sec_cu3pre_cali/ssec_cu3pre_cali_astro.html} for the background computation.}. 
The science observations are obtained by subtracting the bias and the background from the raw observations. The background is dominated by scattered sunlight and the zodiacal light to which the Milky Way makes a significant contribution.

\subsection{Enhancement of the signal-to-noise }
The vast majority of the sources analysed by the pipeline are faint with low signal-to-noise ratio (S/N), with most galaxies being fainter than G=20 (see Figure~\ref{G}). Quasars appear generally brighter than the galaxies in this figure because of their central bright nucleus but their surrounding host galaxy  - when detectable by our pipeline - appears much fainter.

In order to enhance the S/N in the AF data, the nine AF windows of each transit are combined to produce a \textit{median-AF} window. The first step is to align the signal of each window on a common frame. This is performed by first applying a cubic spline interpolation in each observed AF1-9 window and then producing nine corresponding over-sampled (\textit{synthetic AF1-9}) windows (with a pixel size=5 \mas). The signals in the nine synthetic AF1-9 windows are then re-centred respectively using the local plane coordinates (LPCs) \citep{Hobbs2018gdr2.reptE...3H} that place the observed windows onto the sky. The median of the nine aligned synthetic windows is then computed after a three-sigma clipping rejection. This median-AF window is finally re-binned into the original window sampling corresponding to the standard AF2-9 dimensions. The AF1 window contains fewer samples than AF2-9 and therefore requires a specific treatment in order to incorporate its fluxes into the median AF window. For this purpose, AF1 is resized to the AF2-9 dimensions by adding empty samples at its extremities in order to handle it in the same way as other AFs. 

In parallel, the SM windows are binned in the AC direction in order to obtain a 1D window (binned-SM) of (20x1) samples with increased S/N. Similarly to what is done with AF windows, the signal of the SM windows of all transits are aligned via a cubic spline interpolation and the use of the LPCs. 

\subsection{Statistical estimators of the flux in windows}\label{s:estimators}
There are several quantities that characterise the distribution of the flux of the sources in the median-AF and the binned-SM windows. One of them is referred to here as the mean integrated flux in the median-AF or the SM windows and is defined as:  

\begin{equation}\label{int_flux}
\begin{split}
&\overline{flux_{AF}} = \sum_{j=1}^{N_{AF}} \sum_{i=1}^{12} s_{ij} /N_{AF}\\
&\overline{flux_{SM}} = \sum_{j=1}^{N_{SM}} \sum_{i=1}^{20} s_{ij} /N_{SM}\\
\end{split}
,\end{equation}
where s$_{ij}$ is the flux value (in e-/s) of sample $i$ of transit $j$. $N_{AF}$ and $N_{SM}$ are the number of AF and SM transits. These quantities are calculated after the elimination of abnormal sample values using a 3$\sigma$ clipping rejection criterion applied to each sample over all transits. When the central samples or too many samples of a transit are eliminated, the transit is rejected. Only sources with 20 or more remaining transits are kept for processing.

\begin{figure}
        \includegraphics[width=0.48\textwidth]{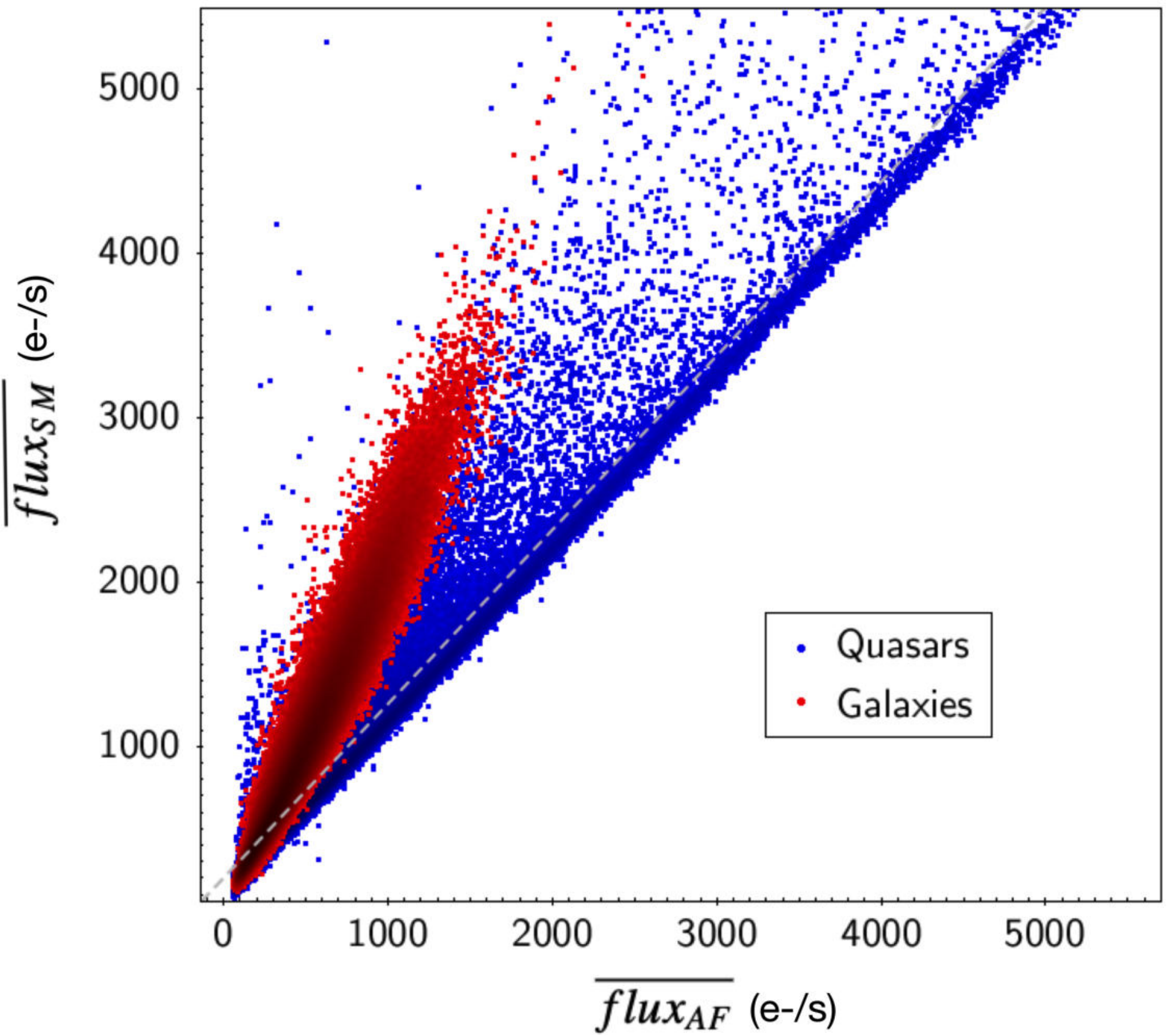}
    \caption {Comparison of the mean integrated fluxes of the quasars (blue) and of the galaxies (red) in the AF window and the SM windows over all transits. The dashed line indicates the limit $\overline{flux_{SM}}$ $< $1.06 $\overline{flux_{AF}}$+200 that isolates point-like objects unsuited for a determination of ellipticity and position angle. }
    \label{AFSM1}
\end{figure}

Figures \ref{AFSM1} presents a comparison of the mean integrated fluxes in the AF detector $\overline{flux_{AF}}$  against the mean integrated fluxes in the SM detector $\overline{flux_{SM}}$ for the lists of quasars and of galaxies. A star has a similar mean integrated flux in both AF and SM windows because it is fully encompassed by both. Sources with a larger flux in the SM window than in the AF window have a significant spatial extension. Most quasars are point-like and lie on the diagonal of this figure. Most galaxies are extended and lie well above the diagonal. Sources with $\overline{flux_{AF}}$ $<$120 e-/s do not  appear to be measurable because of their S/N. These sources are not further processed and considered afterwards as point-like.

\begin{figure}
\includegraphics[height=0.40\textwidth]{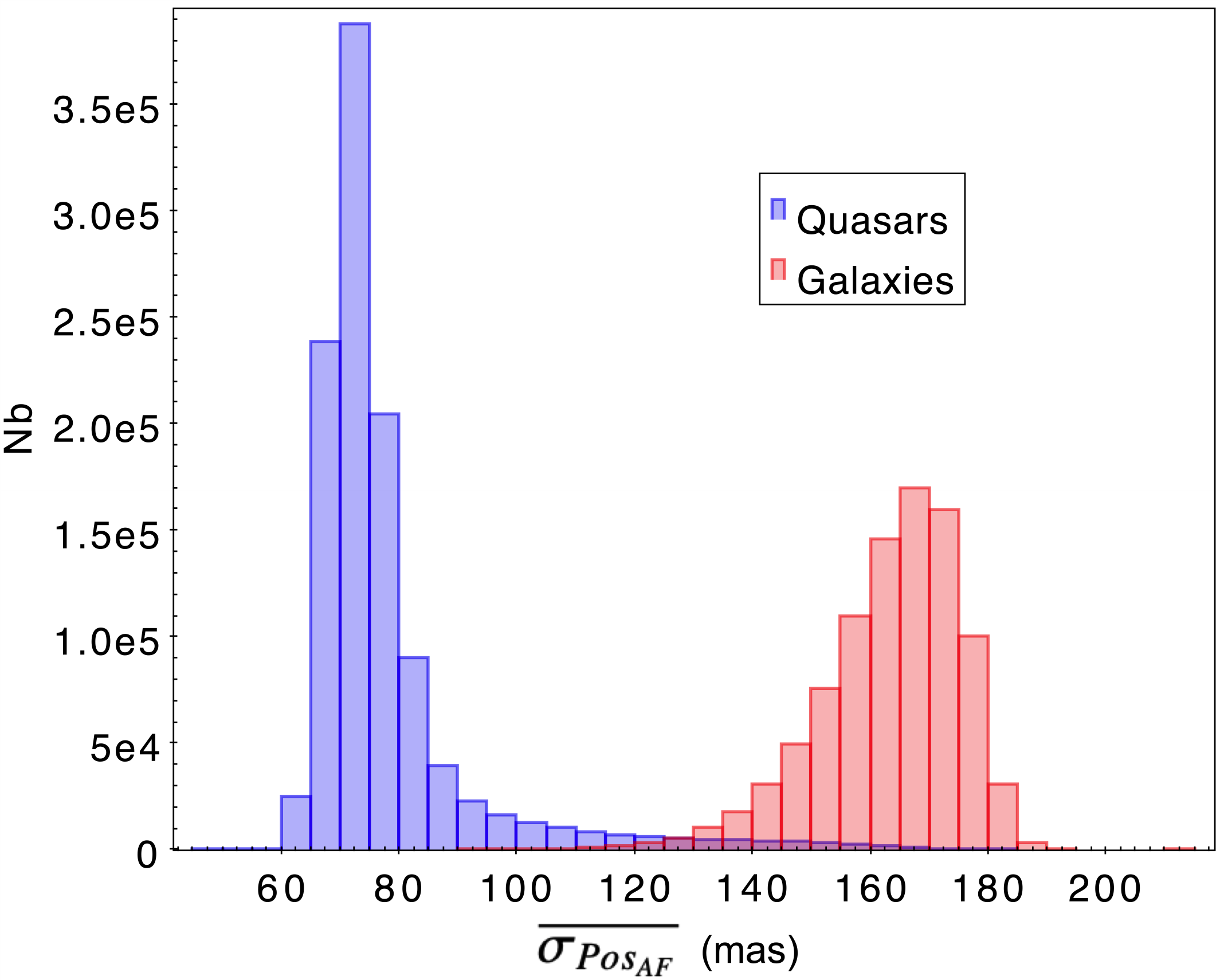}
\caption{Histogram of the mean standard deviation of the distribution of flux ($\overline{\sigma_{Pos_{AF}}}$) of the sources in the AF window over all transits (bin width=5 mas).}
\label{std}
\end{figure}

When analysing the signal in AF windows (highest resolution), several other statistical indicators can be used to investigate the spatial extension of the source: the mean of the flux distribution in each AF window ($Pos_{AF}$) and the standard deviation about this mean position ($\sigma_{Pos_{AF}}$), and finally the mean over all transits of these standards deviations ($\overline{\sigma_{Pos_{AF}}}$), which is a good indicator of the spatial extension of the sources in the AF detector:
 \begin{equation}
 \begin{split}
 &Pos_{AF} = \frac{\sum_{i=1}^{12} i s_{i}}{\sum_{i=1}^{12}s_{i}}\\
 &\sigma_{Pos_{AF}} = \sqrt{\frac{\sum_{i=1}^{12}i^{2} s_{i}-\sum_{i=1}^{12} i s_{i} Pos_{AF}}{\sum_{i=1}^{12} s_{i}}}\\
 &\overline{\sigma_{Pos_{AF}}} = \frac{\sum_{j=1}^{N_{AF}}\sigma_{Pos_{AF}}}{N_{AF}}
\end{split}
.\end{equation}
Figure \ref{std} presents the distribution of $\overline{\sigma_{Pos_{AF}}}$ for the quasars and the galaxies in our lists. It is clear that most quasars extend over 1-1.5 AF samples ($\sim60$-90 mas) while almost all galaxies extend over 2-3 AF samples $(\sim170$ mas). The overlap of both distributions corresponds to quasars for which the host galaxy is detected.

\subsection{Organisation of the observations in the Radon space}
The way \gaia scans the objects in different transit angles is similar to the Radon transform \citep{1986Radon}, which establishes the possibility of reconstructing a real two-variable function (similar to an image) using all of its projections along concurrent straight lines. Specific algorithms have been developed in the medical domain to recover the internal structure of patients from various profiles acquired by tomography. Although very rarely used in astronomy, this technique is perfectly adapted to the observations of \gaia, as demonstrated by \cite{2013Krone-Martins}. 
To analyse the surface brightness profile of a source, we first place its observations in a Radon space (so-called \textit{sinogram}), which organises the fluxes in the observed windows along their transit angles. 

\begin{figure}
        \centering
\includegraphics[height=0.45\textwidth]{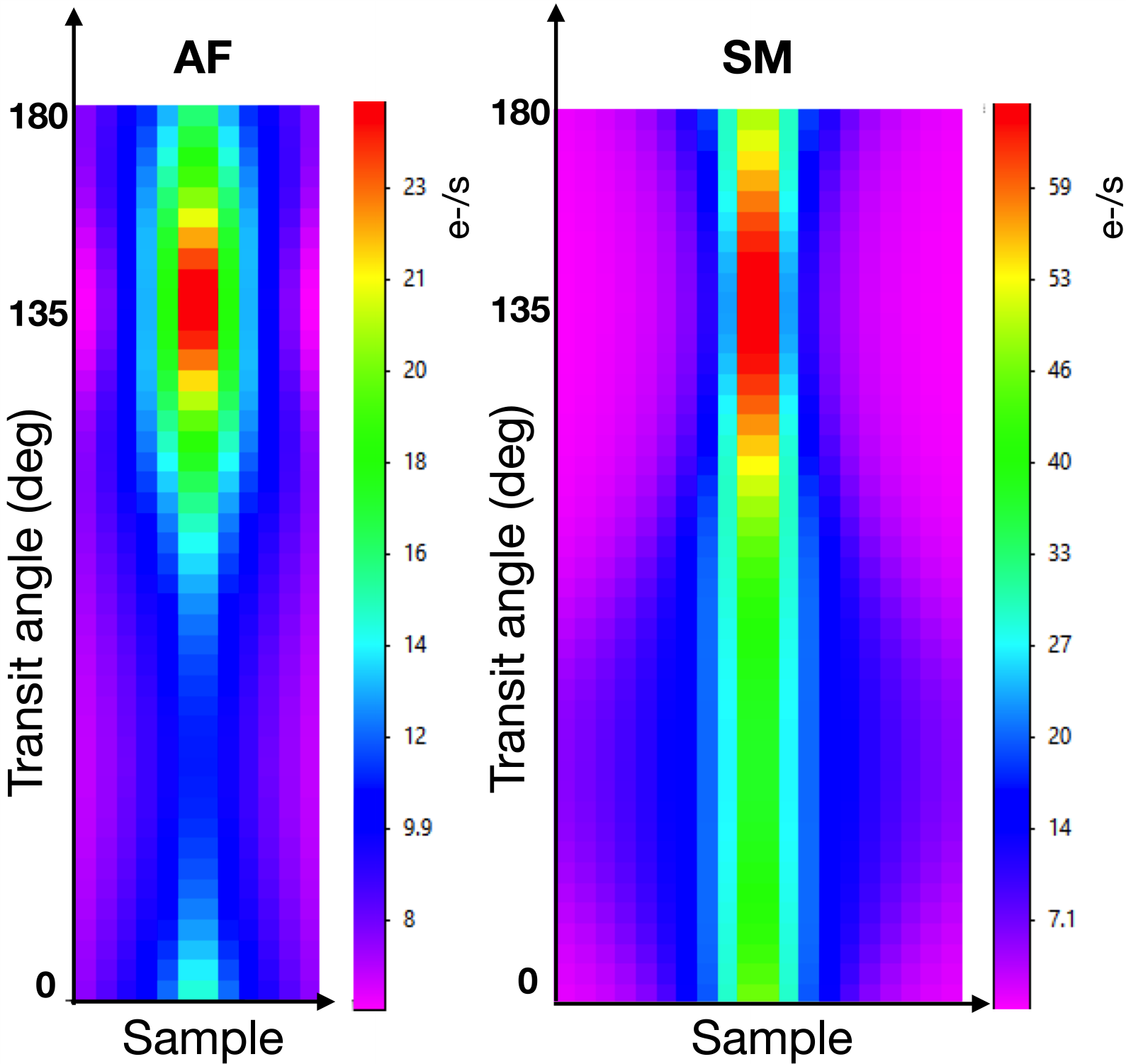} 
 \caption{Sinograms presenting the organisation of median-AF (left) and binned-SM (right) window fluxes of a simulated elliptical galaxy (ellipticity=0.7, position angle=45$\deg$, effective radius=2500 mas and \sersic index=3). Each horizontal line of the sinograms corresponds to one transit. AF windows are 12 samples long, while SM are 20 samples long. The fluxes of the AF and SM windows with transit angles=0$\deg$ lie at the bottom of the sinograms and the ones at 180$\deg$ at the top. The colour scales designates the sample flux values expressed in e-/s.}
    \label{radon}
\end{figure}

Figure~\ref{radon} presents two sinograms (AF left and SM right) of the observations of a simulated galaxy (ellipticity=0.7, position angle=45 $\deg$, effective radius=2500 mas and \sersic index=3) scanned regularly (each 5 $\deg$, from 0$\deg$ to 180$\deg$).

It is easy to note an over-brightness in both sinograms in the upper part around transit angle= 135$\deg$. This over-brightness corresponds to the transit angle where the source was scanned along its minor axis (as explained in Figure~\ref{principle}). The over-brightness corresponds then to a transit angle = galaxy position angle (45 $\deg$)+90$\deg$ = 135$\deg$.
 In real observations by \gaia, transit angles are not regularly spaced and therefore the resulting sinograms are less easily readable.
\section{Fitting surface brightness profiles }\label{fitting}
Fitting surface brightness profiles is achieved by a direct model that tends to reproduce the observed sinograms as best it can. The combination of parameters leading to the closest sinograms is then selected as the fitted profile. This is done via a global iterative strategy based on a direct model with a Bayesian exploration of the parameter space. 

\subsection{Light profiles}
The global iterative algorithm can be applied to both types of objects (quasars and their host galaxies or galaxies) but selects different models to be fitted. 

\subsubsection{Quasars}
In the case of quasars, the model must decompose the structure of the source into two components: the central quasar and a potential surrounding host galaxy. The central quasar is expected to be point-like, its extension being essentially due to the line spread function (LSF) of \gaia \citep{Fabricius2016A&A...595A...3F}. It is modelled by a circular exponential profile with a fixed scale length ($r_{s}=39.4$ mas) which approximately corresponds to the LSF of \gaia. The shape of the host galaxy is known to be spiral for distant quasars but could be bulge-dominated for closer quasars. We adopt a free S\'ersic profile \citep{1963Sersic} to model the host galaxy because of its ability to represent spirals as well as bulges: 
\begin{equation}\label{qso_profile}
\begin{split}
&\mathrm{Exponential\, (quasar):}\;  I_{q}(r) = I_{0}\,\exp\Bigg[\frac{-r}{r_{s}}\Bigg] \\
&\mathrm{Sersic\, (host\,galaxy):}\;  I_{g}(r) = I_{r_{e}}\,\exp\Bigg[-b_n \Bigg(\bigg(\frac{r}{r_{e}}\bigg)^{1/n} - 1\Bigg)\Bigg]\\
\end{split}
,\end{equation}
where $I_{q}(r)$ is the intensity of the central quasar at radius r, $I_{0}$ is the central intensity of the quasar, and $r_{s}$ is the scale length of the quasar (radius where the intensity drops by a factor $e$). $I_{g}(r)$ is the intensity of the host galaxy at radius r, $r_{e}$ is the major-axis effective radius encompassing half of the total flux of the source, $I_{r_{e}}$ is the intensity of the galaxy at effective radius, $n$ is the S\'ersic index, and $b_{n}$ is a function of $n$ such that $\Gamma (2n) = 2\gamma (2n, b_n)$, with $\Gamma$ being  the (complete) gamma function \citep{1991Ciotti} and $\gamma$ the incomplete gamma function. The value of $b_n$ is determined numerically.\\

\begin{figure}
\includegraphics[height=0.44\textwidth]{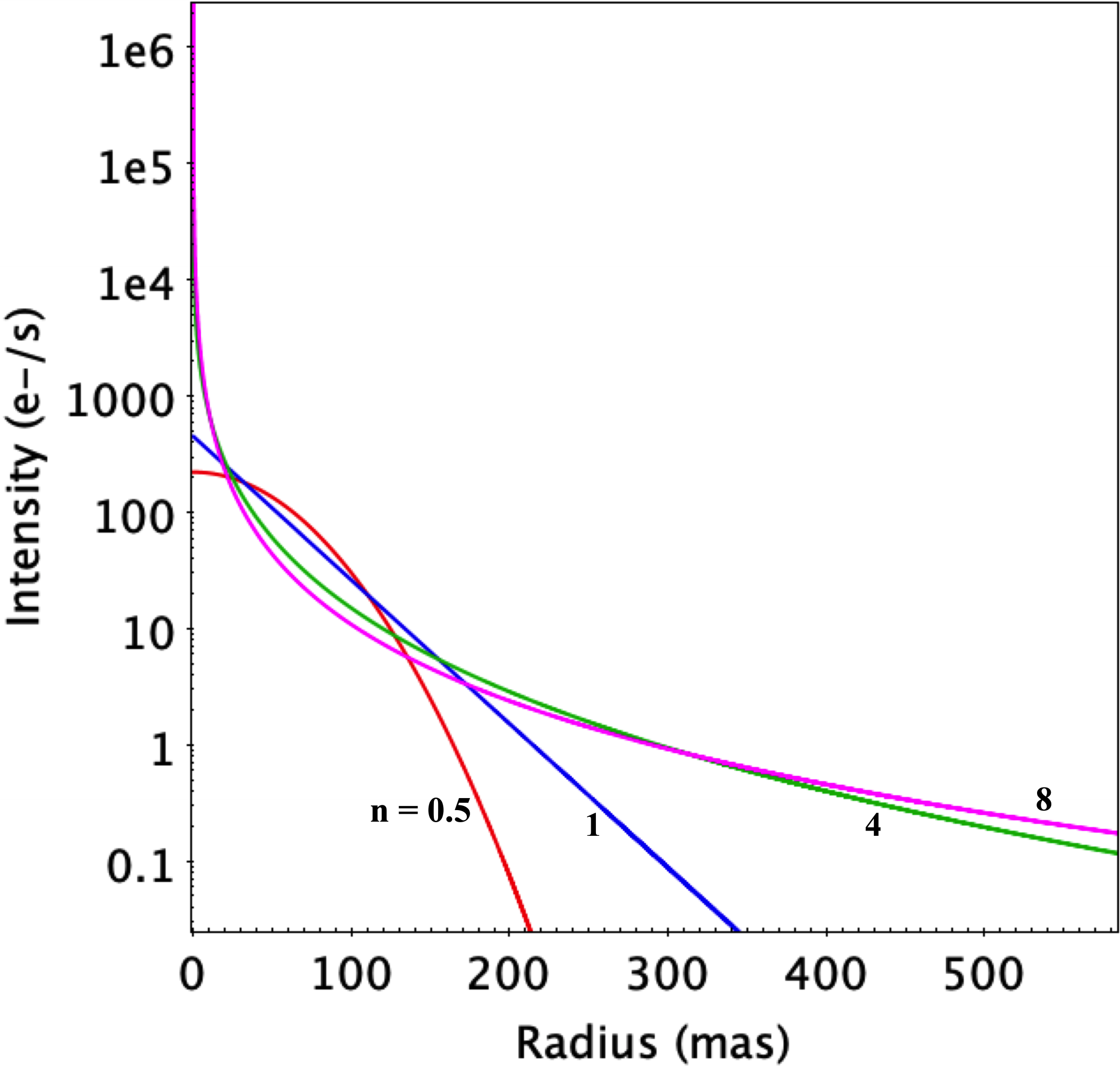}
\caption{\sersic profile where integrated flux (Flux=1000 e-/s) and effective radius ($r_{e}$=58.933 mas) are held fixed. The red line corresponds to the profile with \sersic index n=0.5 (Gaussian profile), blue to n=1 (exponential profile), green to n=4 (de Vaucouleurs), and pink to n=8.}
\label{sersic}
\end{figure}

The typical distribution of the flux of a galaxy along its radius according to the \sersic index is illustrated in Figure~\ref{sersic} which presents the variation of the light profile along the \sersic index for a fixed integrated flux and a fixed effective radius. Index value n=0.5 corresponds to a Gaussian profile, n=1 to an exponential profile, and n=4 to a de Vaucouleurs profile. The larger the index value, the steeper the central core, and the more extended the outer wing. Low \sersic indices have a flatter core and a more sharply truncated wing. Large \sersic indices are very sensitive to uncertainties in the sky background level determination because of the extended wings. The profiles corresponding to indices n>4 are in fact very similar to each other.

Quasars are usually variable and in particular those of our input list coming from the CU7 subset which were selected because  of their photometric variability in the \gaia data. If this variability has a long period then there is no impact in the present analysis. If it is with short-period variations, then it acts as an addition of noise in the data, eventually perturbing the profile fitting. In some cases, it leads to the non-convergence of the fitting. When a host galaxy surrounds the central quasar and is clearly detected by \gaia, the impact of variability on the  parameters of the host galaxy is minor.

\subsubsection{Galaxies}
Spiral and elliptical galaxies have intrinsically different shapes. \gaia  filters out most spiral galaxies and mostly detects elliptical galaxies. The consensus choice in the literature for describing dwarf and ordinary elliptical galaxies is the S\'ersic profile, which is a generalisation of the de Vaucouleurs $R^{1/4}$ model \citep{1948deVaucouleurs, 1953deVaucouleurs}. We successively adjusted these two profiles on our galaxies:   
\begin{equation}\label{gal_profile}
\begin{split}
&\mathrm{Sersic:}\; I_{g}(r)  = I_{r_{e}}\,\exp\Bigg[-b_n \Bigg(\bigg(\frac{r}{r_{e}}\bigg)^{1/n} - 1\Bigg)\Bigg]\\
&\mathrm{de\, Vaucouleurs:}\; I_{g}(r)  = I_{r_{e}}\,\exp\Bigg[-7.6697 \Bigg(\bigg(\frac{r}{r_{e}}\bigg)^{1/4} - 1\Bigg)\Bigg]\\
\end{split}
,\end{equation}
where $I_{g}(r)$ is the intensity of the profiles at radius r, $I_{r_{e}}$ their respective intensity at the effective radius, $r_{e}$ their major-axis effective radius, $n$ is the S\'ersic index, and $b_{n}$ is a function of $n$ (as described above Equation~\ref{qso_profile}).

An exponential profile has also been adjusted on all galaxies but is not published in \gdr{3} (see Section~ \ref{exp} for the discussion).
\\

\subsubsection{Shape parameters}
For both galaxies and galaxies hosting quasars, the shape parameters ellipticity, $\epsilon$, and position angle, $PA$ (from north to east), are also inferred. This is implemented by calculating the intensities given in equations \ref{qso_profile} and \ref{gal_profile} at radius $r$ \citep[e.g][]{1990Athanassoula} were $r$ is defined as
\begin{equation}\label{shape}
\begin{split}
r(x_p, y_p) = \Bigg[x_p^{2} + \bigg(\frac{y_p}{1 - \epsilon}\bigg)^{2}\Bigg]^{1/2} 
\end{split}
,\end{equation}
with $x_p = x  \cos(PA) - y  \sin(PA)$ and $y_p = x \sin(PA) + y \cos(PA)$ and the ellipticity $\epsilon$ = 1-b/a with a and b being the major and minor axis of the elliptical profile, respectively. 

Nevertheless a subset of the sources from the input lists are identified as being too faint for any tentative measurement of shape parameters; in which case they are fitted with circular profiles. To identify these objects, we compare their mean integrated fluxes in the AF and the SM windows ($\overline{flux_{AF}}$, $\overline{flux_{SM}}$) (see Figure \ref{AFSM1}). The extended sources have a larger mean integrated flux in the SM window than in the AF window, indicating that these objects extend beyond the limit of the AF window. It has been empirically determined that the condition\\ $\overline{flux_{SM}}$ $< $1.06 $\overline{flux_{AF}}$+200 isolates objects that are not suited for a determination of ellipticity and position angle, which are then not provided in the catalogue. 

\subsection{Fitting}
The fitting relies on an iterative application of a direct model that produces synthetic sinograms of the selected profile with chosen parameters and compares them to the observed ones.

\subsubsection{Forward model}
The forward model is used to produce synthetic sinograms. It produces a 2D over-sampled synthetic image (with pixel size = 58.9 mas x 58.9 mas) of the selected light profile for a given set of parameters. 
Using the Radon transform, AF-like and SM-like windows are then extracted from the synthetic image that match to the observed transit angles. The synthetic windows are organised into two sinograms and their integrated fluxes are compared to the integrated fluxes in the observed sinograms and the difference is characterised by a weighted sum L2 of two $\ell 2$ norms:
\begin{equation}\label{l2norm}
 L2 = \frac{\sqrt{\sum (SM - SM_{synth})^2}}{N_{SM}} + \frac{\sqrt{\sum (AF - AF_{synth})^2}}{N_{AF}} 
,\end{equation}
where SM and AF correspond to the observed integrated fluxes in the binned-SM and median-AF windows, SM$_{synth}$ and AF$_{synth}$ to the integrated fluxes in the synthetic SM and AF windows, N$_{SM}$ and N$_{AF}$ the number of valid windows for SM and AF corresponding to all transits over the source. 

\subsubsection{Iterative approach of the solution}
The fitting consists in finding the set of parameters from the light profile that minimises the L2 norm defined above (Eq.~\ref{l2norm}). This is done iteratively through a two-step strategy that numerically runs the forward model using different sets of parameters. The algorithm first locates the region of the space of parameters where the minimum L2 norm is found using a global optimiser based on multivariate normal distribution and maximum likelihood estimation and then applies a local optimiser (Matrix Adaptation Evolution Strategy \citep[CMA-ES,][]{hansen2006}) to accurately determine the solution, that is, the set of parameters leading to the lowest L2 norm. 

To locate the region of the parameter space where the smallest L2 norm is likely to be found, the algorithm randomly tries $10\,000$ sets of parameters with uniform distribution within the search domain described in Table~\ref{bounds}. The forward model is run for each set of parameters leading to $10\,000$ L2 norms. The 30 best sets of parameters (those having the lowest L2 norms) are kept. The mean and the covariance matrix of the parameters are computed. At the next iteration, $10\,000$ new sets of parameters are randomly drawn,  this time using a multivariate normal distribution characterised by the covariance matrix derived at the previous iteration. The forward model is evaluated with these new sets and with the 30 best solutions of parameters until the best solution no longer improves for ten consecutive iterations or when 500 iterations are reached.

In order to prevent the optimisation from getting stuck in local minima or on an obviously incorrect location of the parameter space and also to search for parameters in regions where real galaxies are found, we force, at each iteration, the random selection of parameters to follow some local constraints (that are part of the global boundaries presented in Table~\ref{bounds}). For instance, we force the central intensity of the galaxy to be smaller than that of the quasar. Most of these constraints were determined empirically. 

For the quasars, these constraints concern the maximum and the minimum fluxes of the central quasar and of the host galaxy:
\begin{equation}
\begin{split}
    &I_{q}^{min} \geq 2  \frac{flux_{AFmin}}{r_e}\\
    &I_{q}^{max} \leq \frac{1000 + 3  flux_{AFmax}}{r_e}\\
    &I_{g}^{max} \leq \frac{I_q}{\exp(b_n)}
\end{split}
,\end{equation}  

where
\begin{equation}
\begin{split}
    &b_n = 2  n - 1/3.0 + 4 / (405.0  n) + 46.0 / ( 25515.0  n^2)\\
    &r_e^{min} \geq - 0.2669  n^3 - 3.0159  n^2 - 4.7138  n + 112.41
\end{split}
.\end{equation}  

For the galaxies, considering $f_s = 2 \pi  n  \Gamma (2n) / b_n^{2n}$, we can additionally compute the upper and lower boundaries of the effective intensity of the \sersic and de Vaucouleurs profiles using the following equations:
\begin{equation}
\begin{split}
    &I_{Ser}^{min} \geq \frac{\phi_{min}  (58.933 / r_e)^2}{f_s  \exp(b_n)  (1-\epsilon)}\\
    &I_{Ser}^{max} \leq \frac{\phi_{max}  (58.933 / r_e)^2}{f_s  \exp(b_n)  (1-\epsilon)}\\
\end{split}
,\end{equation}
where $\phi^{min}$ and $\phi^{max}$ are empirically defined as
\begin{equation}
\begin{split}
&\phi^{max} = \operatorname{max}\bigg(20.0 \overline{flux_{AF}}, 6.0 \overline{flux_{SM}}\bigg)\\
&\phi^{min} = \operatorname{min}\bigg(1.5 \overline{flux_{AF}}, \overline{flux_{SM}}\bigg)\\
\end{split}
.\end{equation}\label{phimin}

The output of the global optimisation (means of parameters and covariance matrix over the last best solutions) is then used as input of the local optimiser Covariance
Matrix Adaptation Evolution Strategy \citep[CMA-ES,][]{hansen2006}, which is applied to accelerate the convergence and to more efficiently locate  the optimal solution of the problem. CMA-ES is an evolutionary algorithm designed for the optimisation of problems whose input is real. It is a randomised, derivate-free and bounded optimisation method that is considered as state-of-the-art among evolutionary algorithms \citep{Hansen2010Gecco}. This method is also based on multi-variate normal distribution but the covariance matrix of the distribution is incrementally updated such that the likelihood of the previous successful search steps is increased.

A test of convergence stops the iterative process when no improvement to the L2 norm is imposed. A maximum of 150 iterations is set. Objects reaching this limit are probably not well fitted and their parameters should be used with due caution. 
\begin{table*}
\centering
\caption{Search domain of the surface brightness profile parameters. I$_{r_{e}}$ is the intensity [e-/s] of the profile at the effective radius $r_{e}$ [mas], $\epsilon$ is the ellipticity of the source calculated as (1-b/a) where (a,b) are the semi-major and semi-minor axis of the source, $PA$ is the position angle of the source (from north to east), and $n$ the S\'ersic index. $Ic$ is the central intensity [e-/s] of the exponential profile of the central quasar. The scale length of the quasar exponential profile is fixed at $r_{s}$=39.4 mas.}
\label{bounds}
\begin{tabular}{lccccccc}
\hline
\hline
Profile   & Ic & I$_{r_e}$  &  $r_e$& $\epsilon$ & PA & S\'ersic Index n\\
   & $e^-$/s & $e^-$/s  &  mas&  & \deg & \\
\hline
Quasar (exponential)     & 20-2000    &    -      & -  & -      & - & - \\
Host galaxy (S\'ersic)  & -  & 0-200    &  39.4-8000   & 0.01-0.8 & 0-180 & 0.5-6\\
\hline
 Galaxy (S\'ersic) & - &  0-2000   &  30-8000   & 0-0.99 & 0-180 &0.5-8\\
 Galaxy (de Vaucouleurs)  & -        &  0-2000   &  30-8000   & 0-0.99 & 0-180 &\\
\hline
\end{tabular}
\end{table*}


\subsubsection{Correlation matrix}
A correlation matrix as well as internal errors are concurrently provided as auxiliary data product. These quantities are usually extremely small and reflect the final step of the convergence of the iterative process rather than the evaluation of the uncertainties on the quantities. 
During the fit of the parameters, only the 30 best solutions from our $10\,000$ random trials are kept to compute our statistics, such as the mean and covariance matrix. Therefore, the computation of the final covariance matrices is done on the final set of 30 best parameters. This explains the very small values obtained, because at this step, the algorithm should have converged, which should lead to very small differences between the 30 best sets of parameters.



\section{Post-processing}\label{postproc}
After running the pipeline on the two lists of sources, it is necessary to apply a post-processing step to the results to identify the sources that did not converge and to attribute quality flags to the sources. 

\subsection{Quasars}\label{postprocqso}
The model that is fitted on the quasars and their potential host galaxy is complex as it is the combination of two separate models, one for the nucleus and one for the surrounding host galaxy. 
This may lead to non-converged situations as well as situations where some of the fitted parameters converged towards the limit of the search domain, which eventually indicates that the model did not converge properly. All these cases were therefore filtered out. Another important filtering that has been applied to the quasar output is that only host galaxies with a radius of smaller than 2.5\arcsec are published, ensuring that at least one effective radius of the galaxy is encompassed in the SM windows (4\,715 mas AL), consequently preventing it from being in the extrapolation regime of the pipeline.

Two flags, \href{https://gea.esac.esa.int/archive/documentation/GDR3/Gaia_archive/chap_datamodel/sec_dm_extra--galactic_tables/ssec_dm_qso_candidates.html#qso_candidates-host_galaxy_detected}{\textbf{host\_galaxy\_detected}} and \href{https://gea.esac.esa.int/archive/documentation/GDR3/Gaia_archive/chap_datamodel/sec_dm_extra--galactic_tables/ssec_dm_qso_candidates.html#qso_candidates-host_galaxy_flag}{\textbf{host\_galaxy\_flag}}, are attributed to each source to respectively indicate that a host galaxy is detected and to indicate the specificity and quality of the fitted profile. 

The post-processing identified the quasars with a host galaxy detected and raised the flag \href{https://gea.esac.esa.int/archive/documentation/GDR3/Gaia_archive/chap_datamodel/sec_dm_extra--galactic_tables/ssec_dm_qso_candidates.html#qso_candidates-host_galaxy_detected}{\textbf{host\_galaxy\_detected}}\,=`true' when \\
\begin{equation}\label{qso_postproc}
\begin{split}
\overline{flux_{SM}}& > 1.06\,\overline{flux_{AF}}+200\,AND \\ 
\overline{flux_{AF}}& > 120\,e^-/s\,AND \\
r_{e,host}&  > 132\,mas.  
\end{split}
\end{equation}
When set to `false', the \href{https://gea.esac.esa.int/archive/documentation/GDR3/Gaia_archive/chap_datamodel/sec_dm_extra--galactic_tables/ssec_dm_qso_candidates.html#qso_candidates-host_galaxy_detected}{\textbf{host\_galaxy\_detected}} flag recovers two different situations: (i) there is no host (majority of cases), or (ii) there is another source in the immediate neighbourhood of the target in which case it is impossible to determine whether or
not there is a host around the quasar.\\

The values taken by \href{https://gea.esac.esa.int/archive/documentation/GDR3/Gaia_archive/chap_datamodel/sec_dm_extra--galactic_tables/ssec_dm_qso_candidates.html#qso_candidates-host_galaxy_flag}{\textbf{host\_galaxy\_flag}}, their signification, and the reason for their attribution are the following:
\begin{itemize}
\item[1:] Host galaxy well measured with circular S\'ersic profile. 
\item[2:] Host galaxy well measured with elliptical S\'ersic profile. 
\item[3:] No host galaxy detected.
\item[4:] Poor solution measured with elliptical S\'ersic profile. Some parameters are published while the doubtful ones are removed. 
\item[5:] No convergence of the fitting procedure. A host galaxy is detected for some of these sources others not. In all cases, the parameters of the light profile of the host galaxy  are not published.
\item[6:] Doubtful solution due to the presence of a secondary source closer than 5\,\arcsec. For security, for these sources \href{https://gea.esac.esa.int/archive/documentation/GDR3/Gaia_archive/chap_datamodel/sec_dm_extra--galactic_tables/ssec_dm_qso_candidates.html#qso_candidates-host_galaxy_detected}{\textbf{host\_galaxy\_detected}}=`false’, although some sources exhibit a clear host galaxy upon visual inspection. 
\end{itemize}


We refer the reader to Appendix \ref{sec:adql_quasar} for an efficient combination of \href{https://gea.esac.esa.int/archive/documentation/GDR3/Gaia_archive/chap_datamodel/sec_dm_extra--galactic_tables/ssec_dm_qso_candidates.html#qso_candidates-host_galaxy_detected}{\textbf{host\_galaxy\_detected}} and \href{https://gea.esac.esa.int/archive/documentation/GDR3/Gaia_archive/chap_datamodel/sec_dm_extra--galactic_tables/ssec_dm_qso_candidates.html#qso_candidates-host_galaxy_flag}{\textbf{host\_galaxy\_flag}} flags as well as for typical queries of the \href{https://gea.esac.esa.int/archive/documentation/GDR3/Gaia_archive/chap_datamodel/sec_dm_extra--galactic_tables/ssec_dm_qso_candidates.html}{\textbf{qso\_candidates}} table.

\subsection{Galaxies}\label{postprocgal}
The models fitted on the galaxies are not as complex as the model used for the quasars. This is why, in most cases, the fitting procedure converged towards a rather robust solution and the filtering applied to the output is not as severe as for the quasars.

The flags \href{https://gea.esac.esa.int/archive/documentation/GDR3/Gaia_archive/chap_datamodel/sec_dm_extra--galactic_tables/ssec_dm_galaxy_candidates.html#galaxy_candidates-flags_sersic}{\textbf{flag\_sersic}} and \href{https://gea.esac.esa.int/archive/documentation/GDR3/Gaia_archive/chap_datamodel/sec_dm_extra--galactic_tables/ssec_dm_galaxy_candidates.html#galaxy_candidates-flags_de_vaucouleurs}{\textbf{flag\_de\_vaucouleurs}} are given in the output table to indicate the specificity of each of the fitted profiles. The values taken by these flags and their signification are the following:
\begin{itemize}
\item[1:] Elliptical profile fitted, an external source is detected within 2.5" of this source; doubtful solution.
\item[2:] Circular profile fitted, an external source is detected within 2.5" of this source; doubtful solution.
\item[3:] Elliptical profile fitted, PA did not converged and one parameter or more converged towards the limit of the search domain; the solution can be considered as poor.
\item[4:] Elliptical profile fitted, PA did not converge.
\item[5:] Elliptical profile fitted, one parameter or more converged towards the limit of the search domain; poor solution.
\item[6:] Elliptical profile well fitted.
\item[7:] Circular profile fitted, one parameter or more converged towards the limit of the search domain; poor solution.
\item[8:] Circular profile well fitted.
\end{itemize}

We refer the reader to Appendix \ref{sec:adql_galaxy} for typical queries of the \href{https://gea.esac.esa.int/archive/documentation/GDR3/Gaia_archive/chap_datamodel/sec_dm_extra--galactic_tables/ssec_dm_galaxy_candidates.html}{\textbf{galaxy\_candidates}} table.

\section{Results and Validation}\label{results}
Our pipeline has analysed the surface brightness profile of $1\,103\,691$ quasars and $940\,887$ galaxies from the input lists. The results are included in the extragalactic tables \href{https://gea.esac.esa.int/archive/documentation/GDR3/Gaia_archive/chap_datamodel/sec_dm_extra--galactic_tables/ssec_dm_qso_candidates.html}{\textbf{qso\_candidates}} and \href{https://gea.esac.esa.int/archive/documentation/GDR3/Gaia_archive/chap_datamodel/sec_dm_extra--galactic_tables/ssec_dm_galaxy_candidates.html}{\textbf{galaxy\_candidates}} that come along with \gdr{3} and that are presented in \citep{2022Bailer-Jones}. These tables are a compilation of the results from all DPAC processing modules that have classified or analysed extragalactic objects (surface brightness profiles, variability profiles, redshift measurement, source classifications). Concerning the quasars processed by the surface brightness profile module, additional information can be found in the table \href{https://gea.esac.esa.int/archive/documentation/GDR3/Gaia_archive/chap_datamodel/sec_dm_extra--galactic_tables/ssec_dm_qso_catalogue_name.html}{\textbf{qso\_catalogue\_name}} which provides the name of the external catalogues in which the quasars were found (as detailed in Section~\ref{qso_input}). 

\subsection{Quasars}
\begin{figure*}
     \centerline{
     \includegraphics[width=0.33\textwidth]{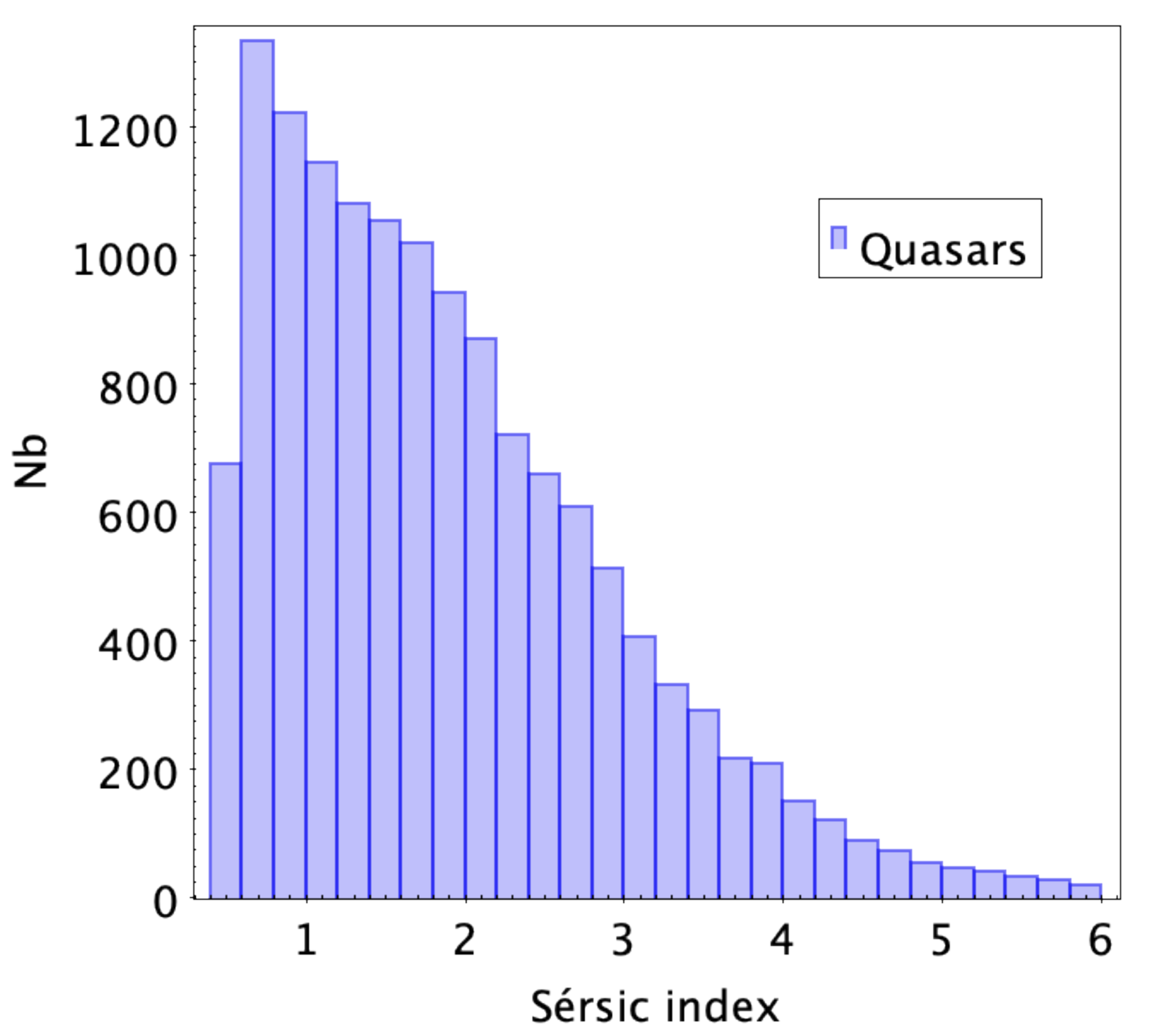} 
     \includegraphics[width=0.33\textwidth]{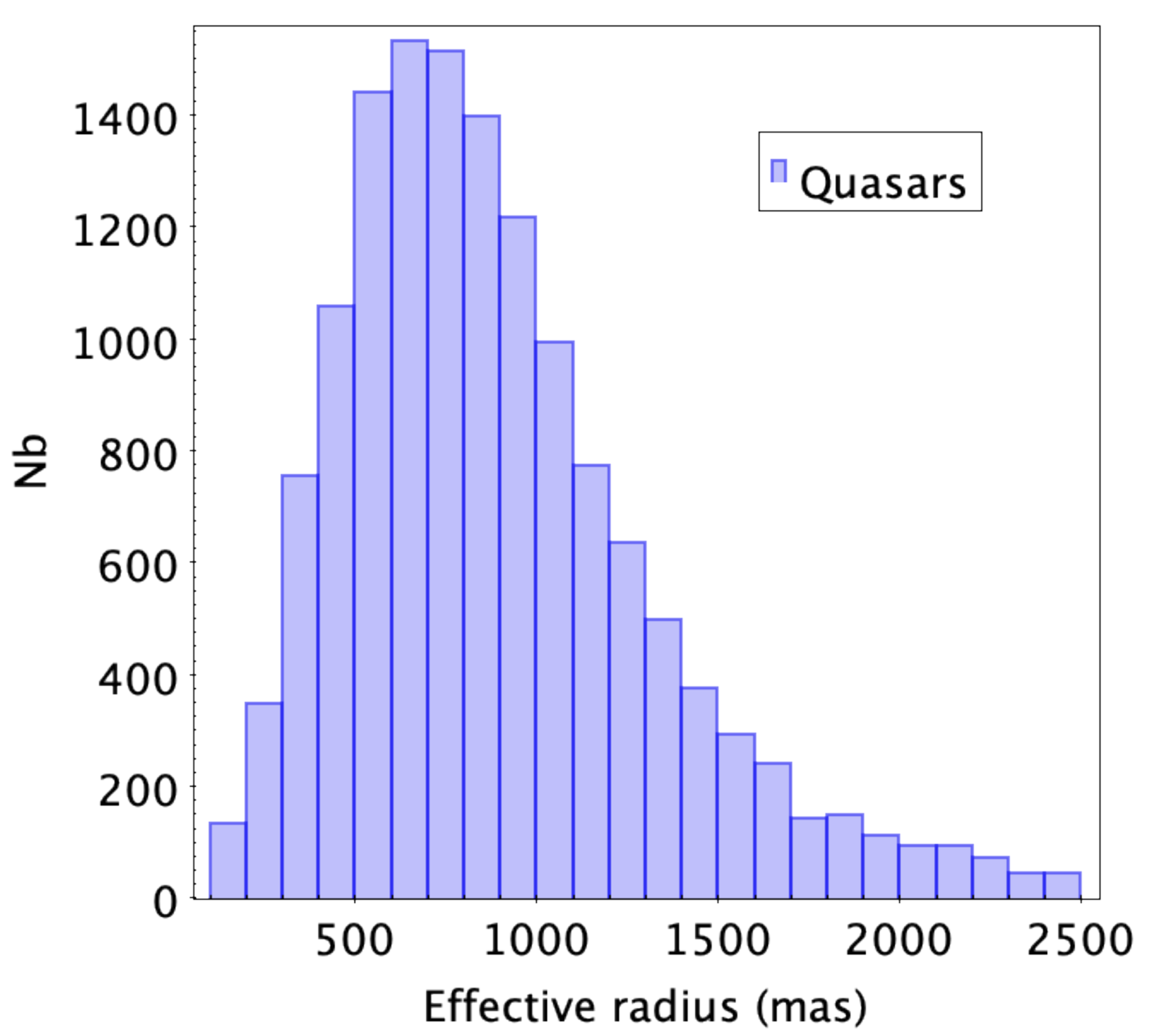}
     \includegraphics[width=0.33\textwidth]{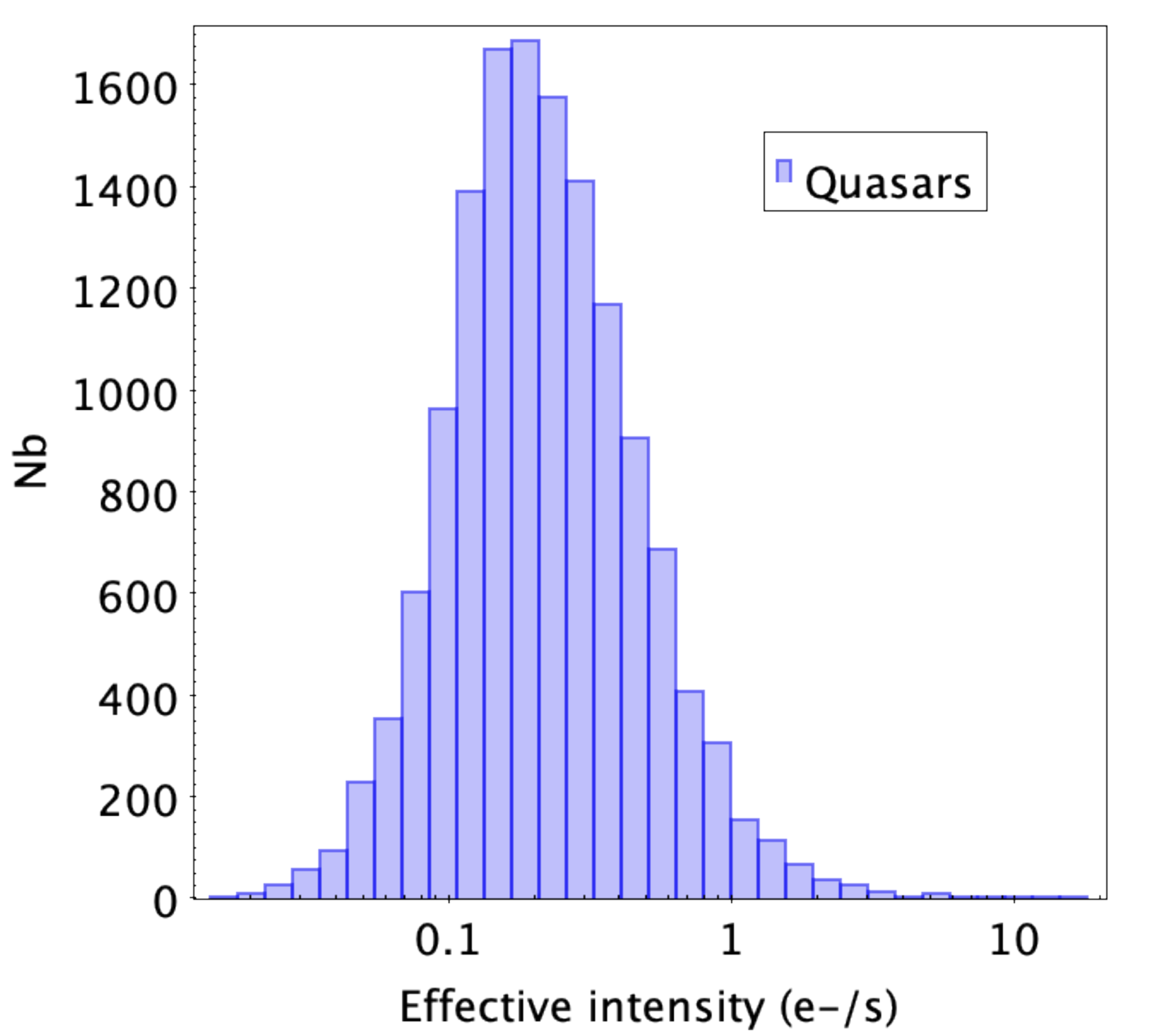} }
     \centerline{ 
     \includegraphics[width=0.33\textwidth]{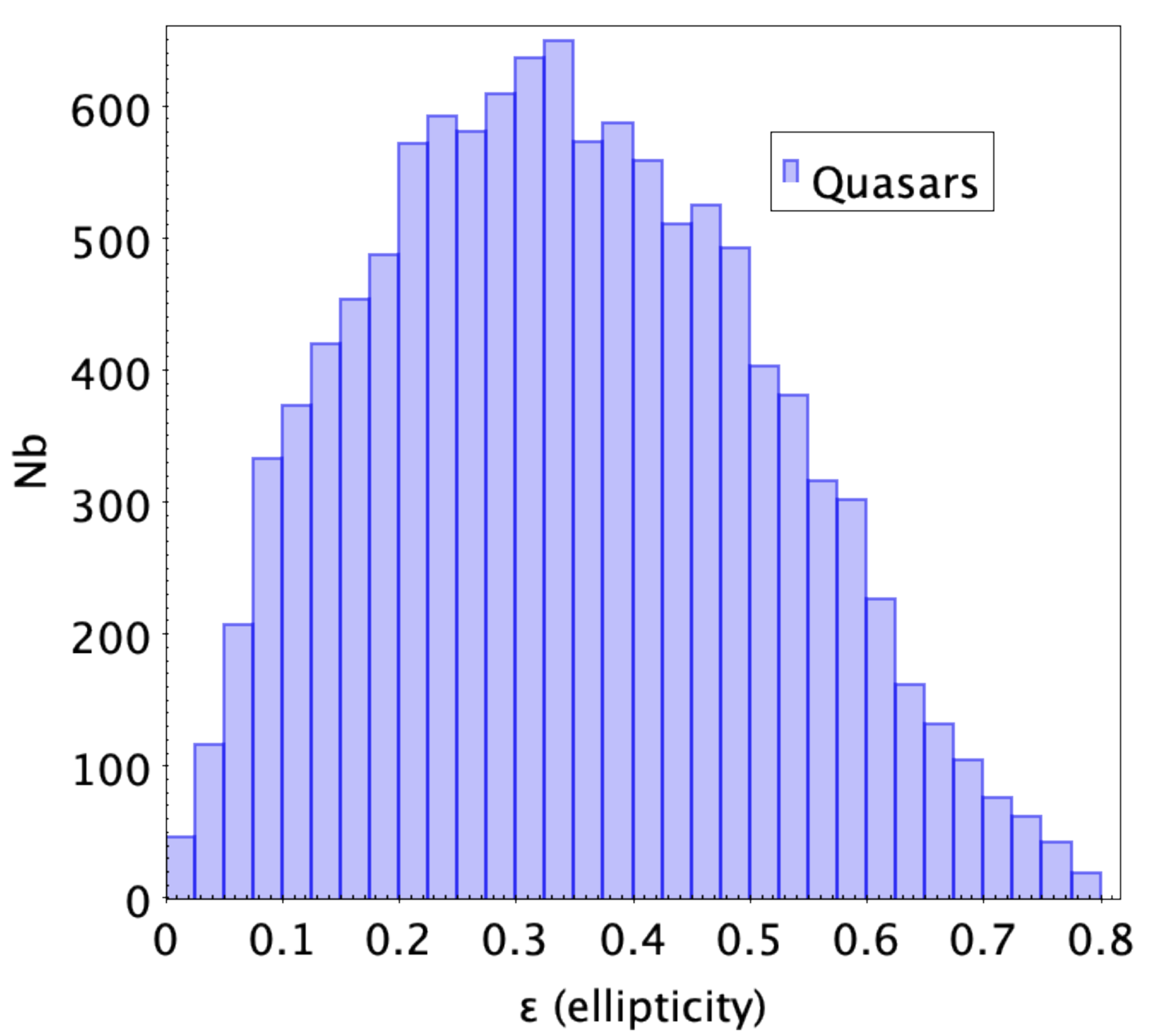} 
     \includegraphics[width=0.33\textwidth]{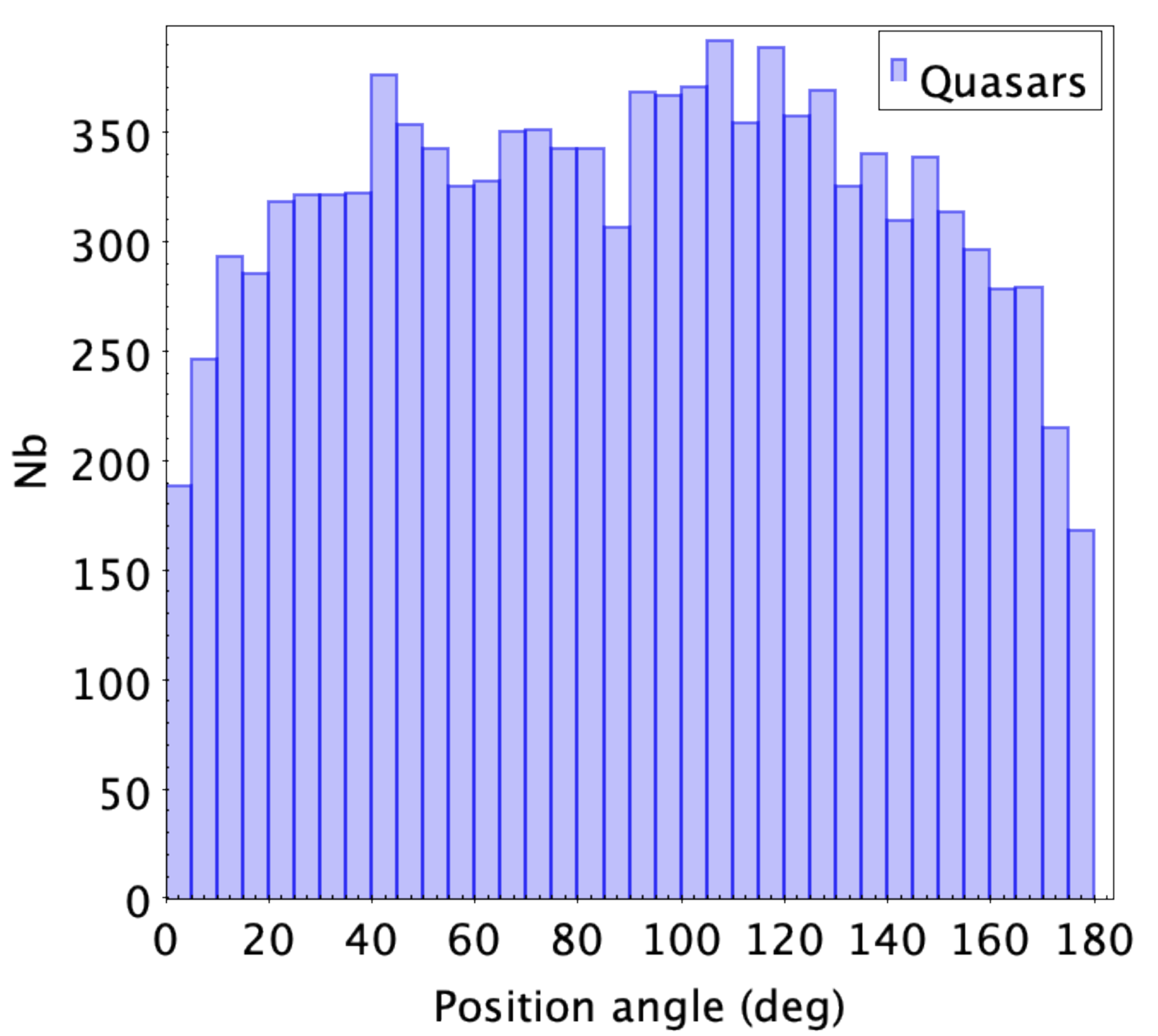}
     }
        \caption{Histograms of the surface brightness profile parameters of the 15\,867 host galaxies of quasars that are published in the \gaia DR3. The bin widths are: 0.2 for \sersic indices, 100 mas for effective radius, variable for effective intensity, 0.025 for ellipticities, and 5 \deg for position angles.} 
     \label{qsodistrib}
  \end{figure*}

From the 1\,103\,691 quasars processed, the vast majority (1\,031\,607) were classified as point-like sources either based on their low integrated flux in the AF data or on the result of the fit. A host galaxy has been detected by our pipeline for 64\,498 sources and for 15\,867 of these, the fitting was satisfying enough to provide all or part of the parameters of the profile in the output table. The distribution of the fitted parameters is given in Figure~\ref{qsodistrib}.

One can observe in this last figure that the \sersic index tends to culminate for n$<$1 which is an indicator that the host galaxies are disc-like. The effective radii culminate around 600-800 mas and are limited to 2\,500 mas. The distribution of position angles is not totally homogeneous, which is unexpected. The pipeline tends to attribute a position angle close to 90$\deg$ when low or no ellipticity is found. The ellipticities culminate around 0.2-0.3 which is what is expected from the projection of random ellipsoids on a sky and what is also observed in an analysis \citep{2022Petit} of the shape of the galaxies from the EAGLE Universe simulation \citep{2015EAGLE}.

\begin{figure}
        \includegraphics[width=0.45\textwidth]{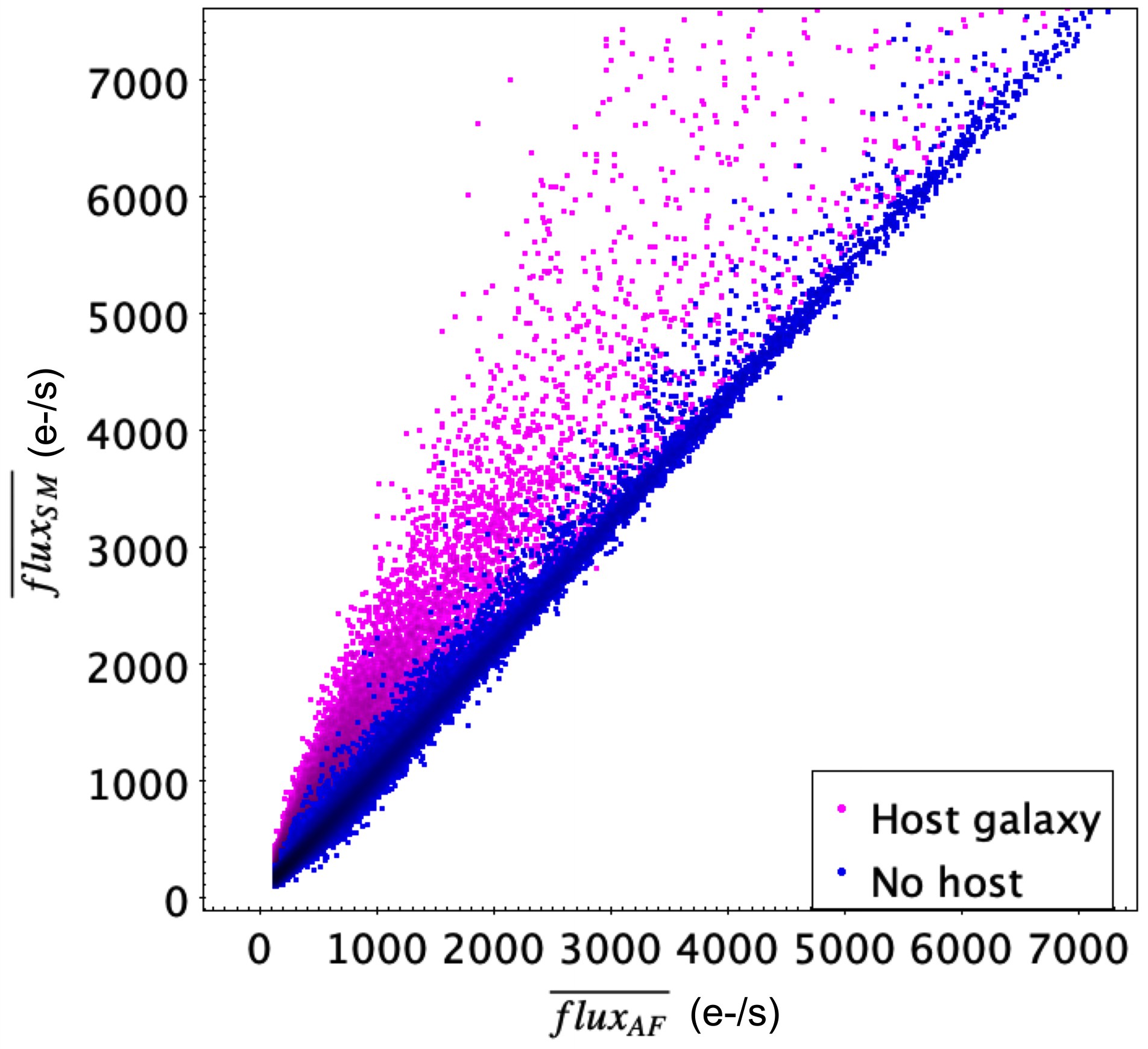}
        \includegraphics[width=0.45\textwidth]{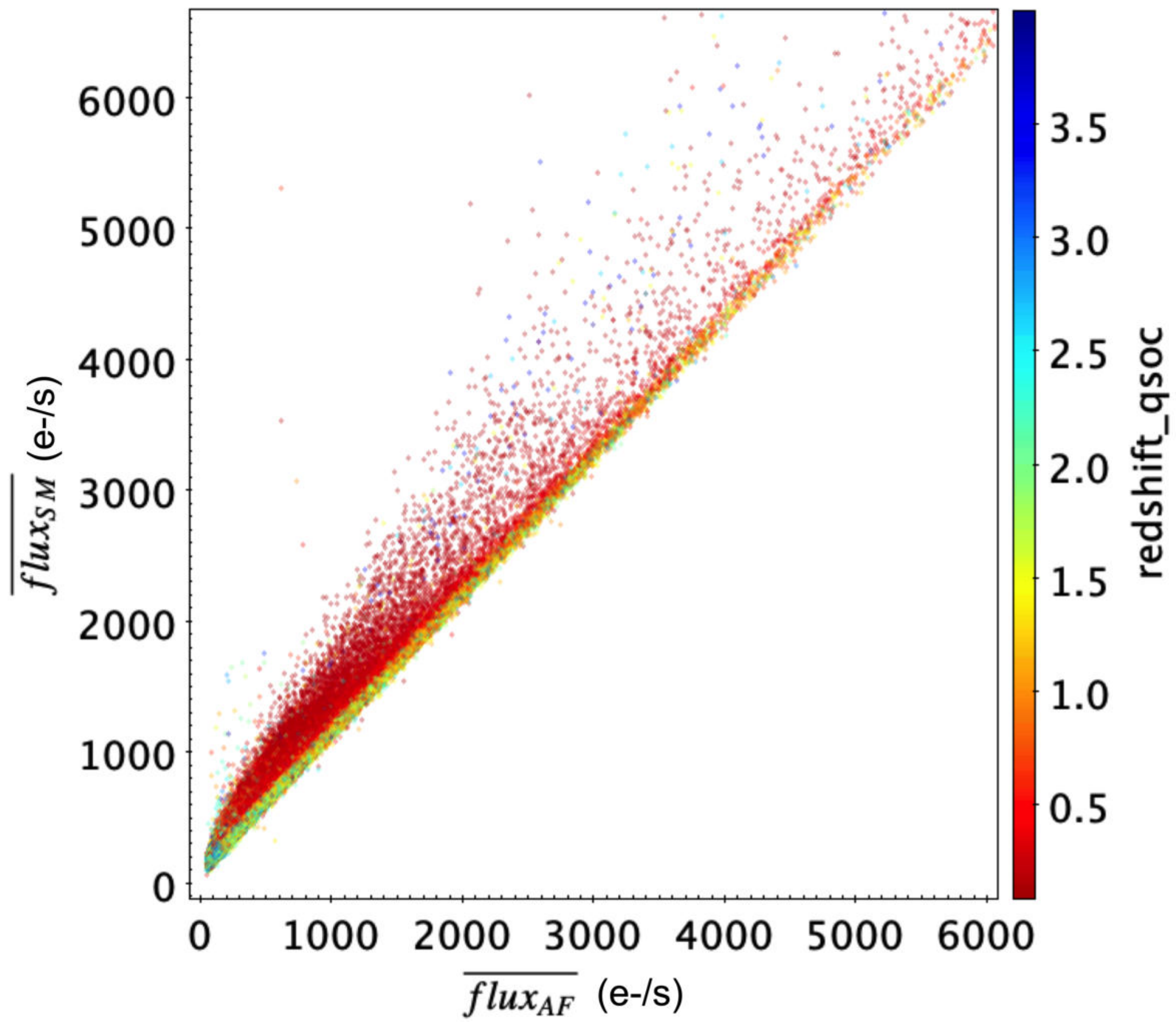}
    \caption {Comparison of the mean integrated flux of the quasars in the AF and the SM windows with indication of the detection of a host galaxy (upper panel) and colour coded with the \gaia (QSOC) redshift (lower panel).} 
    \label{qso_afsm_host}
\end{figure}
\begin{figure}
        \includegraphics[width=0.4\textwidth]{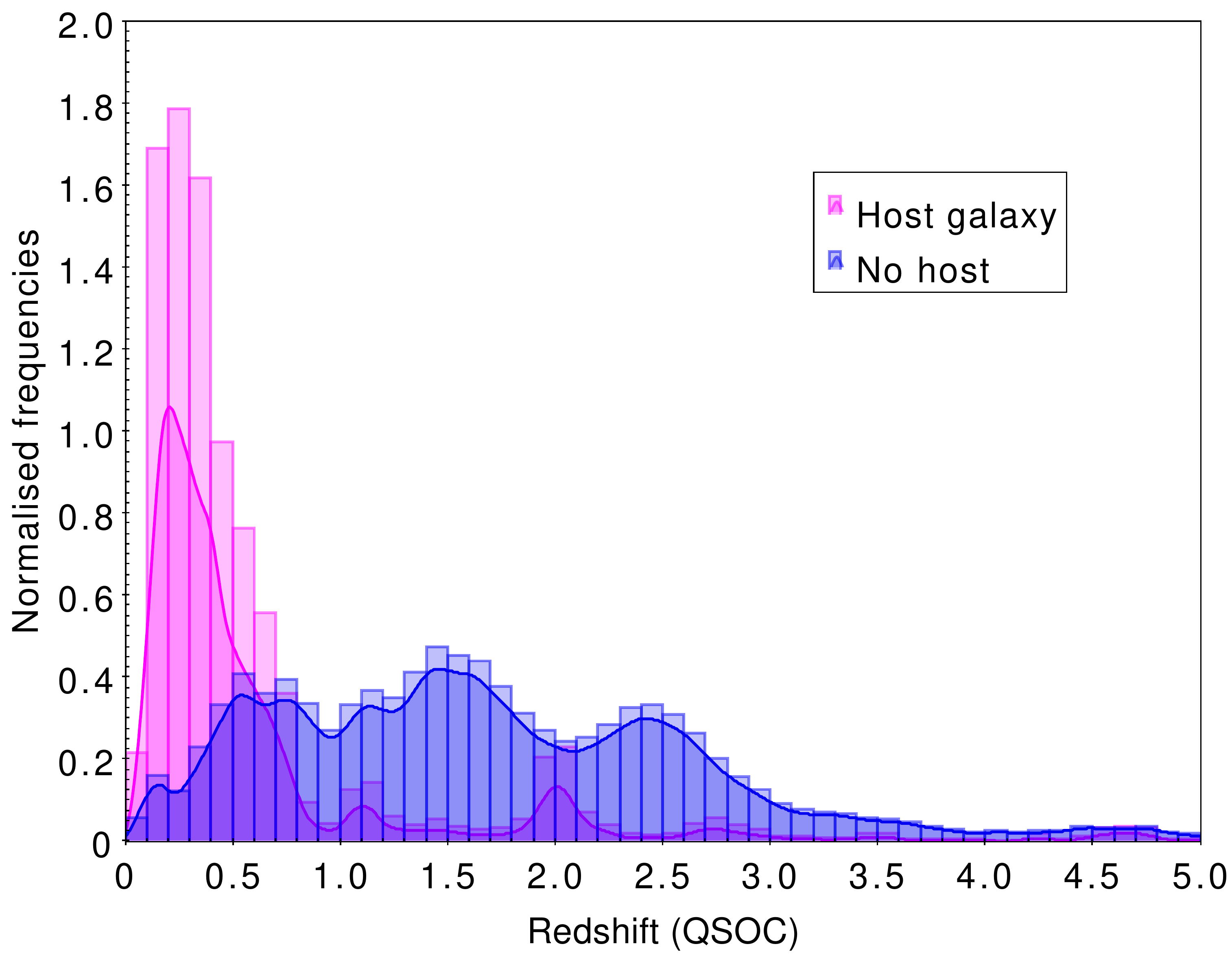}
     \caption {Normalised distribution (by area, bin width=0.1) of the \gaia redshifts (flags\_qsoc=0) for quasars analysed in terms of surface brightness profile.}
     \label{qso_red}
 \end{figure}

One way to validate the results of the quasar analysis is to examine the correlation between the \href{https://gea.esac.esa.int/archive/documentation/GDR3/Gaia_archive/chap_datamodel/sec_dm_extra--galactic_tables/ssec_dm_qso_candidates.html#qso_candidates-host_galaxy_detected}{\textbf{host\_galaxy\_detected}}=`true' flag and the redshift of the sources. One would expect the resolved galaxies hosting quasars to surround nearby quasars while point-like quasars would lie further away. There are $268\,229$ quasars present in our catalogue that benefit from a \gaia Quasar Classifier \citep[QSOC,][]{DelchambreDR3-DPACP-158} redshift with \href{https://gea.esac.esa.int/archive/documentation/GDR3/Gaia_archive/chap_datamodel/sec_dm_extra--galactic_tables/ssec_dm_qso_candidates.html#qso_candidates-flags_qsoc}{\textbf{flags\_qsoc}} = 0.  A host galaxy is detected for 6\,488 of  these objects. In Figure~\ref{qso_afsm_host}, we compare their $\overline{flux_{AF}}$ and $\overline{flux_{SM}}$, colour coded according to the presence or absence of a host galaxy as defined by the pipeline and colour-coded according to the QSOC redshifts. Figure~\ref{qso_red} presents the normalised distributions of the redshifts of the quasars with and without a host galaxy detected.


There is a clear correlation between the two plots of Figure~\ref{qso_afsm_host}: quasars with a host galaxy detected have small redshifts (mean z=0.54) and quasars for which no host galaxy could be resolved have larger redshifts (mean z=1.71), as expected. In very few cases ($\sim$40 sources), the host is detected for larger redshifts. These sources are very faint (\gmag $>$ 20 mag) and suffer either from uncertainties in the light profile fit or in the redshift measurement. 

To the best of our knowledge, there are no HST-based studies analysing the brightness profile of galaxies hosting quasars with objects in common with the \gaia list of quasars processed here. This is essentially due to the different ranges of magnitudes of HST and \gaia which barely overlap. 
We could find comparison data with a ground-based survey adjusting a free S\'ersic profile on sources from the GAMMA survey \citep{2011Robotham}, from the NASA-Sloan-Atlas \citep{2009Maller} (hereafter NASATLAS), and from the work of \cite{2011Simard} which is based on SDSS data. Nevertheless, these surveys considered the quasar with its host galaxy as a whole. In these works, the quasar  drastically influences the fitted profile, preventing any comparison with our analysis. Even the radii cannot be compared because the S\'ersic index is sensitive to the concentration of the light profile and the effective radius is strongly linked to the index of the profile: the smaller the index, the smaller the effective radius.

The distribution of the S\'ersic indices in Figure \ref{qsodistrib} exhibits a peak around an index of $\sim$ 0.8 and a mean value of $\sim$ 1.9 which is consistent with quasars being hosted by galaxies with disc-like profiles. This result is in agreement with a recent study of the sizes of galaxies hosting quasars in the Hyper Suprime-Cam Subaru Strategic Program \citep{2021Li}.

\subsection{Galaxies}
From the $940\,887$ galaxies processed by the pipeline, we were able to derive a valid result with clear convergence for $914\,837$ of them. These are the sources published in the output table.

\begin{figure*}
      \centerline{
      \includegraphics[width=0.33\textwidth]{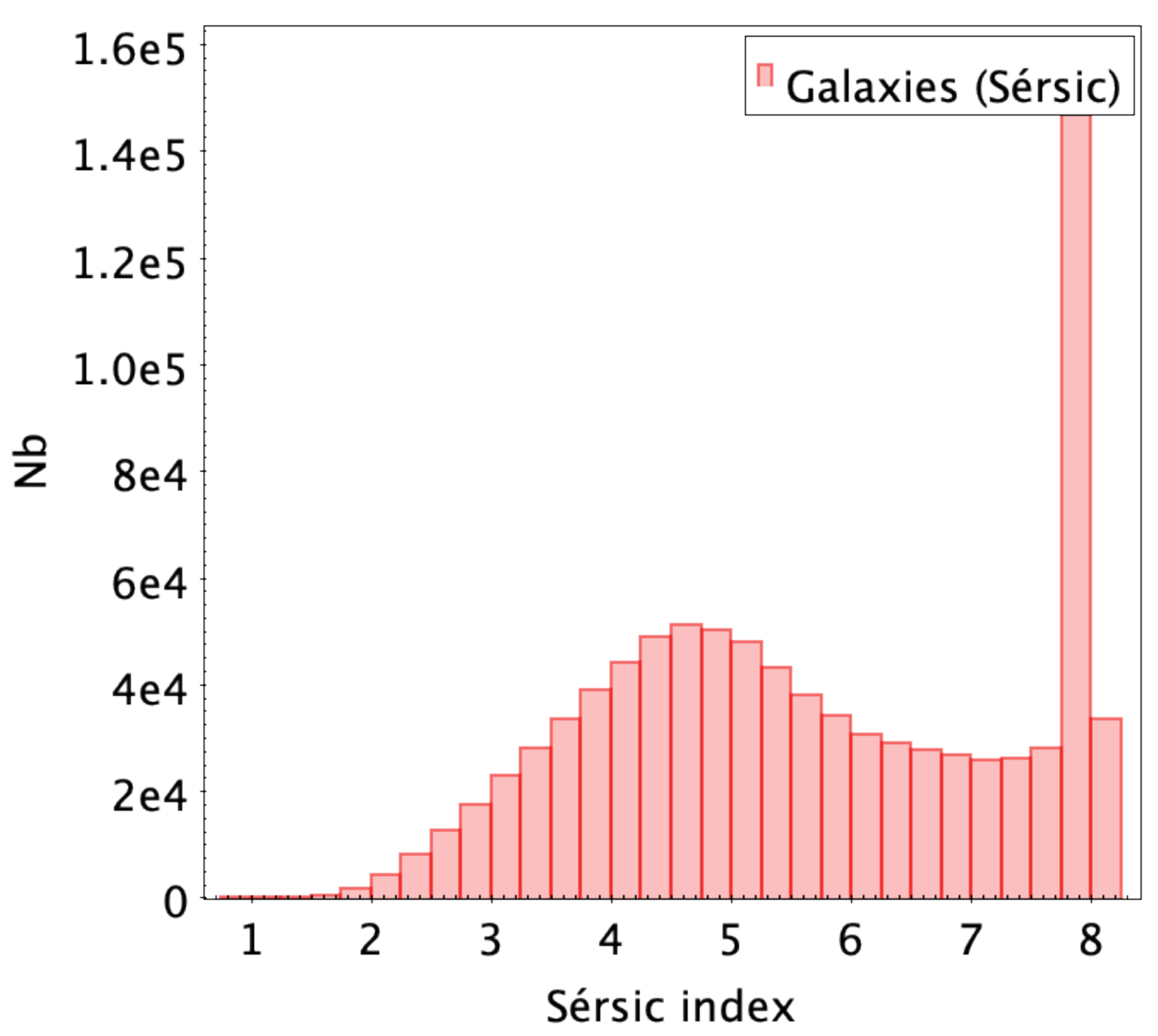} 
      \includegraphics[width=0.33\textwidth]{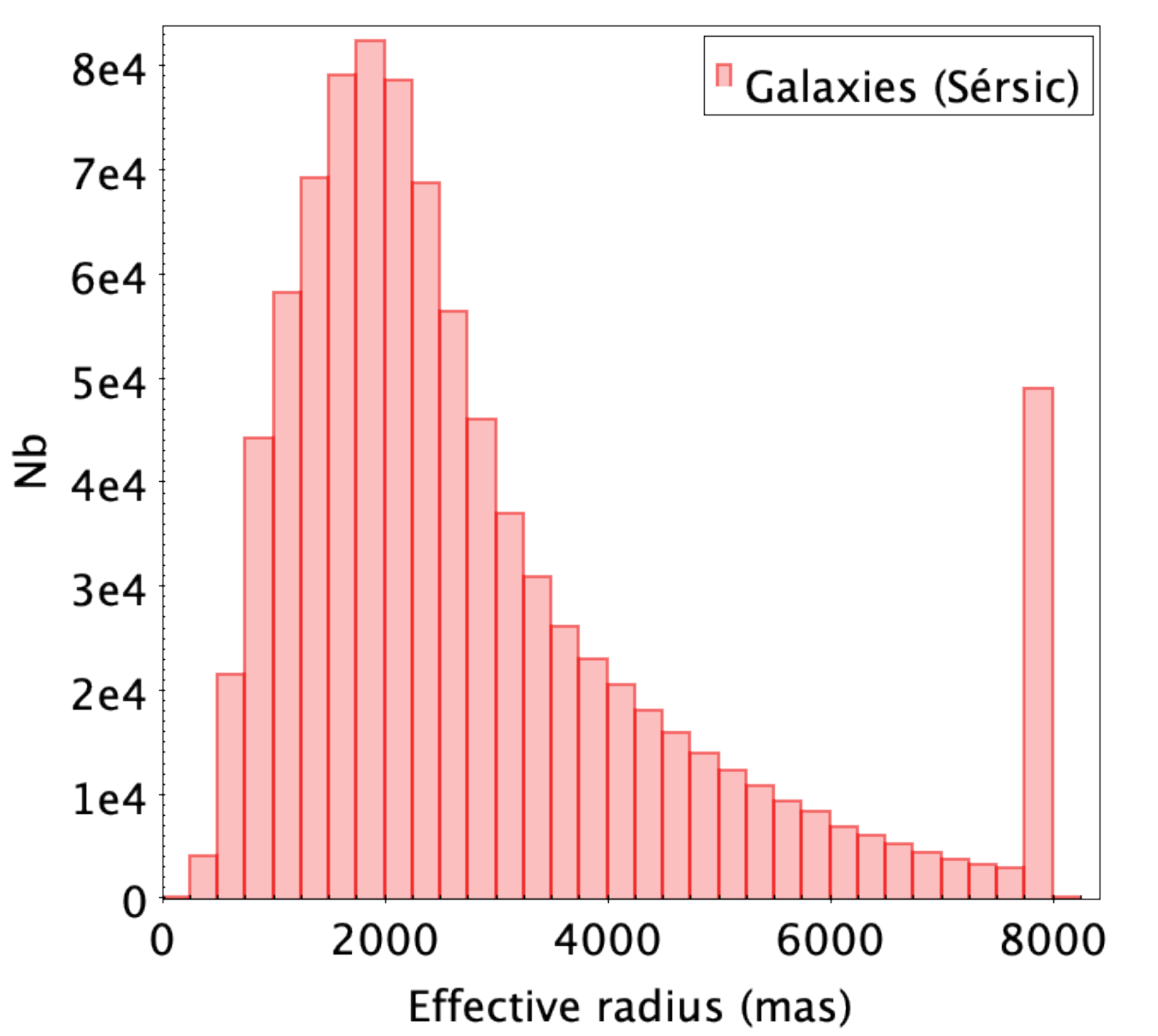} 
      \includegraphics[width=0.33\textwidth]{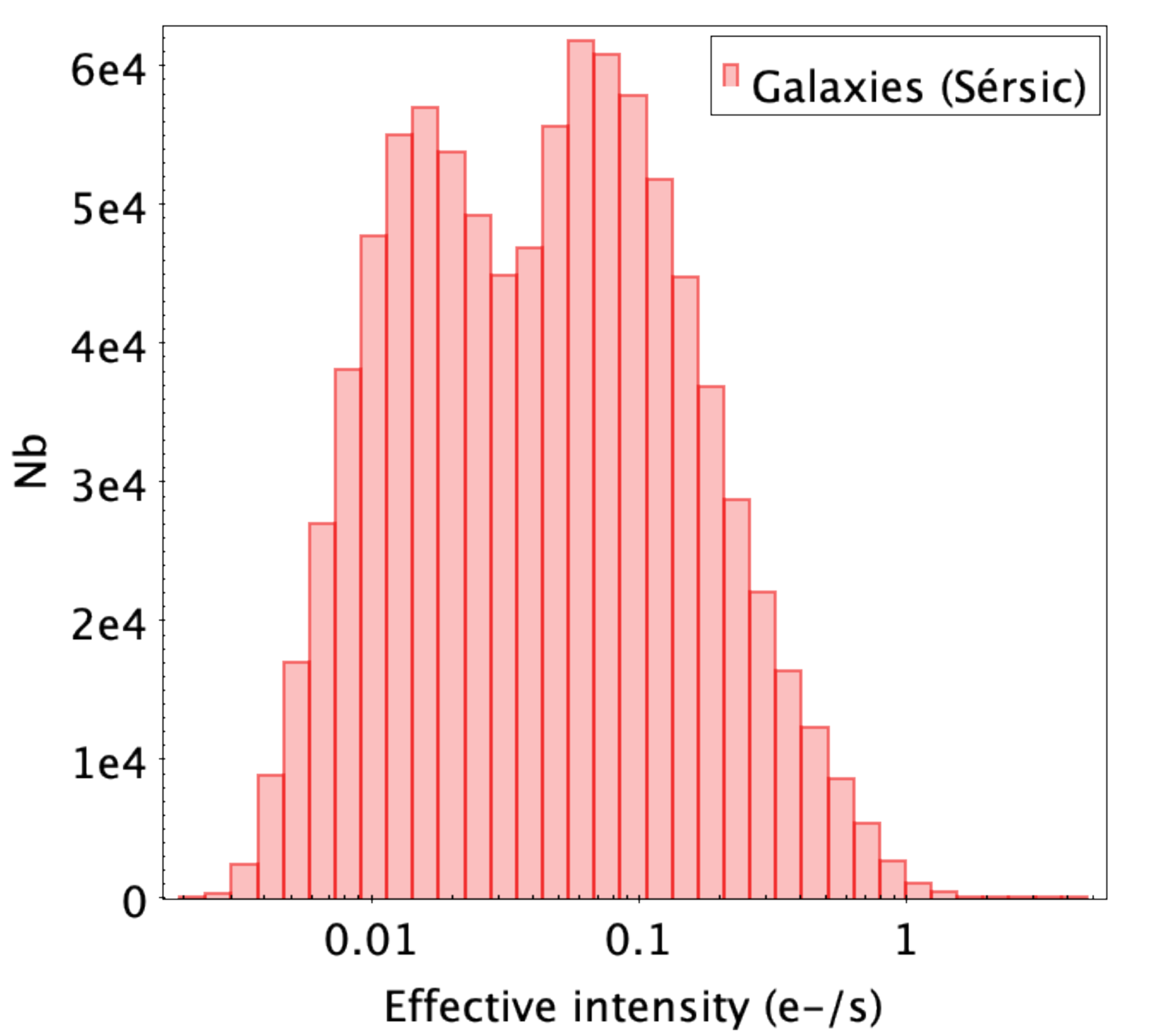}
      }
      \centerline{
      \includegraphics[width=0.33\textwidth]{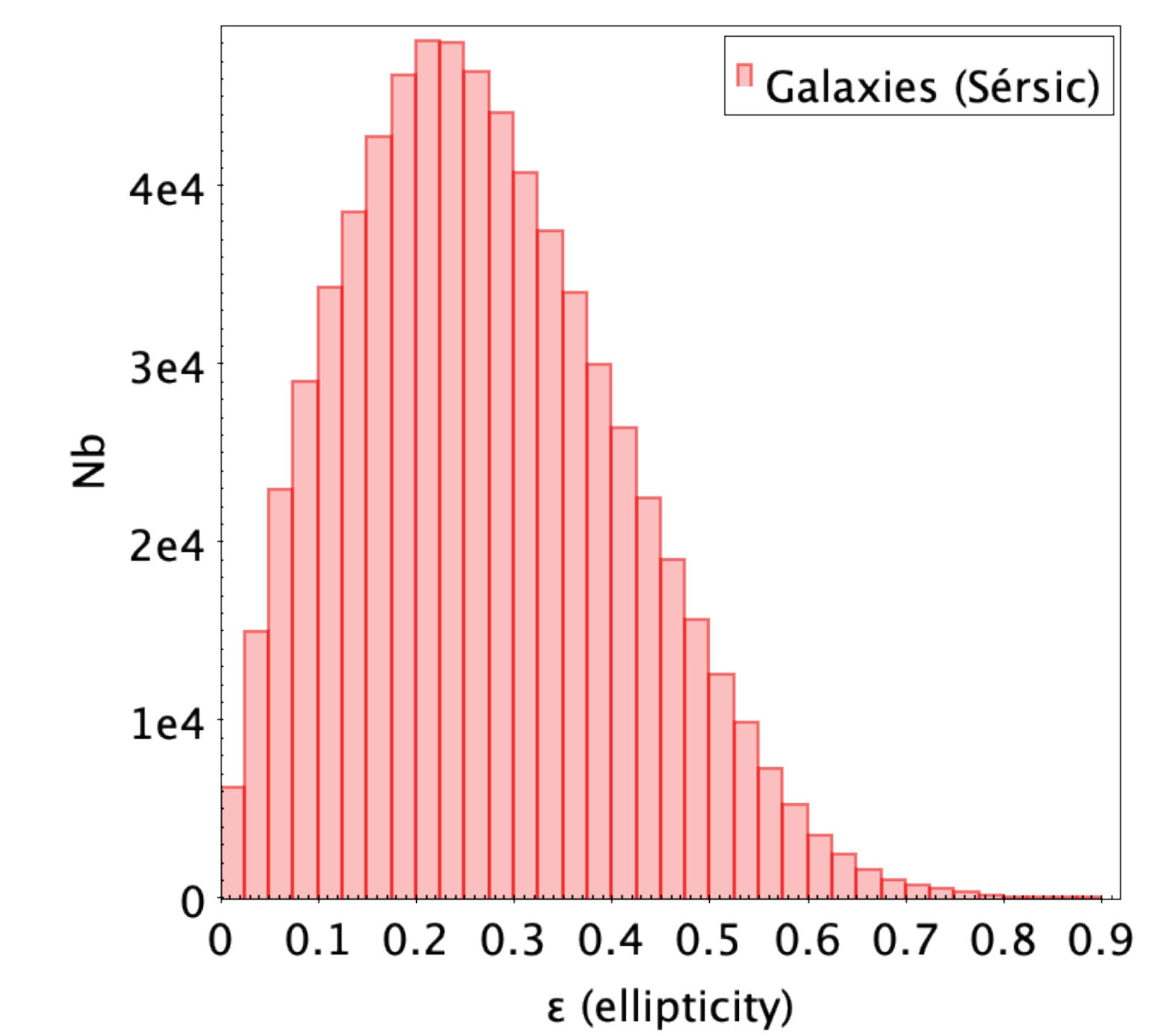}
      \includegraphics[width=0.33\textwidth]{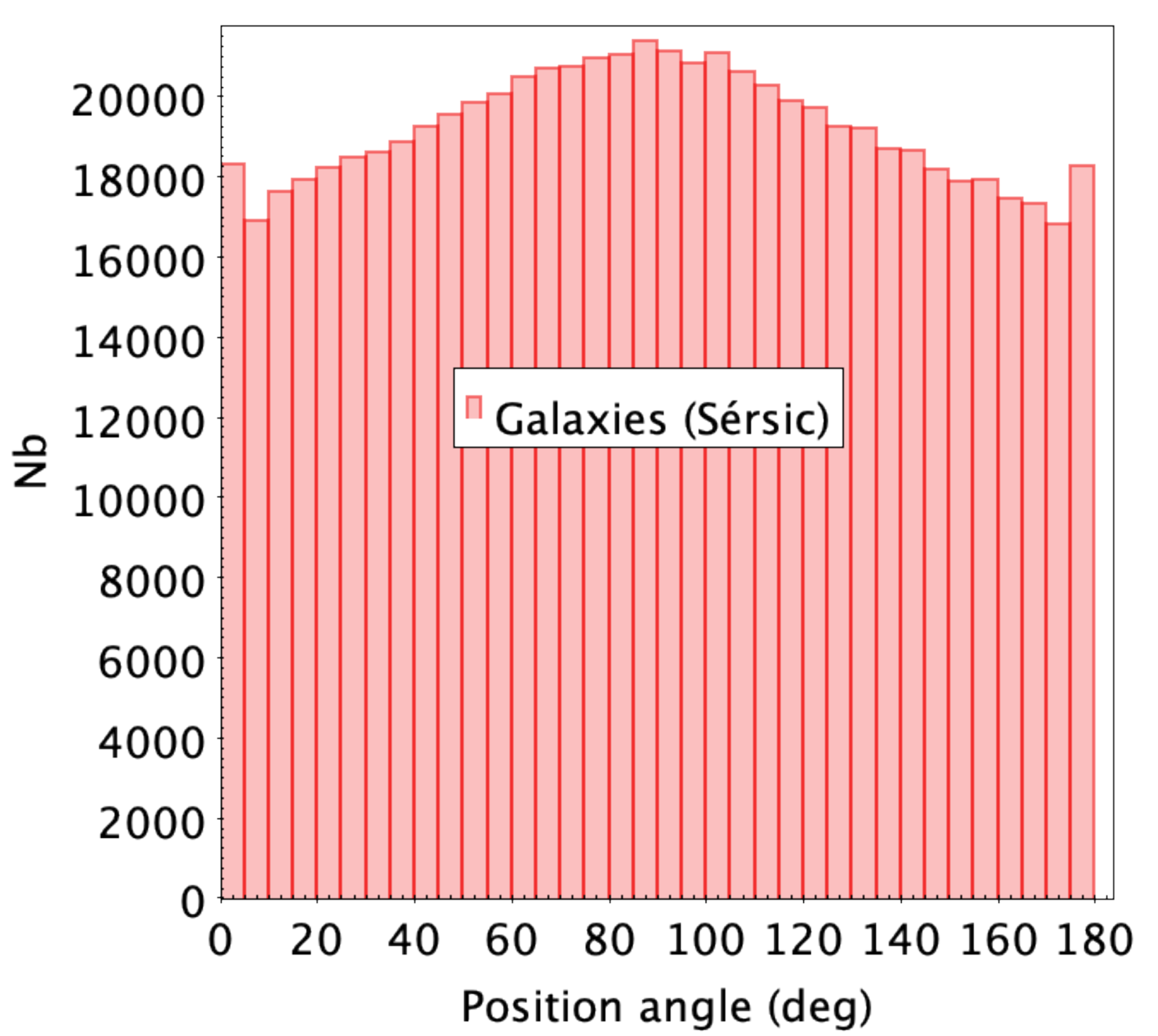} 
      }
      \caption{Distribution of the parameters fitted with a S\'ersic profile on galaxies. The bin widths are: 0.25 for \sersic indices, 250 mas for effective radius, variable for effective intensity, 0.025 for ellipticities, and 5 \deg for position angle.}
      \label{gal_distrib_s}
\end{figure*}

\begin{figure*}
      \centerline{
      \includegraphics[width=0.4\textwidth]{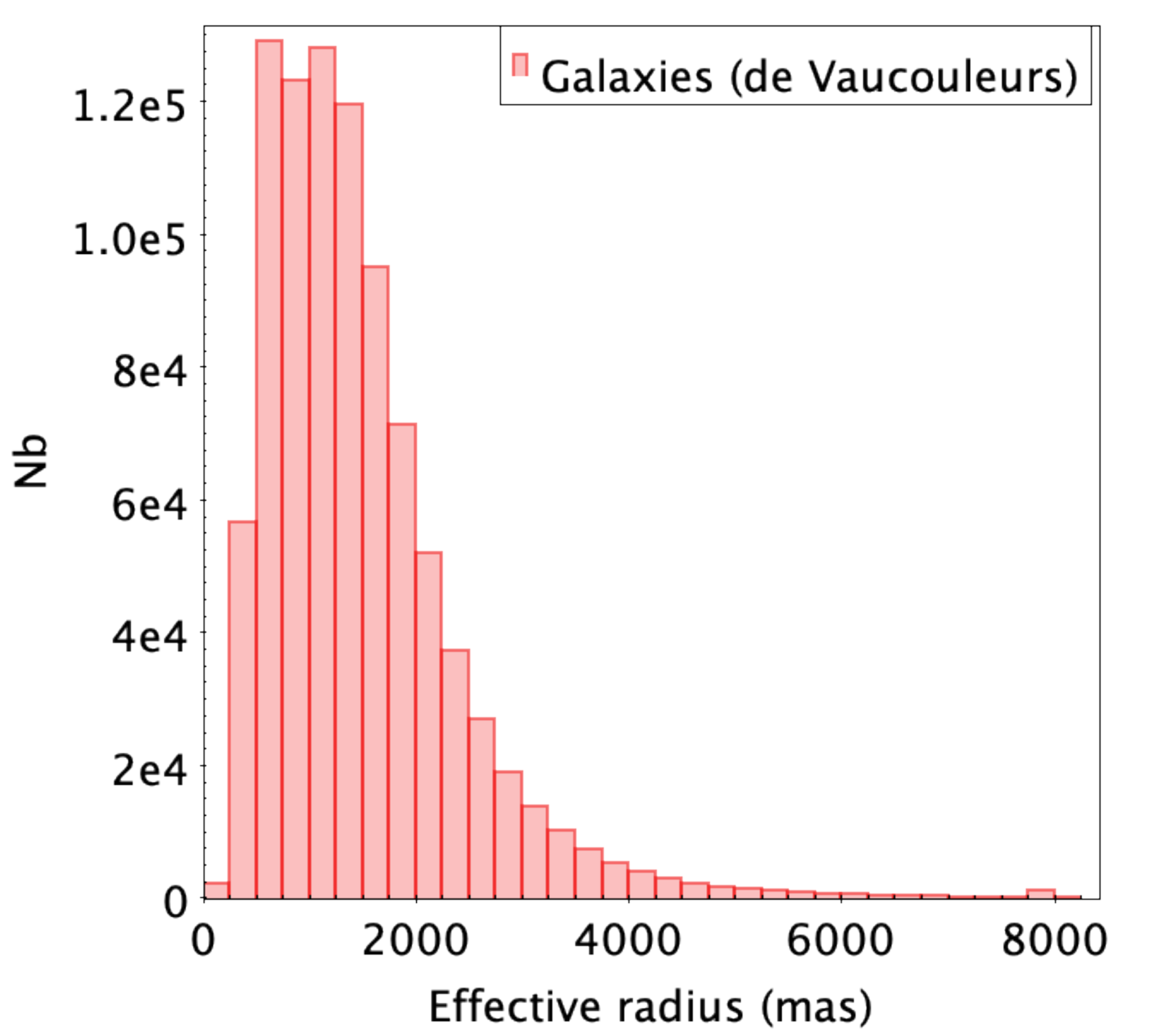} 
      \includegraphics[width=0.4\textwidth]{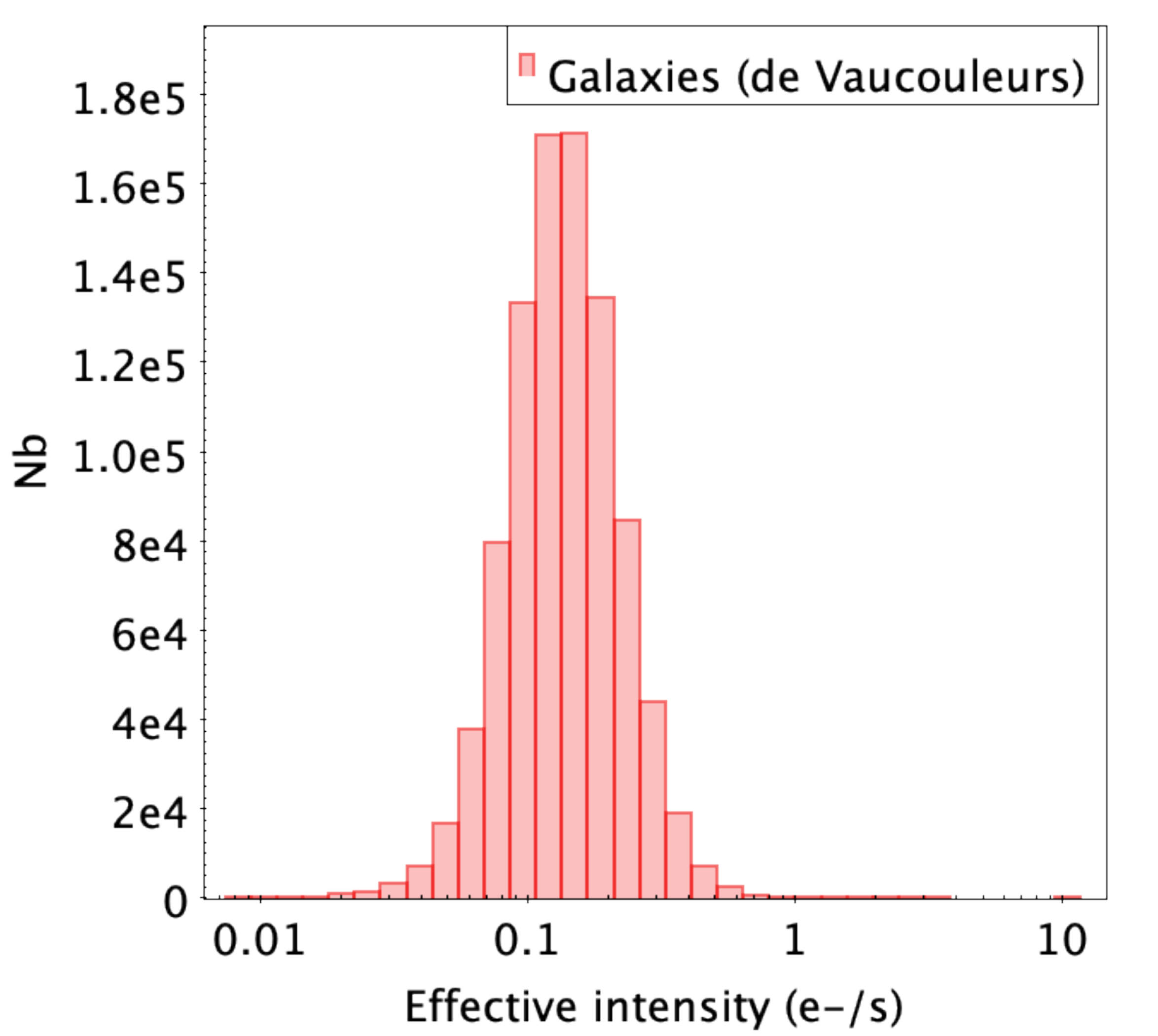}
      }
      \centerline{
      \includegraphics[width=0.4\textwidth]{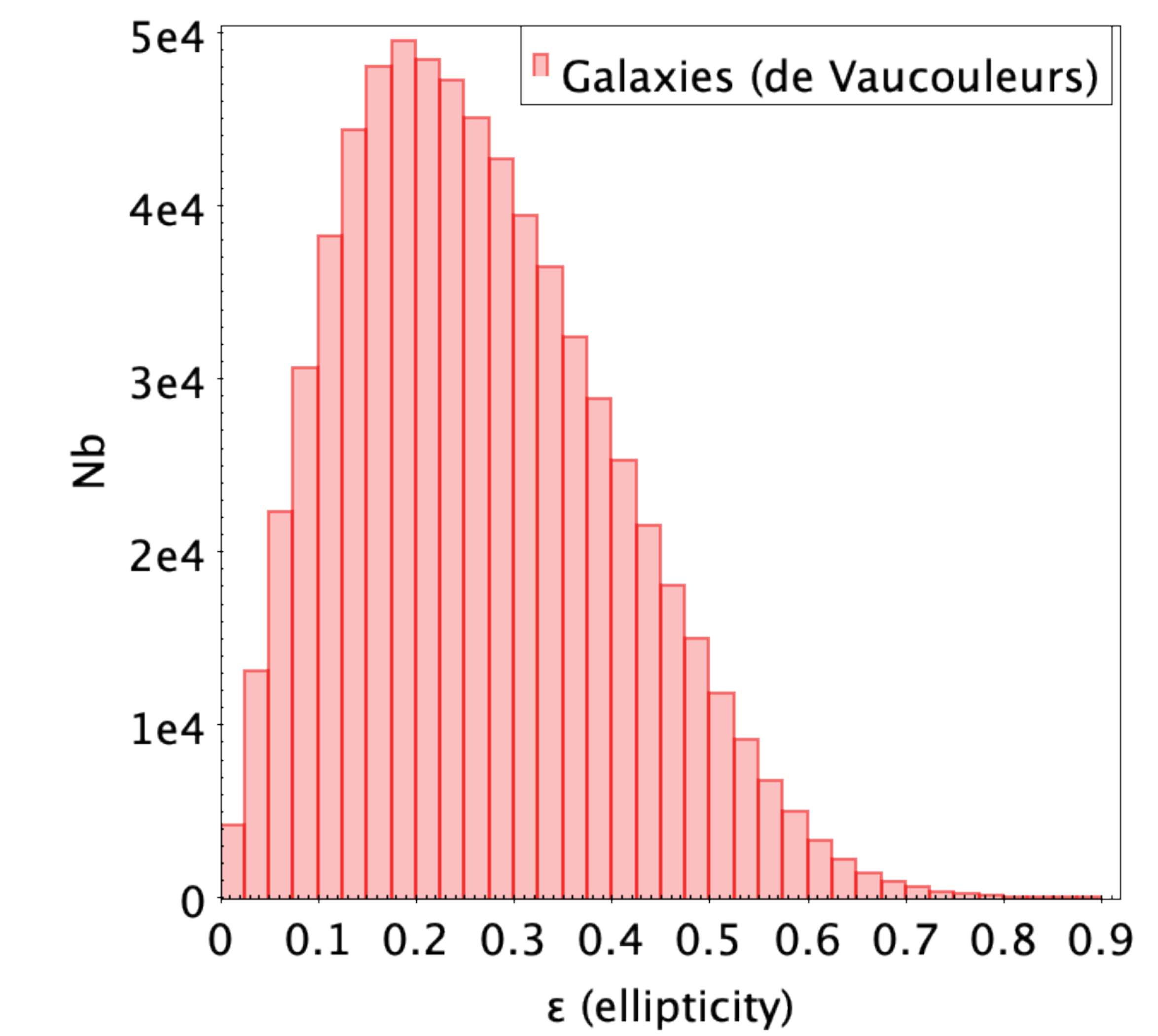}
      \includegraphics[width=0.4\textwidth]{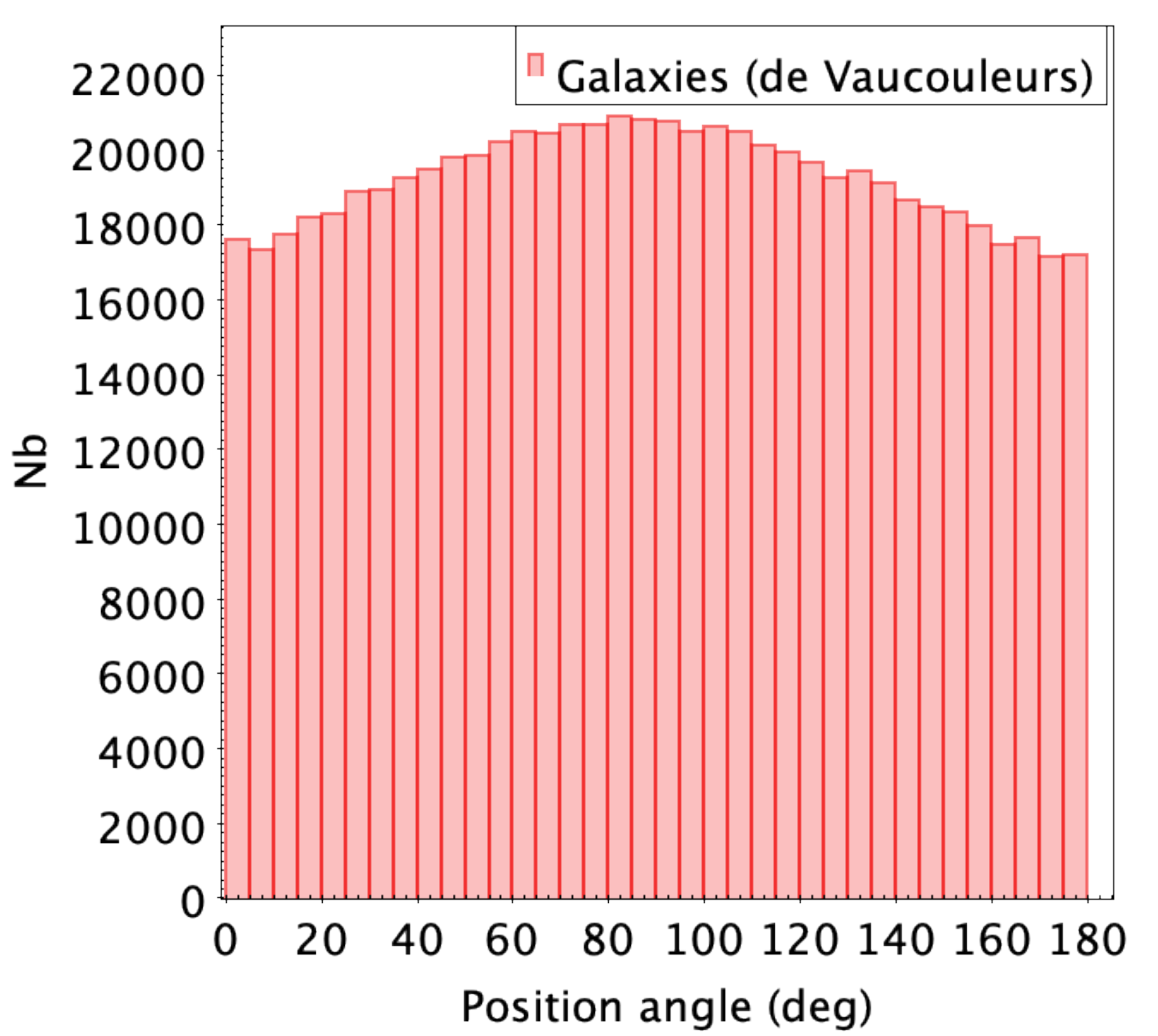} 
      }
      \caption{Distribution of the parameters fitted with a de Vaucouleurs profile on galaxies. The bin widths are: 250 mas for effective radius, variable for effective intensity, 0.025 for ellipticities and, 5\deg\, for position angle.}
      \label{gal_distrib_s4}
 \end{figure*}

The distributions of the fitted parameters are presented in Figure~\ref{gal_distrib_s} for the S\'ersic profile and in Figure~\ref{gal_distrib_s4} for the de Vaucouleurs profile. 

\subsubsection{S\'ersic profile} 
Most galaxies measured have a \sersic index of between 4 and 5, which is typical for elliptical galaxies and is coherent with the theoretical predictions \citep[][]{2014deSouza, 2015deBruijne} that \gaia would filter out disc galaxies. A few thousand galaxies have an index of below 2, indicative of disc galaxies or pseudo-bulge plus disc and only a few hundred of them have an index of below 1.5. The accumulation of indices of around 8 corresponds to small sources that are not well described by a free \sersic profile. Nevertheless, as seen in Figure \ref{sersic}, the light profiles with indices n$=$4 are very similar to profiles with higher indices, which means that the galaxies fitted with n=8 can be considered as elliptical as well, the slight remaining background light being interpreted as wings by the algorithm. The distribution of effective radius of the \sersic profile has a peak value of around 2000 mas. These radii are not comparable from source to source, or with other models because the radius is linked to the \sersic index: the larger the index, the larger the radius. There is an accumulation of radii at 8000 mas which is our upper limit of investigation, meaning that these galaxies are eventually larger. The effective intensity exhibits a bimodal distribution which corresponds to the circular and the elliptical fitting of the sources. This is because the intensity is calculated at the major-axis effective radius for elliptical profiles. 

\begin{figure}
    \includegraphics[width=0.45\textwidth]{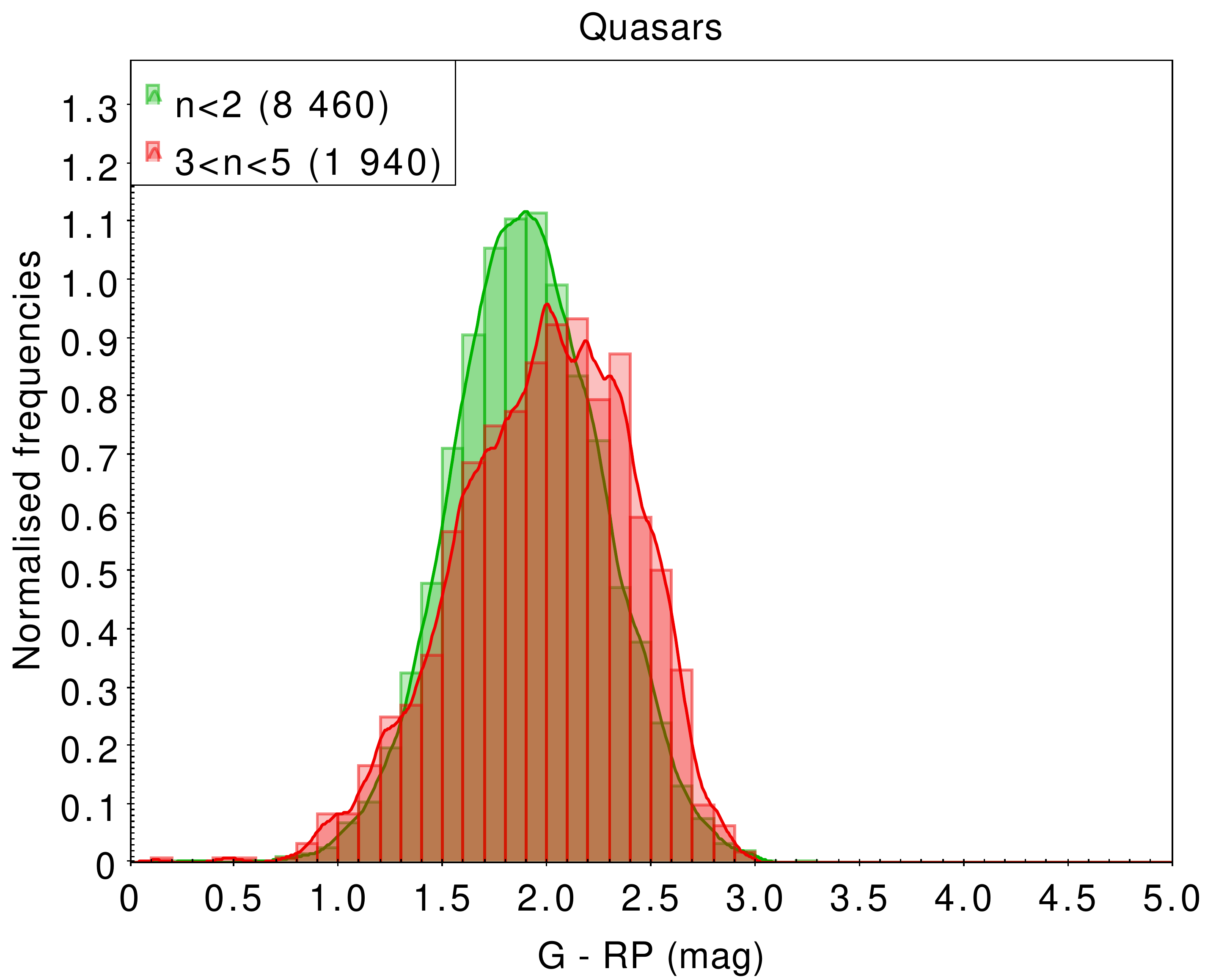}
    \includegraphics[width=0.45\textwidth]{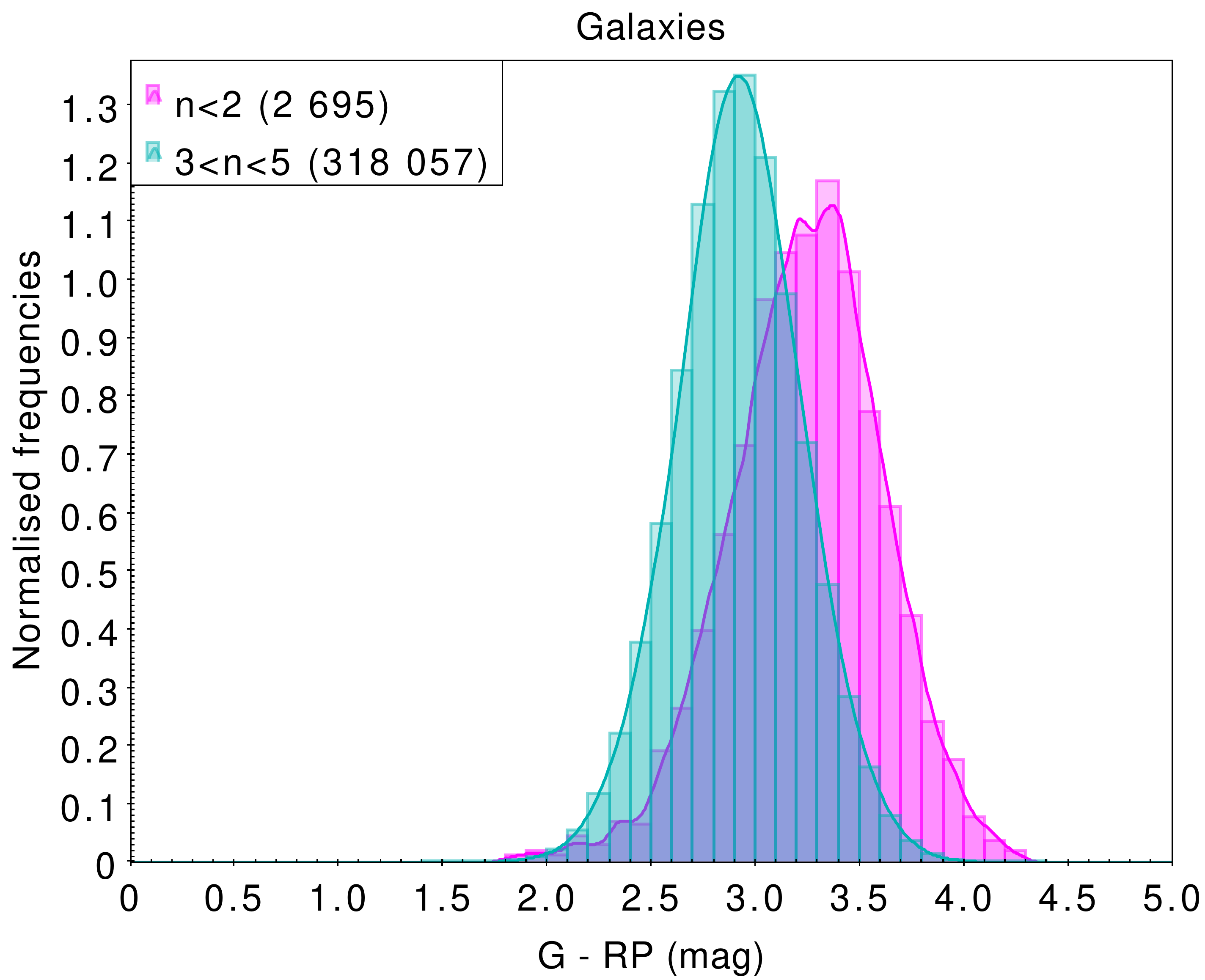} 
    \caption{Normalised distribution (by area) of the colours G-RP (phot\_g\_mean\_mag - phot\_rp\_mean\_mag) of the host galaxies of quasars and of galaxies for sources selected based on their \sersic index (bin width=0.1 mag). Sources with n$<$2 are expected to be disc-like while those with n$\sim$4 are expected to be elliptical.}
    \label{colour}
\end{figure}

Figure~\ref{colour} presents the distribution of colours G-RP of galaxies hosting quasars and of galaxies for sources selected based on their \sersic index: n$<$2 for disc galaxies and n$\sim$4 for elliptical galaxies. We observe a dependence of the \sersic index along the colour G-RP of the sources. Disc-like galaxies that are well represented by index$<$2 appear bluer than elliptical galaxies (index$\sim$4) as expected from the ongoing star formation in discs. This effect also concerns galaxies hosting quasars. 

\begin{figure}
    \includegraphics[width=0.45\textwidth]{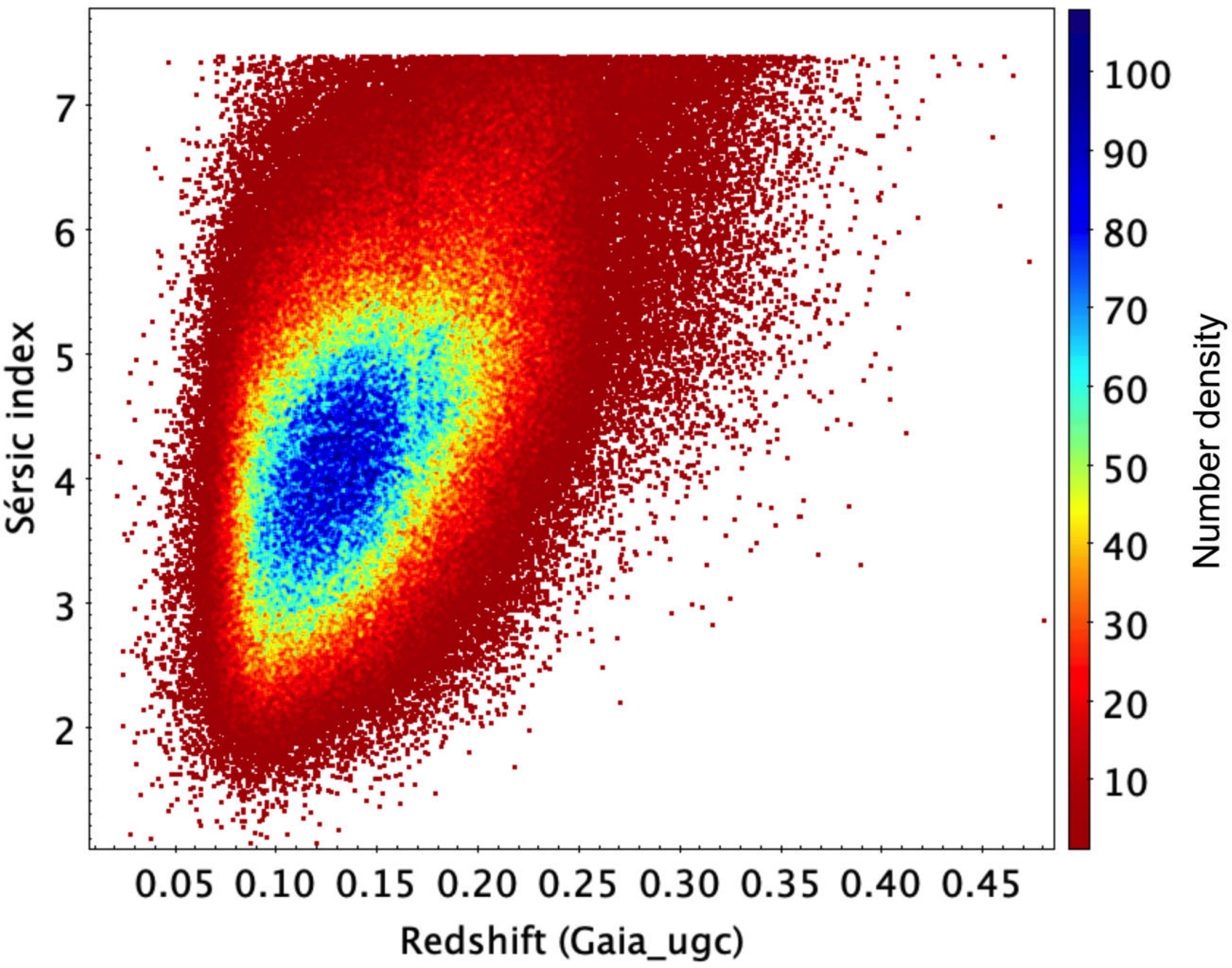} 
    \caption{Density plot of the distribution of the \sersic indices of  fitted galaxy  profiles as a function of the redshift measurement from \gaia (UGC).}
    \label{gal_n_red}
\end{figure}

From the $914\,837$ galaxies successfully processed, $388\,552$ benefit from a redshift measurement from the \gaia Unresolved Galaxy Classifier \citep[UGC,][]{DelchambreDR3-DPACP-158}. 
Figure~\ref{gal_n_red} presents the distribution of \sersic indices fitted as a function of the \gaia redshifts.
There is a clear dependence of the \sersic index on redshift. This is not completely expected but results from the apparent size of the sources: the most distant sources appear more compact and are fitted with large indices, while the closest galaxies are better represented by various indices corresponding to the difference in light concentration in their inner part.

\begin{figure*}
    \includegraphics[width=0.33\textwidth]{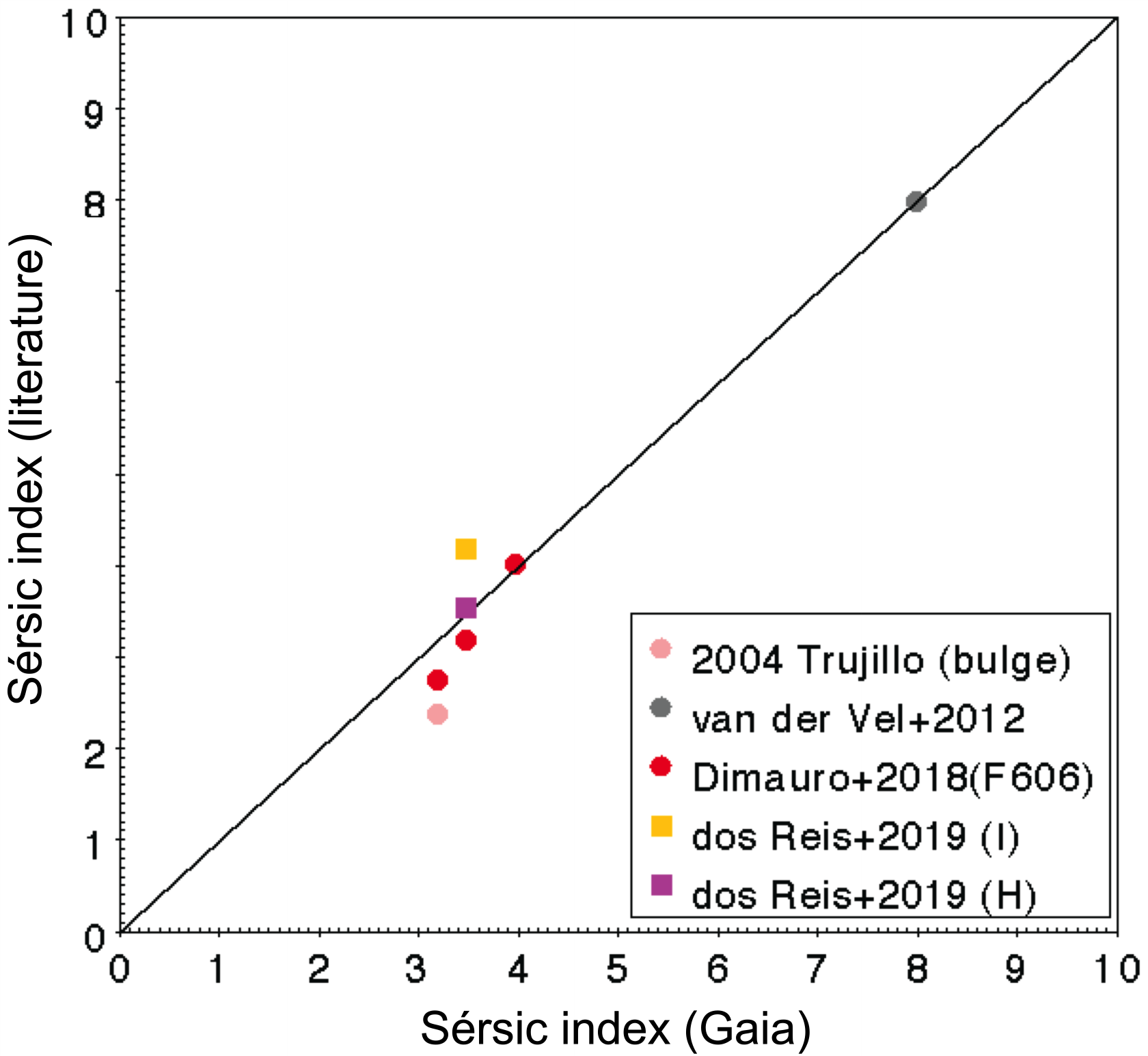} 
    \includegraphics[width=0.33\textwidth]{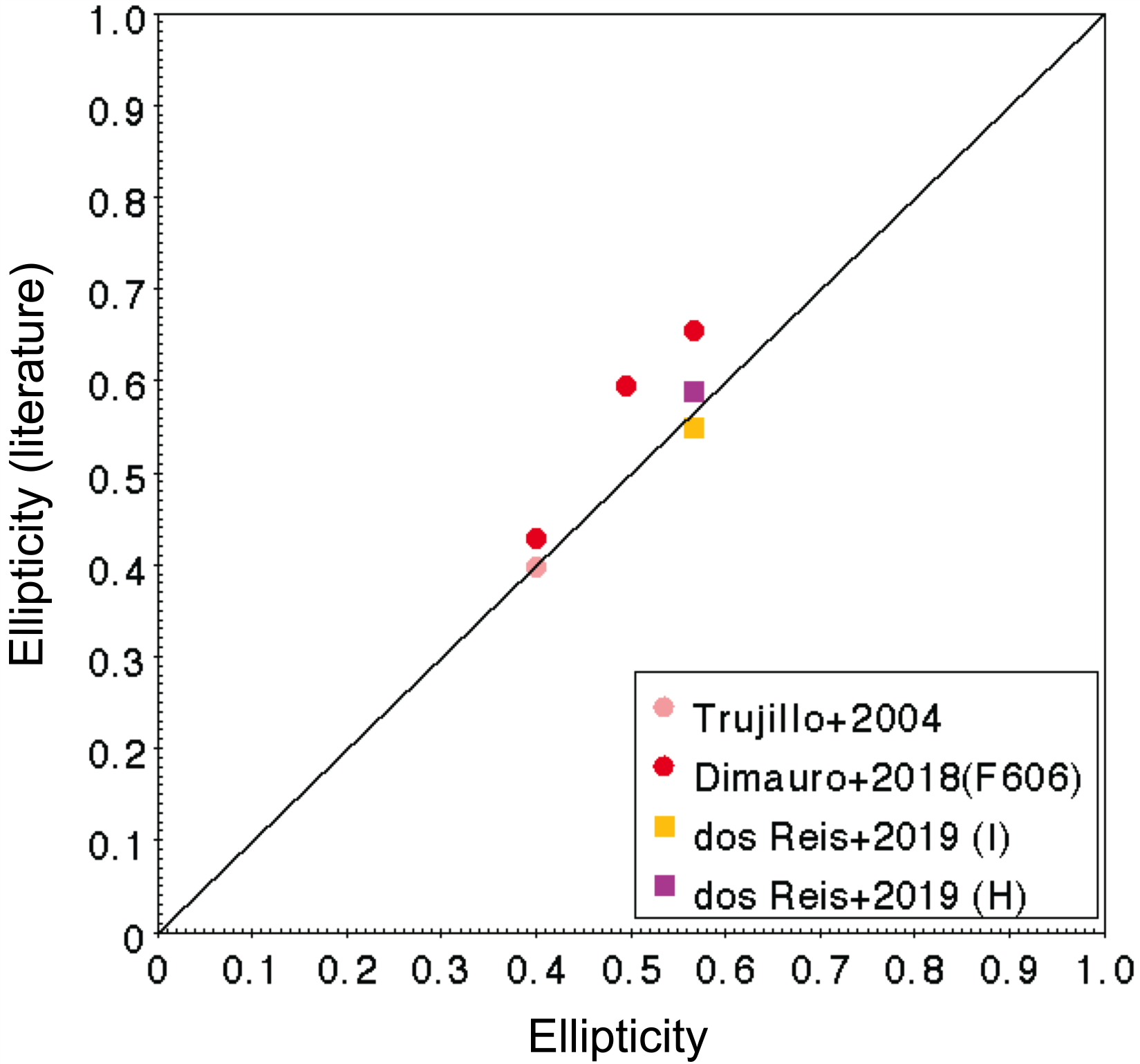}
    \includegraphics[width=0.33\textwidth]{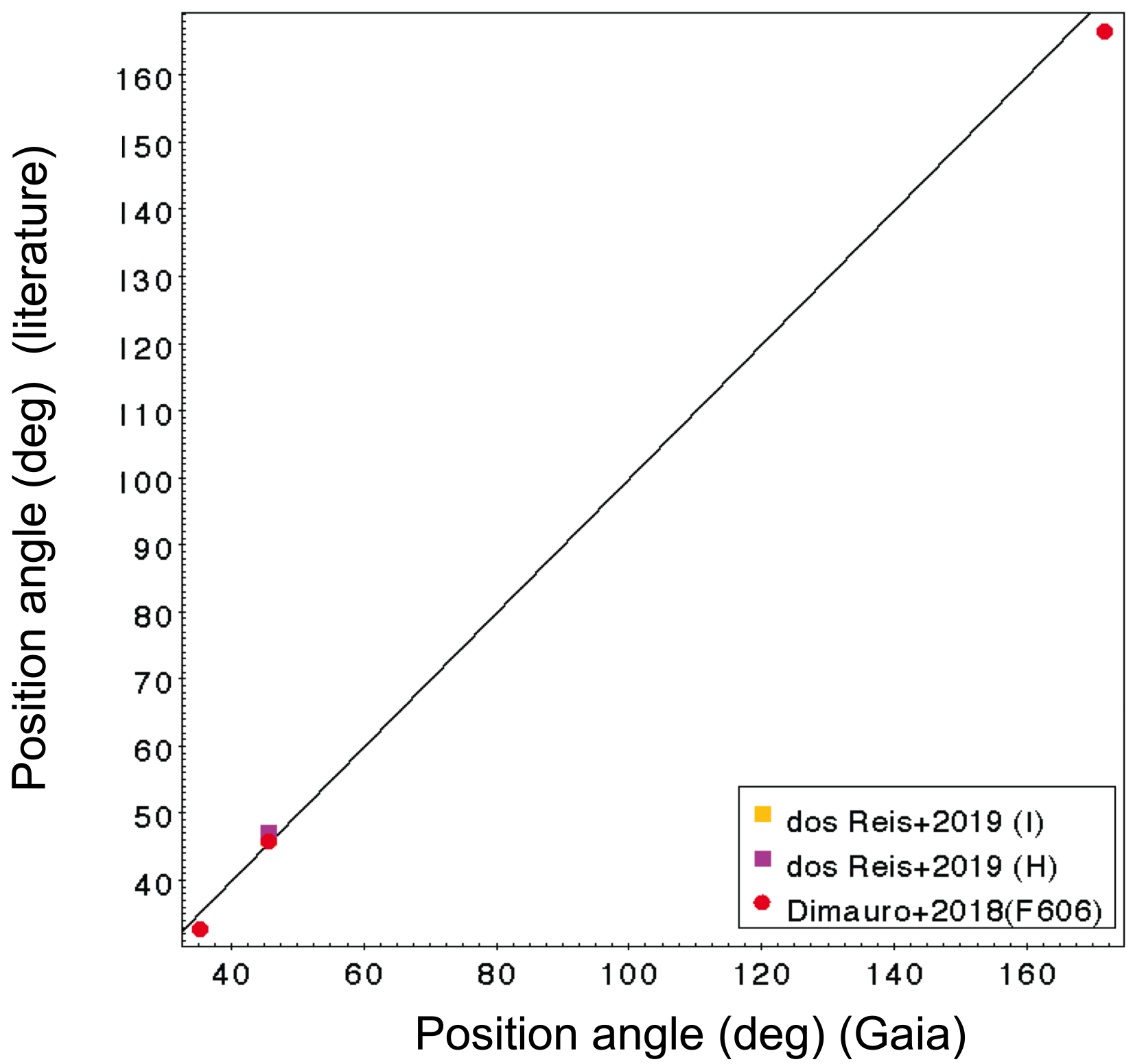}
    \caption{Comparison of the fitted S\'ersic indices, ellipticities, and position angles of galaxies (S\'ersic profile) with works based on HST observations, while fitting S\'ersic profiles.}
    \label{gal_hst}
\end{figure*}

To perform an external validation of the fitted profiles, it would be desirable to have space-based studies for comparison.
While analysing the space-based studies of galaxy morphology that fit a S\'ersic profile and use HST data, only five galaxies were found in common with our list of processed sources: one in common with \cite{2004Trujillo}, one with \cite{2012vanDerWel}, two with \cite{2018Dimauro}, and one with \cite{2020dosReis}. The comparison of the fitted S\'ersic indices, ellipticities, and position angles with these studies is given in Figure~\ref{gal_hst}. This comparison, although very limited because of the small number of objects, shows good agreement between these studies and the \gaia parameters.

\begin{figure}
    \includegraphics[width=0.45\textwidth] {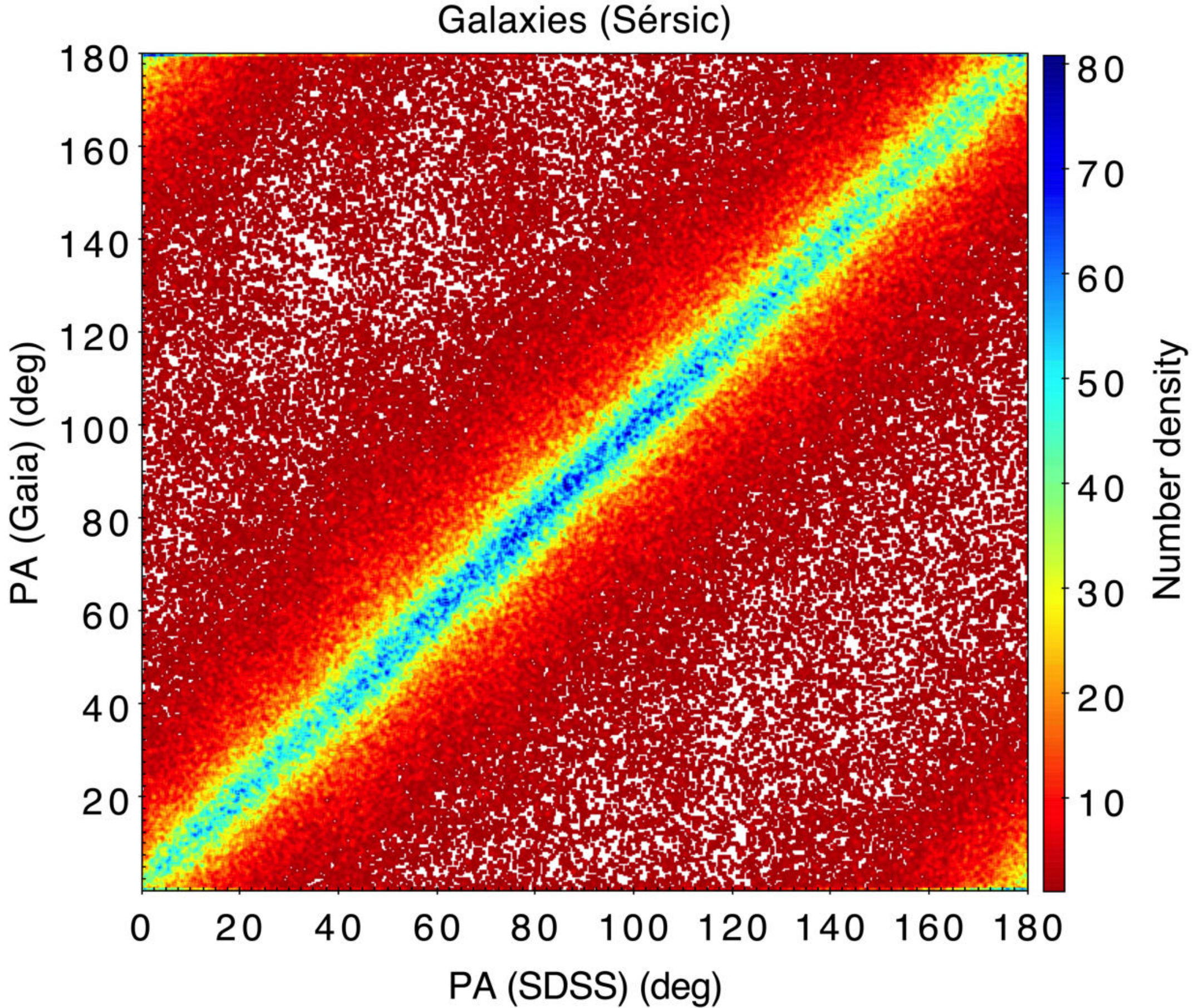} 
    \includegraphics[width=0.45\textwidth] {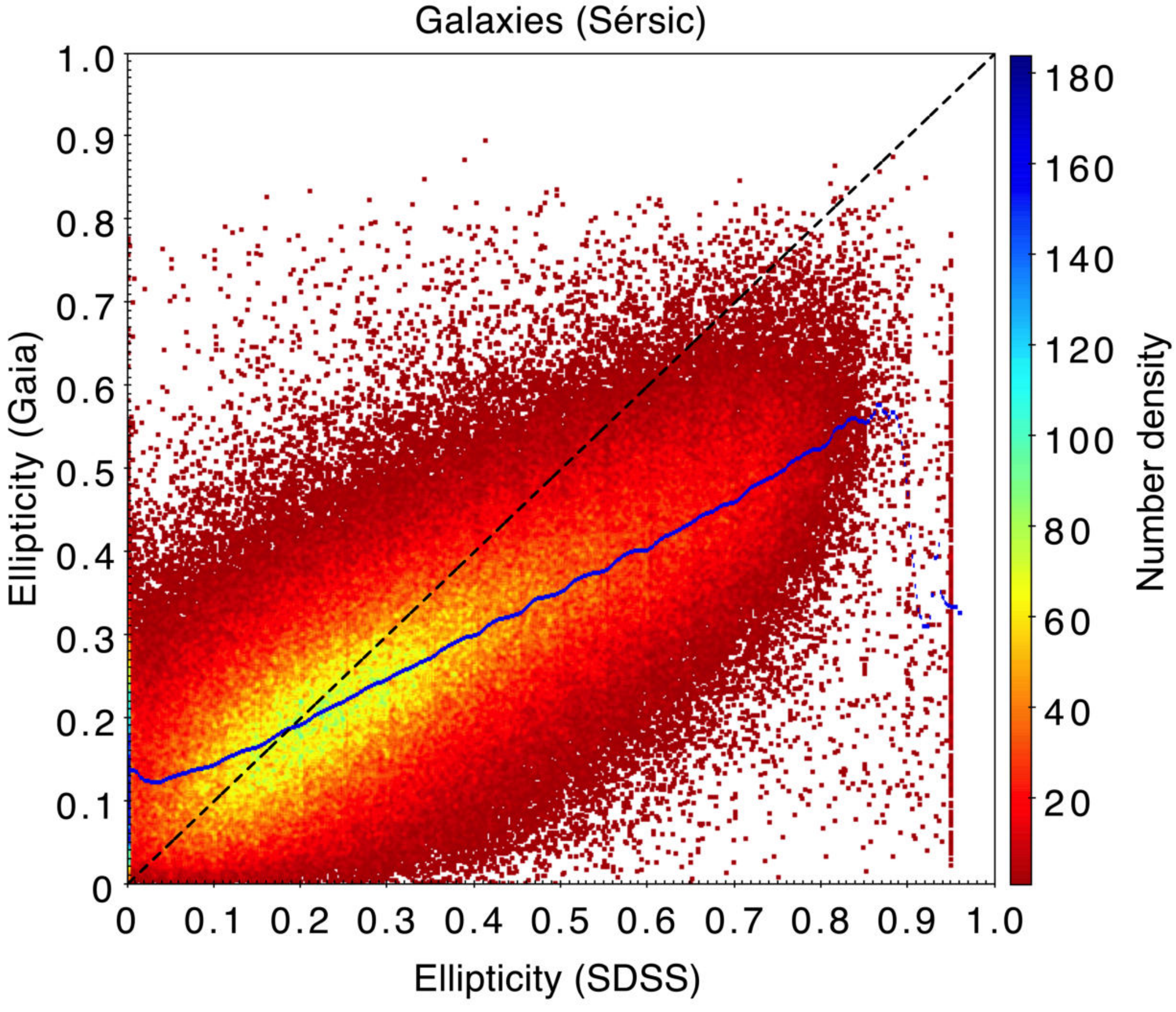}         \caption{Comparison of the fitted position angles and ellipticities of the S\'ersic profile of galaxies with SDSS DR16 de Vaucouleurs profile parameters. The blue line indicates the median of the distribution of ellipticities.}
    \label{gal_S_dr16}
\end{figure}

There are a few ground-based surveys that make use of a free S\'ersic profile \citep[such as GAMMA, NASATLAS or the work from][]{2011Simard}. The values of S\'ersic indices derived by our pipeline and those given in these surveys are not similar. One reason of this disagreement is the atmospheric seeing, which modifies the inner light profile of the sources \citep[see][for similar analysis]{2003Balcells, 2004Trujillo} leading to smaller \sersic indices when observed from the ground. Nonetheless, the shape parameters (ellipticity and position angle) of the galaxies derived from space and from the ground should be globally comparable. We present a comparison for the galaxies of the \gaia shape parameters of S\'ersic profile with the SDSS DR16 \citep{2020Ahumada} de Vaucouleurs profile (no S\'ersic profile has been adjusted by SDSS) in Figure \ref{gal_S_dr16}. 

The comparison of position angles is excellent; the sources that depart from the diagonal usually exhibit a small ellipticity, for which the position angle parameter is meaningless.

The comparison of ellipticities reveals a systematic trend: our pipeline tends to find galaxies to be rounder than SDSS. This trend is also observed as a function of the effective radius: the larger the galaxies, the larger the difference in ellipticity between Gaia and SDSS. \gaia tends to observe the inner parts of large objects. The fitted ellipticities are therefore influenced by the bulge shape, as ellipticities vary along the radius where they are measured \citep[][]{2004Ferrari}.

\begin{figure}
    \includegraphics[width=0.45\textwidth]{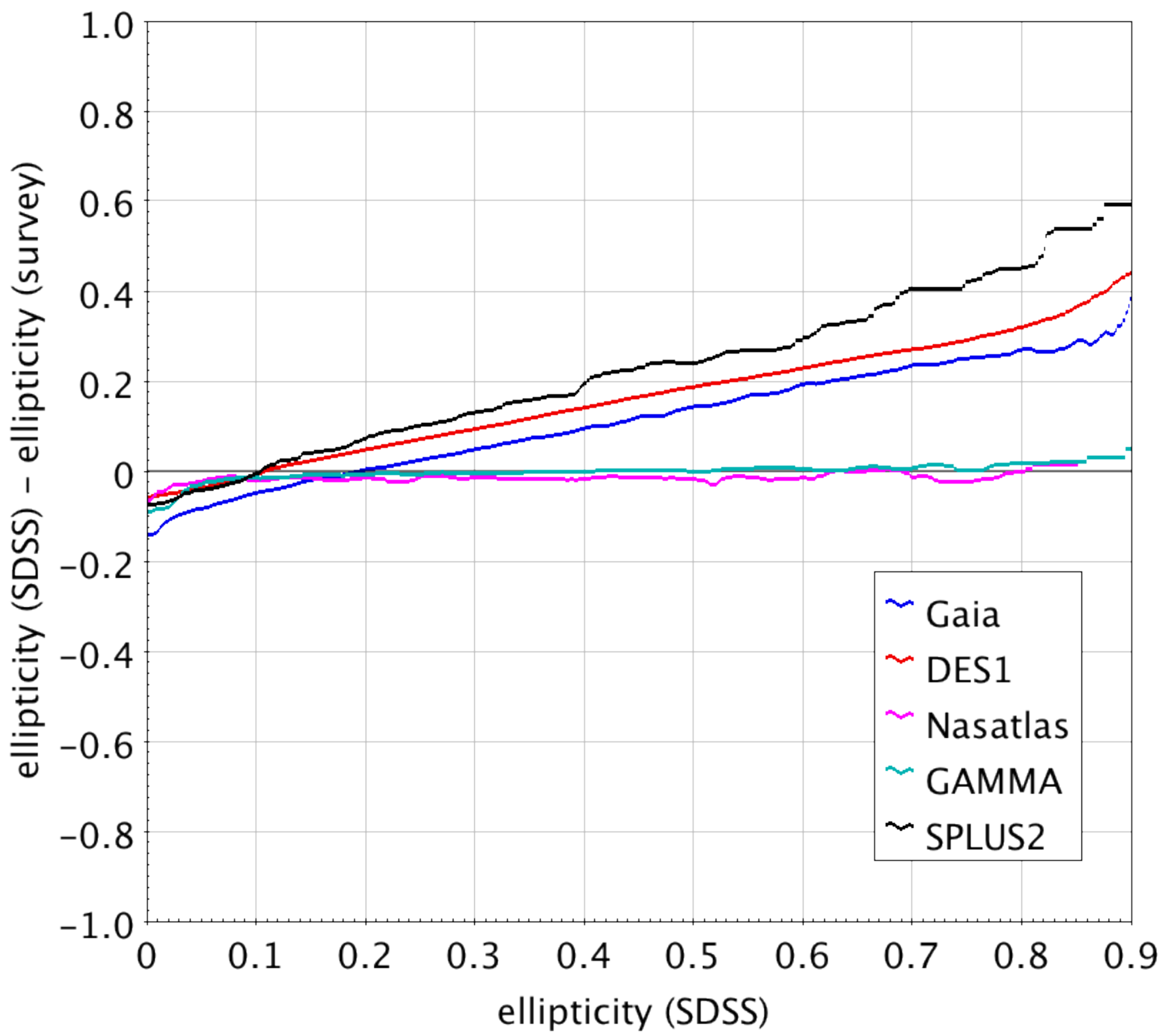}
    \caption{Comparison of SDSS DR16 ellipticities (de Vaucouleurs profile) with several surveys. Coloured lines indicate the median of the distributions.}
    \label{gal_de_dr16}
\end{figure}

To further investigate the systematic difference in ellipticity between our fit and the SDSS DR16, we confronted the SDSS values with several other surveys (Figure~\ref{gal_de_dr16}). There is a systematic shift between SDSS and DES1 \citep{2018Tarsitano} or SPLUS \citep{2019MendesdeOliveira} surveys, and good agreement is observed when comparing SDSS with GAMMA and NASATLAS surveys. It is therefore very difficult to conclude which survey is at the origin of the observed shifts. 

\subsubsection{de Vaucouleurs profile} 
\begin{figure*}
    \includegraphics[width=0.33\textwidth]{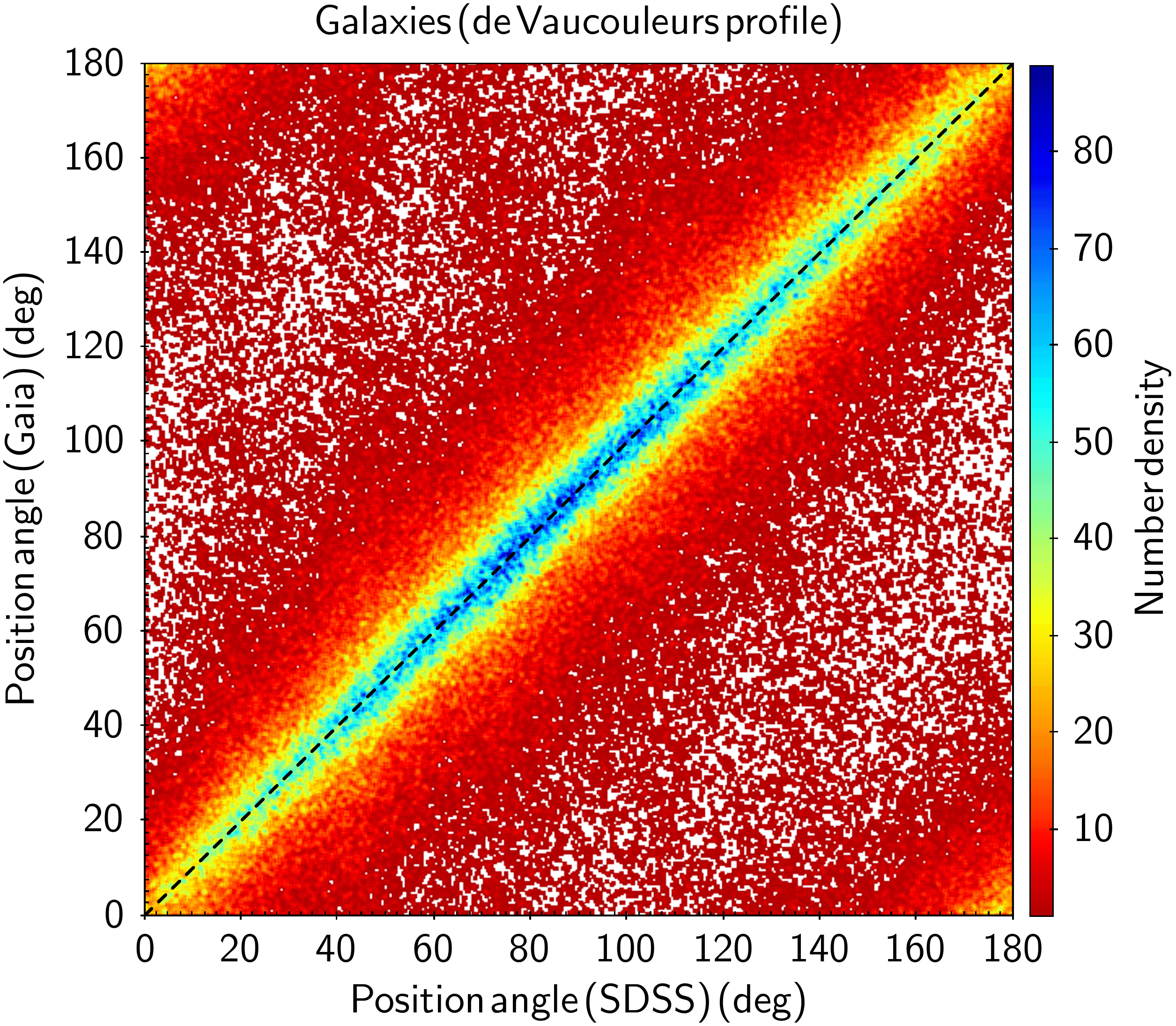} 
    \includegraphics[width=0.33\textwidth]{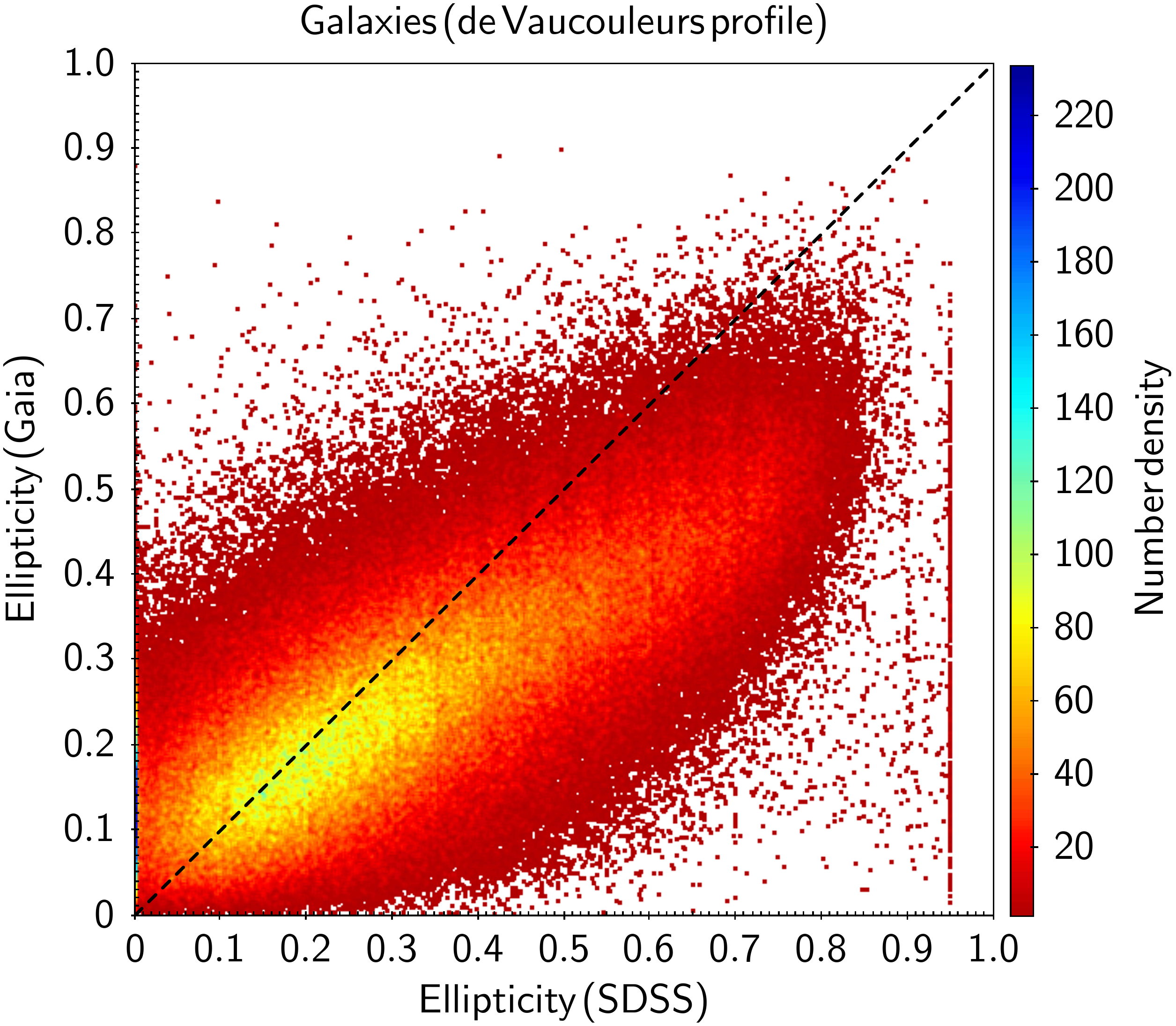} 
    \includegraphics[width=0.33\textwidth]{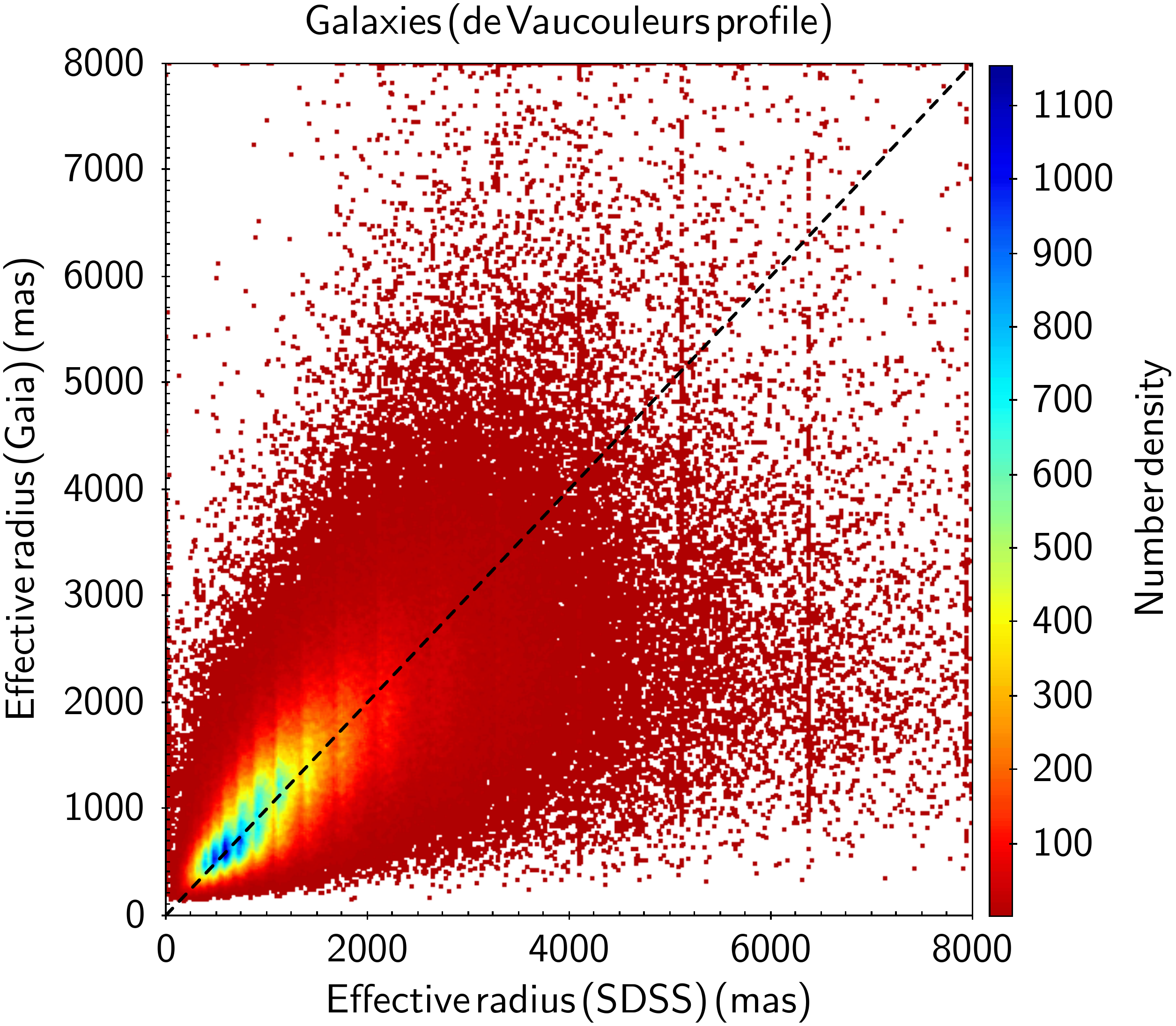} 
    \caption{Comparison for galaxies of the fitted parameters of the de Vaucouleurs profile with the SDSS DR16 de Vaucouleurs profile parameters.}
    \label{gal_S4_dr16}
\end{figure*}

To the best of our knowledge, there is also no space-based survey that adjusts a de Vaucouleurs profile on galaxies. The SDSS DR16 ground-based survey provides a de Vaucouleurs profile for $390\,615$ objects in common with the list of galaxies used in this study. The problem of the seeing modifying the shape of the light profile is still present and limits the comparison that can be done. We present the comparison of the parameters of the fit with the SDSS in Figure~\ref{gal_S4_dr16}. The accordance between both radii is reasonable, with no systematic effect except for the very large objects for which \gaia is in the extrapolation regime and tends to underestimate the radii. The position angles are in very good agreement with those provided by the S\'ersic model. The same systematic effect as seen with the S\'ersic profile affects the ellipticities of the de Vaucouleurs profile.

\subsubsection{Internal coherence: \sersic\,versus de Vaucouleurs}
\begin{figure*}
    \includegraphics[width=0.33\textwidth]{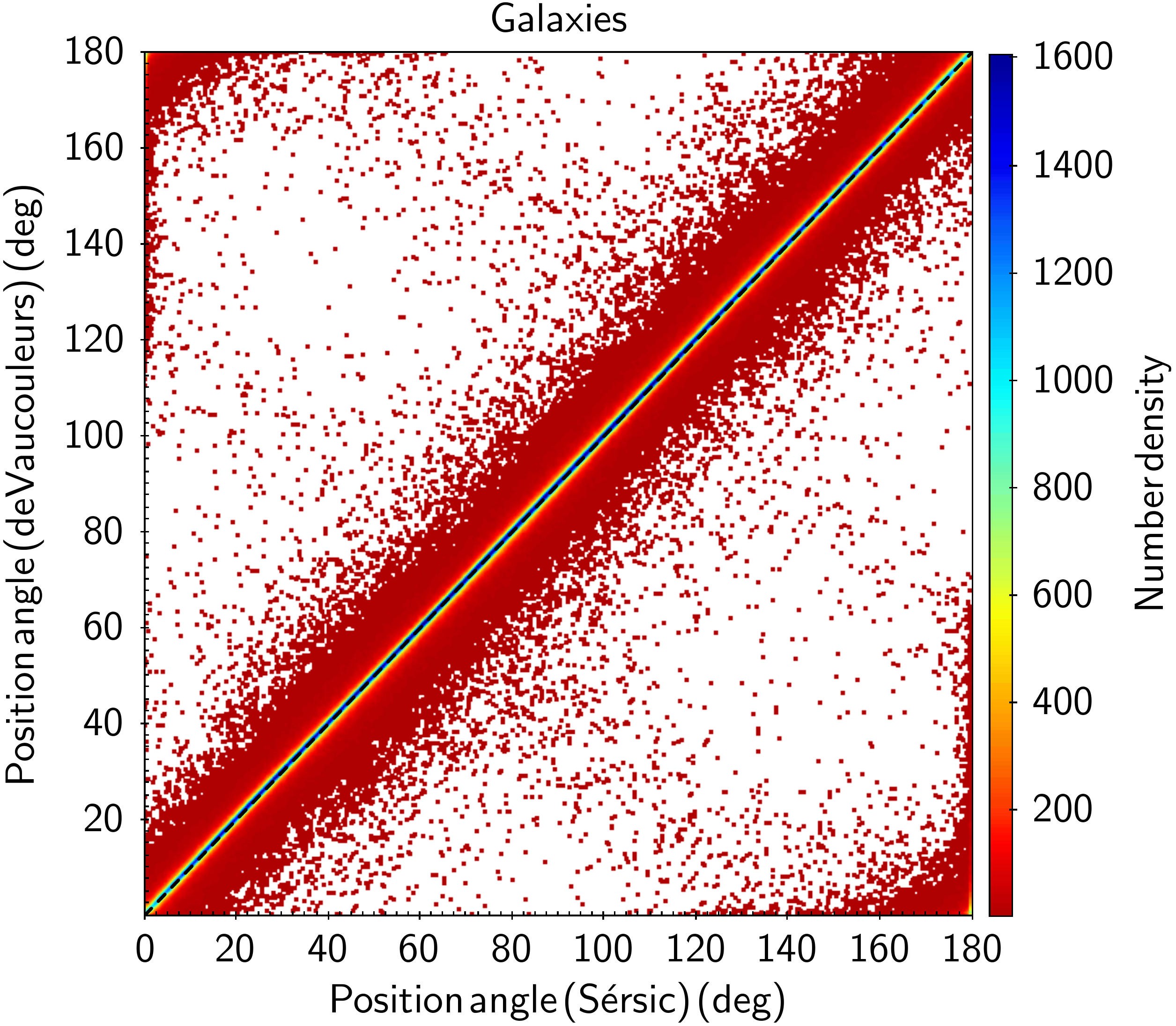}
    \includegraphics[width=0.33\textwidth]{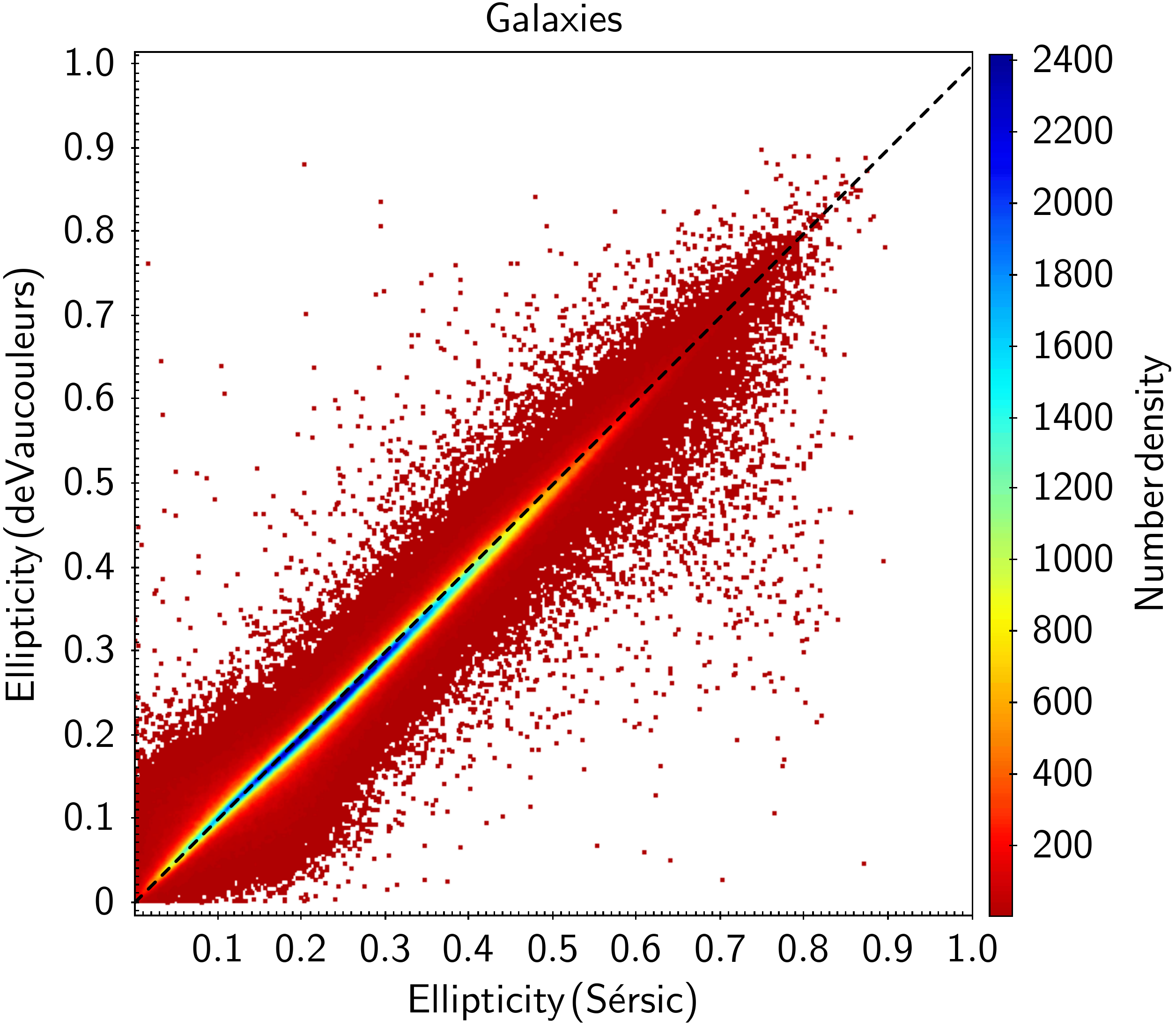} 
    \includegraphics[width=0.33\textwidth]{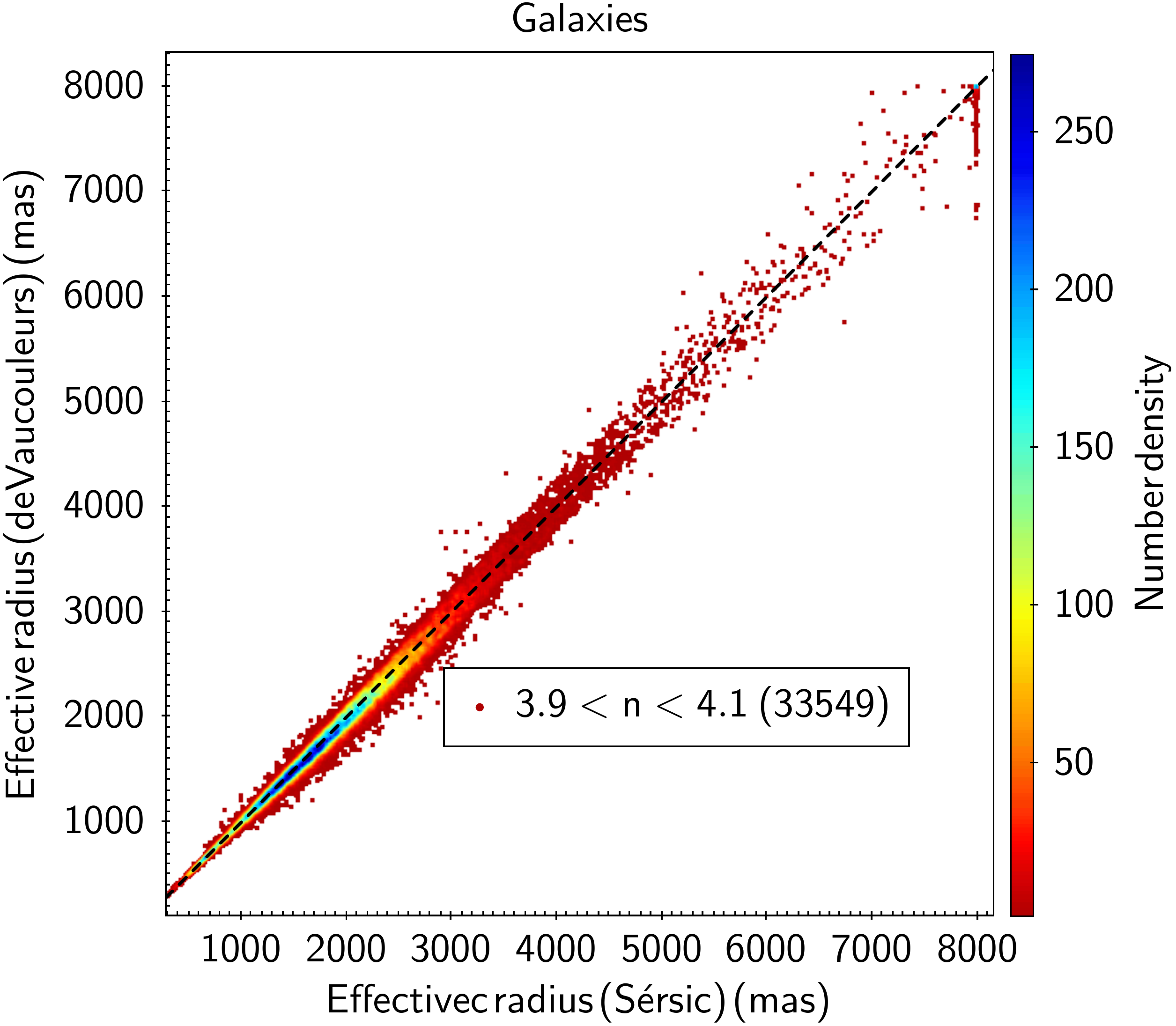}
    \caption{Internal comparison of the parameters obtained for galaxies with the S\'ersic profile and with the de Vaucouleurs profile. (Left and middle panels) Internal comparison of the position angles and ellipticities of galaxies obtained with the S\'ersic profile and with the de Vaucouleurs profile. (Right panel)  Comparison of effective radii from the S\'ersic profile and from the de Vaucouleurs profile, for a selection
of sources that have a S\'ersic index of close to 4, corresponding to typical
elliptical galaxies well represented by a de Vaucouleurs profile,}
    \label{gal_S_S4}
\end{figure*}

To some extent, the comparison of the two profiles fitted for the galaxies allows some internal assessment of the validity of the results.
We compare the position angles and ellipticities derived by the S\'ersic and the de Vaucouleurs models in Figure~\ref{gal_S_S4}. The comparison of effective radii is also presented for a selection of sources. To compare the effective radii given by the two models, we selected the sources with a S\'ersic index of close to 4 that have profiles similar to a de Vaucouleurs profile and for which the comparison of the two models is meaningful.

There is excellent agreement between the parameters fitted by the two models which indicates that the models are coherent. 

\subsubsection{Exponential profile}\label{exp}
The theoretical analyses from \citet{2014deSouza} and \citet{ 2015deBruijne} indicate that the on-board video processing unit would filter almost all disc galaxies that are typically well modelled by an exponential profile. This is confirmed by the results of our \sersic profile fitting (see Figure~\ref{gal_distrib_s}) which converges towards small values of the \sersic index ($< 1.5$), that is, those typical of discs, for only a few hundred objects in the entire analysed sample of galaxies.

Despite this, we also adjusted an exponential profile on all galaxies as $I(r) = I_{0} \exp(-r/r_{s})$. A comparison of the results with the SDSS DR16 exponential profile shows that very few objects are well described by our exponential profile. The fitted radii are systematically smaller than those given by the SDSS.
The size of the differences appears to correlate with the angular size of the objects: the larger the size, the larger the difference. These differences are mostly attributable to the following factors: (i) The real shape of most \gaia galaxies does not follow an exponential profile as the disc galaxies have been filtered out by the VPA algorithm; and (ii) \gaia observes the bulge of large galaxies that are better represented by a de Vaucouleurs profile. Our fitting tends to extrapolate the radius though it is systematically underestimated, while SDSS considers images convolved by the atmosphere and of the entire objects, thus accounting for a larger influence of the disc. 

In our simulations of bulge plus disc galaxies, the fitting of an exponential profile leads to radii following mostly the bulge characteristics.


\subsection{Known issues}
There are limitations to the efficiency of the CU4-\textit{Surface brightness profile} fitting applied to extragalactic sources. The first is the angular size of the field of view of \gaia. The algorithm extrapolates the solution for sources with an effective radius of larger than 2.5\arcsec, working with less than half of their total flux and the solutions are therefore less reliable. It is observed that, for these sources, the pipeline analyses the bulge properties instead of the galaxy as a whole. 

Most sources with a disc-like light distribution are filtered out by the detection algorithm of \gaia unless they encompass a bright compact bulge. For this reason, the resulting table of galaxies contains a majority of elliptical galaxies and almost no disc galaxies and can therefore not be used in its current state for statistical analysis of the local environment of the Milky Way. 

A systematic effect between the ellipticities determined by the pipeline and ground-based catalogues is observed. The origin of the problem may concern large sources for which the algorithm analyses the bulge ellipticity when the surveys measure the objects as a whole, but may also be related to the effect of atmosphere in the ground-based measurements.

The fitting of galaxies hosting quasars appears to be more difficult than fitting other galaxies due to the complexity of the combined model (quasar plus host galaxy). The filtering applied during the post processing is strict, removing all host galaxies with an effective radius of larger than 2.5\arcsec. Several modifications to the model are being tested in order to improve its robustness in view of \gaia DR4.
\section{Catalogue overview}\label{catalogue}
The pipeline delivers the surface brightness light profile parameters  of the sources analysed, including the shape parameters (position angle and ellipticity). These are stored accordingly in the \href{https://gea.esac.esa.int/archive/documentation/GDR3/Gaia_archive/chap_datamodel/sec_dm_extra--galactic_tables/ssec_dm_qso_candidates.html}{\textbf{qso\_candidates}} and \href{https://gea.esac.esa.int/archive/documentation/GDR3/Gaia_archive/chap_datamodel/sec_dm_extra--galactic_tables/ssec_dm_galaxy_candidates.html}{\textbf{galaxy\_candidates}} tables. A flag of quality and specificity of the fitting is given in both tables: \href{https://gea.esac.esa.int/archive/documentation/GDR3/Gaia_archive/chap_datamodel/sec_dm_extra--galactic_tables/ssec_dm_qso_candidates.html#qso_candidates-host_galaxy_flag}{\textbf{host\_galaxy\_flag}} for quasars and \href{https://gea.esac.esa.int/archive/documentation/GDR3/Gaia_archive/chap_datamodel/sec_dm_extra--galactic_tables/ssec_dm_galaxy_candidates.html#galaxy_candidates-flags_sersic}{\textbf{flags\_sersic}}, \href{https://gea.esac.esa.int/archive/documentation/GDR3/Gaia_archive/chap_datamodel/sec_dm_extra--galactic_tables/ssec_dm_galaxy_candidates.html#galaxy_candidates-flags_de_vaucouleurs}{\textbf{flags\_de\_vaucouleurs}} for the galaxies that complement the light profiles. The flag \href{https://gea.esac.esa.int/archive/documentation/GDR3/Gaia_archive/chap_datamodel/sec_dm_extra--galactic_tables/ssec_dm_qso_candidates.html#qso_candidates-host_galaxy_detected}{\textbf{host\_galaxy\_detected}} is also provided in the \href{https://gea.esac.esa.int/archive/documentation/GDR3/Gaia_archive/chap_datamodel/sec_dm_extra--galactic_tables/ssec_dm_qso_candidates.html}{\textbf{qso\_candidates}} table to indicate when the host galaxy around the central quasar is detected by \gaia. 

The sources analysed by our pipeline are a fraction of the sources provided in the tables because we analysed external lists set up by ourselves and not the sources classified as quasars or galaxies by the groups working on classification within DPAC. The complete content of these tables is presented in \cite{2022Bailer-Jones}. 

\subsection{Quasars}\label{product_qso}
 Here is the list of surface brightness profile parameters provided in the \href{https://gea.esac.esa.int/archive/documentation/GDR3/Gaia_archive/chap_datamodel/sec_dm_extra--galactic_tables/ssec_dm_qso_candidates.html#qso_candidates-host_galaxy_flag}{\textbf{qso\_candidates}} table. 
 
 \begin{itemize}
 \item  n\_transits: number of transits used for the fit.
 \item  intensity\_quasar: Intensity ($I_0$) of the quasar at the centre (e-/s) obtained with the exponential profile using a fixed scale length (39.4 mas).
 \item  intensity\_quasar\_error: Uncertainty on the intensity of the quasar  [e-/s].
 \item  intensity\_hostgalaxy: Intensity ($I_{re}$) of the host galaxy at the fitted effective radius [e-/s].
 \item  intensity\_hostgalaxy\_error: Uncertainty on intensity of the host galaxy [e-/s].
 \item  radius\_hostgalaxy: Effective radius  ($r_e$) of the S\'ersic profile containing half of the total luminosity of the host galaxy [mas]. 
 \item  radius\_hostgalaxy\_error: Uncertainty of the effective radius of the S\'ersic profile [mas].
 \item  sersic\_index: S\'ersic index (n) of the S\'ersic profile describing the host galaxy.
 \item  sersic\_index\_error: Uncertainty on the S\'ersic index  of the S\'ersic profile. 
 \item  ellipticity\_hostgalaxy: Ellipticity ($\epsilon$ defined as 1-($b/a$) where $b/a$ is the axis ratio) of the host galaxy fitted with the S\'ersic profile. This parameter is not provided for extremely faint objects for which its determination was not possible.
 \item  ellipticity\_hostgalaxy\_error: Uncertainty in the ellipticity of the host galaxy fitted with the S\'ersic profile.
 \item  posangle\_hostgalaxy: Position angle of the host galaxy for the fitted S\'ersic profile (from north to east) [\deg]. This parameter is not provided for extremely faint objects for which its determination was not possible.
 \item  posangle\_hostgalaxy\_error: Uncertainty in the position angle for the S\'ersic profile [\deg]. 
 \item  host\_galaxy\_detected: flag indicating the presence (TRUE) or absence (FALSE) of a host galaxy as detected by the pipeline. 
 \item  l2\_norm: L2 norm for the combined exponential and S\'ersic profiles. This value represents the mean squared error between the integrated flux of all observed samples (from the SM and AF) and the integrated flux of synthetic samples produced with the fitted profile.
 \item  morph\_params\_corr\_vec: Upper triangular part of the correlation matrix of the fitted profile parameters, as obtained by morphological fitting of an exponential profile combined with a S\'ersic profile.
 \item  host\_galaxy\_flag: This flag provides information about the processing or scientific quality of the results of the Galaxy morphology analysis chain for the de Vaucouleurs profile.
 \end{itemize}

Two flags, namely \href{https://gea.esac.esa.int/archive/documentation/GDR3/Gaia_archive/chap_datamodel/sec_dm_extra--galactic_tables/ssec_dm_qso_candidates.html#qso_candidates-host_galaxy_detected}{\textbf{host\_galaxy\_detected}} and \href{https://gea.esac.esa.int/archive/documentation/GDR3/Gaia_archive/chap_datamodel/sec_dm_extra--galactic_tables/ssec_dm_qso_candidates.html#qso_candidates-host_galaxy_flag}{\textbf{host\_galaxy\_flag,}} are given in the table to indicate (i) if a host galaxy is detected by \gaia (\href{https://gea.esac.esa.int/archive/documentation/GDR3/Gaia_archive/chap_datamodel/sec_dm_extra--galactic_tables/ssec_dm_qso_candidates.html#qso_candidates-host_galaxy_detected}{\textbf{host\_galaxy\_detected}}=`true') and (ii) to indicate the specificity of the profile fitted (\href{https://gea.esac.esa.int/archive/documentation/GDR3/Gaia_archive/chap_datamodel/sec_dm_extra--galactic_tables/ssec_dm_qso_candidates.html#qso_candidates-host_galaxy_flag}{\textbf{host\_galaxy\_flag}}). The values taken by the flags as well as their meaning are presented in Section~\ref{postprocqso}. Typical ADQL queries based on the combination of these flags are given in Appendix \ref{sec:adql_quasar}.

\subsection{Galaxies}\label{product_gal}
The list of surface brightness profiles parameters given in table \href{https://gea.esac.esa.int/archive/documentation/GDR3/Gaia_archive/chap_datamodel/sec_dm_extra--galactic_tables/ssec_dm_galaxy_candidates.html}{\textbf{galaxy\_candidates}} is the following:\\

 \begin{itemize}
 \item n\_transits: Number of transits used for the fit. 
 \item posangle\_sersic: Position angle (pa) for the fitted S\'ersic profile (from north to east) [\deg].
 \item posangle\_sersic\_error: Uncertainty in the position angle for the S\'ersic profile [\deg].
 \item intensity\_sersic: Intensity ($I_{re}$) at the fitted effective radius for the S\'ersic profile [e-/s].
 \item intensity\_sersic\_error: Uncertainty of intensity\_sersic for the S\'ersic profile [e-/s].
 \item radius\_sersic: Effective radius ($r_e$) containing half of the total luminosity as obtained by fitting a S\'ersic profile. [mas]     
 \item radius\_sersic\_error: Uncertainty in the effective radius for the S\'ersic profile [mas].
 \item ellipticity\_sersic: Ellipticity ($\epsilon$ defined as 1-($b/a$) where $b/a$ is the axis ratio) for the S\'ersic profile. This parameter is not provided for extremely faint objects for which its determination was not possible.
 \item ellipticity\_sersic\_error: Uncertainty in the ellipticity for the S\'ersic profile.
 \item l2\_sersic: L2 norm for the S\'ersic Profile. This value represents the mean squared error between the integrated flux of all observed samples (from the SM and AF) and the integrated flux of synthetic samples produced with the fitted profile.
 \item morph\_params\_corr\_vec\_sersic: Vector form of the correlation matrix of the fitted profile parameters, as obtained by morphological fitting for the S\'ersic profile.
 \item flags\_sersic: This flag provides information about the processing or scientific quality of the results for the S\'ersic Profile.
 \item n\_sersic: S\'ersic index of the S\'ersic profile.
 \item n\_sersic\_error: Uncertainty on the S\'ersic index of the S\'ersic profile.
 \item posangle\_de\_vaucouleurs: Position angle (pa) for the fitted de Vaucouleurs profile (from north to east) [\deg].
 \item posangle\_de\_vaucouleurs\_error: Uncertainty in the position angle [\deg].
 \item intensity\_de\_vaucouleurs: Intensity ($I_{re}$) at the fitted effective radius for the de Vaucouleurs profile [e-/s].
 \item intensity\_de\_vaucouleurs\_error: Uncertainty of the light intensity at the fitted effective radius for the de Vaucouleurs profile [e-/s].
 \item radius\_de\_vaucouleurs: Effective radius ($r_e$) containing half of the total luminosity of the source as obtained by fitting a de Vaucouleurs profile. [mas]  
 \item radius\_de\_vaucouleurs\_error: Uncertainty in the effective radius for the de Vaucouleurs profile [mas].
 \item ellipticity\_de\_vaucouleurs: Ellipticity ($\epsilon$ defined as 1-($b/a$) where $b/a$ is the axis ratio) for the de Vaucouleurs profile. This parameter is not provided for extremely faint objects for which its determination was not possible.
 \item ellipticity\_de\_vaucouleurs\_error: Uncertainty in the ellipticity for the de Vaucouleurs profile.
 \item l2\_de\_vaucouleurs: L2 norm for the de Vaucouleurs Profile. This value represents the mean squared error between the integrated flux of all observed samples (from the SM and AF) and the integrated flux of synthetic samples produced with the fitted profile.
 \item morph\_params\_corr\_vec\_de\_vaucouleurs: Vector form of the correlation matrix of the fitted profile parameters, as obtained by morphological fitting for the de Vaucouleurs profile.
 \item flags\_de\_vaucouleurs: This flag provides information about the processing or scientific quality of the results of the de Vaucouleurs profile.
 \end{itemize}

The flags \href{https://gea.esac.esa.int/archive/documentation/GDR3/Gaia_archive/chap_datamodel/sec_dm_extra--galactic_tables/ssec_dm_galaxy_candidates.html#galaxy_candidates-flags_sersic}{\textbf{flags\_sersic}} and \href{https://gea.esac.esa.int/archive/documentation/GDR3/Gaia_archive/chap_datamodel/sec_dm_extra--galactic_tables/ssec_dm_galaxy_candidates.html#galaxy_candidates-flags_de_vaucouleurs}{\textbf{flags\_de\_vaucouleurs}} are given in the table to indicate the specificity of each of the fitted profiles. The values taken by these flags and their significations are presented in Section~\ref{postprocgal}. Typical ADQL queries based on these flags are given in Appendix \ref{sec:adql_galaxy}.

\section{Conclusions}\label{conclusion}
We present the \gaia DPAC CU4-\textit{Surface brightness profile} pipeline, which we used to analyse the light profile of galaxies from the local Universe (z$<$0.45) and of quasars with their host galaxy.

The pre-defined lists of extragalactic sources that were analysed have been previously established. For quasars, several major catalogues of quasars and candidates were compiled. For galaxies, we used a machine learning analysis of \gaia DR2 combined with the WISE survey to identify extended sources. Both lists favour purity at the expense of completeness but are not completely free of contamination. Of these lists, we only retained the sources that have at least 25 \gaia observations that together cover at least 86\% of the surface area of the source.

The pipeline has processed the data collected during the first three years of operations of the satellite which represents $\sim$116 million transits. 

The surface brightness profile fitting consists in a global iterative strategy based on a direct model with a Bayesian exploration of the parameter space that tends to best reproduce the AF and SM observations of \gaia through simulations. The combination of parameters leading to the lowest difference between the observations and the simulations is then selected as the fitted profile. We analysed quasars with a two-component profiles: the central quasar described by an exponential and the host galaxy represented by a \sersic profile. All the galaxies were analysed with two separate profiles: a \sersic and a de Vaucouleurs profile.

A post-processing step was applied to the results of the pipeline in order to flag the sources according to the outcome of the fitting process. All host galaxies with an effective radius of larger than 2.5" were removed from the catalogue in order to avoid the extrapolation regime of the pipeline which relates to the limited size of the field of view of \gaia. 

The pipeline identified 64\,498 host galaxies around quasars whereas 1\,031\,607 quasars appear as point-like sources to our analysis. The distribution of the \sersic indices of the host galaxies indicates that most of them are disc-like galaxies. Regarding galaxies, 914\,837 were successfully analysed and two profiles are published (a \sersic and a de Vaucouleurs profiles). The distribution of their \sersic indices indicates that most are ellipticals and confirms that the \gaia detection system is filtering out most disc galaxies unless they host a bright bulge. Most of the sources analysed in this work are compact with effective radii of smaller than 2\arcsec, and have never been resolved from the ground. 

The results are released in the \gdr{3} associated tables \href{https://gea.esac.esa.int/archive/documentation/GDR3/Gaia_archive/chap_datamodel/sec_dm_extra--galactic_tables/ssec_dm_qso_candidates.html}{\textbf{qso\_candidates}}, 
\href{https://gea.esac.esa.int/archive/documentation/GDR3/Gaia_archive/chap_datamodel/sec_dm_extra--galactic_tables/ssec_dm_qso_catalogue_name.html}{\textbf{qso\_catalogue\_name}} 
and \href{https://gea.esac.esa.int/archive/documentation/GDR3/Gaia_archive/chap_datamodel/sec_dm_extra--galactic_tables/ssec_dm_galaxy_candidates.html}{\textbf{galaxy\_candidates}} 
and offer for the first time an all-sky, space-based catalogue of the morphology of galaxies and of galaxies hosting quasars in the visible wavelengths derived from exceptional data. 

For \gdr{4}, several improvements are currently foreseen for the treatment of extragalactic sources: for example, (i) a finer sampling of the simulated images, which will ease the convergence of the fitting, and (ii) a search for an offset between the central quasar and the host galaxy. Concerning the galaxies, a composite model (bulge plus disc) should be fitted in addition to the \sersic and de Vaucouleurs models in order to better represent the true profile of the sources and to avoid the windowing effect observed with the exponential model.

The number of transits over the sources will double for \gdr{4} compared to the data used for \gdr{3}. There will therefore be an improved
angular coverage for these  sources. As a consequence, less sources will be discarded during the filtering. We estimate that the lists of sources will be at least twice as large as the present ones. The input list will also improve, complemented by the findings of the DPAC classification together with newer published catalogues.

\begin{acknowledgements}
Over the ten years that it took to develop the modules of Surface Brightness Profile fitting, we benefited from the contributions of Ulrich Bastian, Jose Hernandez, Michael Davidson, Nigel Hambly and Jordi Portel, we wish to thank them for it. We also want to thank B. Frezouls, G. Walmsley, G. Prat and other CNES colleagues, for their commitment to improve the design of the chains.\\
We would like to warmly thank D. Despois for helpful discussions and suggestions on the validation of the solutions.\\ 
This work is part of the reduction of the \gaia satellite observations (https://www.cosmos.esa.int/gaia).
The \gaia space mission is operated by the European Space Agency, and the data are being processed by the Gaia Data
Processing and Analysis Consortium (DPAC, https://www.cosmos.esa.int/web/gaia/dpac/consortium)). The \gaia archive
website is https://archives.esac.esa.int/gaia. Funding for the DPAC is provided by national institutions, in particular the
institutions participating in the Gaia Multi Lateral Agreement (MLA).\\ 
We acknowledge the french “Centre National d’Etudes Spatiales” (CNES), the french national program PN-GRAM and Action Sp\'ecifique Gaia as well as Observatoire Aquitain des Sciences de l'Univers (OASU) for financial support along the years. We also acknowledge funding from the Brazilian Fapesp institution as well as from Brazilian-French cooperation institution CAPES/COFECUB. \\
Our work was eased by the use of the data handling and visualisation software TOPCAT \citep{2005Taylor}.
This research has made use of "Aladin sky atlas" developed at CDS, Strasbourg Observatory, France \citep{Aladin2014ASPC..485..277B, Aladin2000A&AS..143...33B}.
This research has made use of the VizieR catalogue access tool, CDS, Strasbourg, France.

\end{acknowledgements}

\bibliographystyle{aa}
\bibliography{cu4_bibliography}

\begin{appendix} 
\section{ADQL queries}
\subsection{Queries on the quasar catalogue}\label{sec:adql_quasar}
1- The following query returns the list of {\tt source\_id} of all the quasars having a host galaxy detected by \gaia. It selects 64\,498 sources.

{\small
\begin{verbatim}
SELECT source_id 
FROM gaiadr3.qso_candidates
WHERE host_galaxy_detected='true'
\end{verbatim}
}

2- The following query returns the list of {\tt source\_id} of the quasars that have a host galaxy detected by \gaia and a morphological profile at least partially measured. It selects 15\,867 sources. 

{\small
\begin{verbatim}
SELECT source_id 
FROM gaiadr3.qso_candidates
WHERE host_galaxy_detected='true' 
AND  (host_galaxy_flag = 1 
          OR host_galaxy_flag = 2
          OR host_galaxy_flag = 4)
\end{verbatim}
}

3- The following query returns the list of {\tt source\_id} of the quasars that have a host galaxy detected by \gaia with no morphological profile published. It selects 48\,631 sources. 

{\small
\begin{verbatim}
SELECT source_id 
FROM gaiadr3.qso_candidates
WHERE host_galaxy_detected='true' 
        AND  host_galaxy_flag = 5 
\end{verbatim}
}

4- The following query returns the list of {\tt source\_id} of all the quasars that have no host galaxy detected by \gaia. It selects 1\,031\,607 sources.

{\small
\begin{verbatim}
SELECT source_id 
FROM gaiadr3.qso_candidates
WHERE host_galaxy_detected='false'
AND  host_galaxy_flag < 6 
\end{verbatim}
}

5- The following query returns the list of {\tt source\_id} of all the quasars belonging to the ICRF2 catalogue, being processed by our pipeline and producing results.

{\small
\begin{verbatim}
SELECT source_id 
FROM gaiadr3.qso_catalogue_name
WHERE catalogue_id=2
\end{verbatim}
}

\subsection{Queries on the galaxy catalogue}\label{sec:adql_galaxy}
1- The following query returns the list of {\tt source\_id} of galaxies that have a measured effective radius of larger than 2\,000 mas regardless of the profile used. It selects 606\,331 sources.

{\small
\begin{verbatim}
SELECT source_id 
FROM gaiadr3.galaxy_candidates
WHERE  (radius_sersic > 2000 
        OR radius_de_vaucouleurs > 2000)
\end{verbatim}
}

2- The following query returns the list of {\tt source\_id} of galaxies that have a measured ellipticity larger than 0.7 regardless of the profile used. It selects 2\,255 sources.

{\small
\begin{verbatim}
SELECT source_id 
FROM gaiadr3.galaxy_candidates
WHERE  (ellipticity_sersic > 0.7 
                OR ellipticity_de_vaucouleurs > 0.7)
\end{verbatim}
}

3- The following query returns the list of {\tt source\_id} of galaxies with a \sersic elliptical profile well fitted. It selects 569\,382 sources.

{\small
\begin{verbatim}
SELECT source_id 
FROM gaiadr3.galaxy_candidates
WHERE flag_sersic =6 
\end{verbatim}
}

4- The following query returns the list of {\tt source\_id} of galaxies fainter than magnitude 18 and having a morphological profile measured. It selects 914\,837 sources.

{\small
\begin{verbatim}
SELECT source_id 
FROM gaiadr3.galaxy_candidates as galaxy
JOIN gaiadr3.gaia_source as gaia
USING (source_id)
WHERE  gaia.phot_g_mean_mag > 18 
AND galaxy.n_transits > 0
\end{verbatim}
}

\end{appendix}
\end{document}